\documentclass[entropy,article,accept,oneauthor,pdftex]{Definitions/mdpi}
\pdfoutput=1

\firstpage{1}
\pubvolume{xx}
\issuenum{1}
\articlenumber{5}
\pubyear{2020}
\copyrightyear{2020}
\history{Received: 14 February 2020; Accepted: 22 February 2020; Published: date}

\usepackage{epigraph}
\newcommand{\D}{{\mathrm{d}}}

\Title{Universal Gorban's Entropies: Geometric Case Study}

\Author{Evgeny M. Mirkes $^{1,2}$\orcidA{}}

\AuthorNames{Evgeny Mirkes}

\address{%
$^{1}$ \quad School of Mathematics and Actuarial Science, University of Leicester, Leicester  LE1 7HR, UK; em322@le.ac.uk\\
$^{2}$ \quad Laboratory of advanced methods for high-dimensional data analysis,  Lobachevsky State University, 603105, Nizhny Novgorod, Russia
}

\corres{Correspondence: em322@le.ac.uk}

\abstract{Recently, A.N. Gorban presented a rich family of universal Lyapunov functions for any linear or non-linear reaction network with detailed or complex balance. Two main elements of the construction algorithm are partial equilibria of reactions and convex envelopes of families of functions. These new functions aimed to resolve  ``the mystery'' about the difference between the rich family of Lyapunov functions ($f$-divergences) for linear kinetics and a limited collection of Lyapunov functions for non-linear networks in thermodynamic conditions. The lack of examples did not allow to evaluate the difference between Gorban's entropies and the classical Boltzmann--Gibbs--Shannon entropy despite obvious difference in their construction. In this paper, Gorban's results are briefly reviewed, and these functions are analysed and compared for several mechanisms of chemical reactions. The level sets and dynamics along the kinetic trajectories are analysed. The most pronounced difference between the new and classical thermodynamic Lyapunov functions was found far from the partial equilibria, whereas when some fast elementary reactions became close to equilibrium then this difference decreased and vanished in partial equilibria.}


\DeclareMathOperator*{\argmin}{argmin}

\begin{document}


\section{Introduction}

\subsection{Classical Entropic Lyapunov Functions for General Kinetics}

The classical example of the Lyapunov functional in kinetics was provided by Boltzmann in 1872~\cite{Boltzmann1872} (twenty years before the famous Lyapunov thesis):
\begin{equation}\label{BoltzEntr}
H(f)=\int f(x,v)\ln(f(x,v)) \, d^3 v \, d^3x
\end{equation}
where $f(x,v)$ is an one-particle distribution function in space ($x$) and velocity ($v$).

The analogue of this functional for chemical reaction was known already for Gibbs \cite{Gibbs1879}:
\begin{equation}\label{LyapFreeEN}
H=\sum_{i=1}^n c_i \left(\ln\left(\frac{c_i}{c_i^{\rm eq}}\right)-1 \right)
\end{equation}
where $c_i\geq 0$ is the concentration of the $i$th component $A_i$ and $c_i^{\rm eq}>0$ is an equilibrium concentration of $A_i$ (under the standard convention that $x\ln x=0$ for $x=0$). This is the thermodynamic potential for systems under constant temperature and volume (up to a constant factor).

In 1938, Zeldovich \cite{Zeld} used convexity of function (\ref{LyapFreeEN}) and  logarithmic singularity of its derivatives at zeros  for his proof of uniqueness of positive chemical equilibrium for given values of linear balances. In the 1960s, this approach was applied for many systems under different conditions and became standard \cite{ShapiroShapley1965}.

For systems with detailed balance, the time derivative of $H$ is the  sum (or integral, for continua of elementary processes) of the terms:
\begin{equation}\label{ElementEntProd}
-(w^+-w^-)\ln\left(\frac{w^+}{w^-}\right)\leq 0
\end{equation}
where $w^+$ and $w^-$ are the rates of the direct and reverse elementary process, respectively, and the term  $(w^+-w^-)\ln(w^+/w^-)\geq 0$ is   the entropy production in an elementary process.

Boltzmann used principle of detailed balance in the proof of his $H$-theorem in 1872, but in 1887 he invented a remarkable generalization of his theorem (after criticisms by Lorentz) \cite{Boltzmann1887}. 
His new sufficient condition for $H$-theorem, the cyclic balance or the semidetailed balance, was several times rediscovered later on. In chemical kinetics, it is called `the complex balance' \cite{HornJackson1972}. For linear kinetics, the generalisation  from detailed balance to complex balance is equivalent to the generalisation from the reversible Markov chains to general Markov chains (with positive equilibrium). For non-linear kinetics this condition seems to be more restrictive (it will be discussed below in more detail).

Shannon proved an analogue of the $H$-theorem for general random manipulation with information (for Markov chains, essentially). This is the information processing lemma  \cite{Shannon1948}.

The classical Lyapunov functions (\ref{BoltzEntr}), (\ref{LyapFreeEN}) have an important property, {universality}: they do not depend directly on the  collision and reaction mechanisms and kinetic constants but on the equilibrium distributions (concentration and the detailed or complex balance condition in general form \cite{Gorban2019}). This universality can be considered as a manifestation of the universality of thermodynamics that does not depend on the microscopic details directly.

\subsection{General Lyapunov Functions for Linear Kinetics}

In 1960, an extremely rich family of Lyapunov functions was discovered for general Markov chains. R\'{e}nyi \cite{Renyi1961} proved that the following functions  ($f$-divergences) are the  Lyapunov functions for general linear kinetics (Markov chains) with positive equilibrium $c_i^{\rm eq}$:
\begin{equation}\label{F-div}
H_f(c|c^{\rm eq})=\sum_i c_i^{\rm eq}f\left(\frac{c_i}{c_i^{\rm eq}}\right)
\end{equation}
where $f$ is an arbitrary convex function on the positive semi-axis.

Moreover, $H_f$ are not just Lyapunov functions but divergences:
$$H_f(c^1(t)|c^2(t))$$
is monotonically non-increasing function of time $t$ for any two kinetic curves $c^1(t)$ and $c^2(t)$ with the same value of $\sum_i c_i$.

This discovery attracted less immediate attention than the  R\'{e}nyi entropy
$$H_{\alpha }(P)=\frac{1}{1-\alpha } \ln \left(\sum_{i=1}^n p_i^{\alpha }\right)$$
proposed in the same paper. (Here, $P$ here is a vector of probability distribution with coordinates $p_i$.) Nevertheless, a bit later the theory of $f$-divergences was recognised as an important instrument of information theory and kinetics \cite{Csiszar1963,Morimoto1963}. In 2003, P.A. Gorban proved in that all universal Lyapunov functions for Markov kinetics can be produced by monotonic transformations of $f$-divergences \cite{ENTR3}. In 2009, Amari \cite{Amari2009} got a similar result.

\subsection{Conditionally Universal Lyapunov Functions for General Kinetics}

The $f$-divergences (\ref{F-div}) are universal Lyapunov functions as they do not depend on kinetic constants directly but on the equilibrium only. Nevertheless, their universality is weaker than the universality of the classical thermodynamic Lyapunov functions like \ref{LyapFreeEN}  because the classical thermodynamic potentials change monotonically in time for any reaction mechanism, linear or non-linear, under conditions of detailed or complex balance, whereas $f$-divergences are defined for linear kinetics only: for the sets of elementary processes like $A_i \rightleftharpoons A_j$, where $A_i$ are the components (or states). If a function changes monotonically in time  for a given reaction mechanism under conditions of detailed or complex balance, then we call it a {conditionally universal}  Lyapunov function for this reaction mechanism  \cite{Gorban2019}.

For linear reaction mechanisms a rich family of conditional Lyapunov functions (\ref{F-div}) is known since 1960 \cite{Renyi1961}. Nevertheless, there were no general constructions of conditionally universal Lyapunov functions for non-linear reaction mechanisms till the series of works \cite{Gorban2012arXiv,Gorban2014,Gorban2019}, where new conditionally universal Lyapunov functions were constructed for an arbitrary reaction mechanism under detailed or complex balance condition. These functions differ from the classical thermodynamic potentials, by construction. Nevertheless, it could be important to analyse how different they are. For this purpose, in this paper we compare the level sets of these functions and their changes over time for several typical chemical reaction examples.

\subsection{Structure of the Paper}

The basic notions and generalised mass action law equations are systematically introduced in Section~\ref{sec:Kinur}. The time derivative of the thermodynamic Lyapunov functions is calculated explicitly for systems with detailed balance. The  conditionally universal Lyapunov functions are characterised in this section implicitly, through their geometric properties. An extension of the general results to systems with complex balance is given in Section~\ref{SectionComplex}.  The explicit construction and algorithm for calculation of Gorban's Lyapunov functions are described in Section~\ref{GorLya}. Section~\ref{SectionCase} is devoted to the case studies and comparative analysis of the level sets and dynamics of classical and Gorban's Lyapunov functions for several reaction kinetic systems.  The results and outlooks are summarized in Conclusion.

\section{Kinetic Equations and General $H$-theorem \label{sec:Kinur}}

\subsection{Generalised Mass Action Law}

The construction of the Generalised Mass Action Law (GMAL) kinetic equations uses several basic elements:
\begin{itemize}
\item The list of components that is a finite set of symbols $A_1,\ldots , A_m$;
\item For each $A_i$ a non-negative variable $N_i$ (`the amount of $A_i$') is defined; the vector $N$ with coordinates $N_i$ is `the composition vector';
\item The list of elementary reactions (the reaction mechanism) that is  a finite set of the { stoichiometric equations}
\begin{equation}
\label{stoichiometricequation}
\sum_i\alpha_{\rho i}A_i \to \sum_i \beta_{\rho i} A_i \,,
\end{equation}
where $\rho =1, \ldots, m$ is the reaction number and the { stoichiometric
coefficients} $\alpha_{\rho i}$, $\beta_{\rho i}$ are nonnegative real numbers;
\item A dimensionless free entropy $S(N)$ that is a concave function in $\mathbb{R}_{\geq 0}^n$.
\end{itemize}

We use the following notations:
$\alpha_{\rho }$, $\beta_{\rho}$ are
the vectors with coordinates $\alpha_{\rho i}$, $\beta_{\rho i}$, respectively, $\gamma_{\rho}=\beta_{\rho }-\alpha_{\rho}$ is the stoichiometric vector of the reaction (\ref{stoichiometricequation}) (the `gain minus loss' vector).

The definition of the dimensionless free entropy function for a physico-chemical system depends on the conditions. For isolated systems it is just the thermodynamic entropy divided by the gas constant $R$. For isothermal isochoric  conditions $S=-F/(RT)$, where $F$ is the Helmholtz free energy, $T$ is the temperature, for isothermal isobaric conditions $S=-G/(RT)$, where $G$ is the Gibbs energy (free enthalpy), etc. \cite{G1,Yab1991,Hangos2010}. Introduction in the theory of thermodynamic potentials including free entropies (Massieu--Plank functions) is given by Callen \cite{Callen1985}.

For the general GMAL construction, $S$ is just a concave function. For the sake of generality, the value $S=-\infty$ is also allowed. The function $H=-S$ is assumed to be a closed convex function, this means that the its sublevel set $ \{ N \in \mathbb{R}_{\geq 0}^n | H(N) \leq a \} $
is a closed set for any real $a$. It is also assumed that $H$ takes finite values on a convex domain $U \subset \mathbb{R}_{\geq 0}^n $ with non-empty interior. $H$ is twice differentiable almost everywhere in $U$ (A.D. Alexandrov theorem \cite{Alexandrov1939,VodopGoldshtResh1979}). Following Boltzmann's tradition, we will use further the $H$-function $H=-S$.

A non-negative quantity, the reaction rate $r_{\rho}$ is defined by GMAL almost everywhere in $U$ for every elementary reaction (\ref{stoichiometricequation}) \cite{GorbanShahzad2011,Gorban2014} (compare to the thermodynamic GMAL presentations of reaction rates in earlier works \cite{Feinberg1972,G1,Grmela1993,GiovangigliMatus2012}):
\begin{equation}\label{GMAL}
r_{\rho}=\varphi_{\rho} \exp\left(\sum_{i=1}^n \alpha_{\rho i}\frac{\partial H(N)}{\partial N_i}\right)\, ,
\end{equation}
where the {\em kinetic factor} $\varphi_{\rho}\geq 0$ is an intensive quantity.

Here and below, all the equalities and inequalities with gradients of $H$ are considered `almost everywhere' in $U$ if the convex function $H$ is not everywhere continuously differentiable.

For the perfect isothermal isochoric mixtures $H$-function has the form
\begin{equation}\label{MALLyapFreeEN}
H=\sum_{i=1}^n N_i \left(\ln\left(\frac{c_i}{c_i^{\rm eq}}\right)-1 \right)\, ,
\end{equation}
where $c_i=N_i/V$ and $c_i^*=const$.

For such systems, GMAL (\ref{GMAL}) becomes the standard mass action law:
\begin{equation}\label{MAL}
r_{\rho}=\varphi_{\rho} \prod_{i=1}^n \left( \frac{c_i}{c_i^{\rm eq}}\right)^{\alpha_{\rho i}}\, .
\end{equation}

The corresponding GMAL kinetic equation is
\begin{equation}\label{KinUrChem}
\frac{\D N}{\D t}=V \sum_{\rho=1}^m r_{\rho} \gamma_{\rho}\, ,
\end{equation}
where $V>0$ is a positive extensive variable (volume). It can also change with time and its dynamic is defined by the equation of state and by  the conditions of the process.

The structure of kinetic equations (\ref{KinUrChem}) and GMAL formula for reaction rates (\ref{GMAL}) allow the elegant expression for $\D H/ \D t$. Let an auxiliary function of real variable  $\theta(\lambda)$  be given by the following expression for a given composition vector $N$  \cite{G1,GorbanShahzad2011,OrlovRozonoer1984}:
\begin{equation}\label{auxtheta}
\theta(\lambda)=\sum_{\rho}\varphi_{\rho}\exp\left[\sum_{i=1}^n (\lambda\alpha_{\rho i}+ (1-\lambda)\beta_{\rho i})\frac{\partial H(N)}{\partial N_i}\right]
\end{equation}
Function $\theta(\lambda)$ is convex. With this function, $\D {H}/ \D t$  has a very simple form:
\begin{equation}\label{EntropProdtheta}
\frac{\D H}{\D t}=-V\left.\frac{\D \theta(\lambda)}{\D
\lambda}\right|_{\lambda=1} \, .
\end{equation}
Convexity of $\theta(\lambda)$ implies the following sufficient condition of non-positivity $\D H/ \D t$.
\begin{Proposition}\label{Prop1}
If $\theta(1) \geq \theta(0)$ then $\D H/ \D t \leq 0$.
\end{Proposition}

General kinetic Equations (\ref{KinUrChem}) with GMAL reaction rate (\ref{GMAL}) can describe arbitrarily complex dynamics and approximate any dynamical system in $U$ even for perfect mixtures and constant kinetic factors \cite{Ocherki1986}. The specific thermodynamic properties of kinetic equations are based on special relations between kinetic factors  $\varphi_{\rho}$ that are detailed balance and complex balance.

\subsection{Detailed Balance}

The principle of detailed balance is a special symmetry between direct and reverse elementary reactions caused by the so-called microreversibilty (invariance of the equations of microscopic dynamics with respect to time reversal). In the GMAL formalism, the principle of detailed balance has a simple form: kinetic factors of direct and reverse elementary reaction coincide. In such situations, it is convenient to rearrange the list of elementary reactions (\ref{stoichiometricequation}), join the reactions with their reverse reactions in a shorter list of pairs of reactions:
\begin{equation}\label{reversibleMechanism}
\sum_i\alpha_{\rho i}A_i \rightleftharpoons \sum_i \beta_{\rho i} A_i\, .
\end{equation}

If the reverse reaction does not exist in the original reaction mechanism (\ref{stoichiometricequation}) then we can, nevertheless, add the reverse reaction formally, with zero kinetic factor. For the reaction mechanism in the reversible form (\ref{reversibleMechanism}), we use the superscripts $+$ and $-$ for the reaction rates and kinetic factors of the direct and reverse reactions, respectively:
\begin{equation}\label{GMALrev}
\begin{split}
&r^+_{\rho}=\varphi_{\rho}^+ \exp\left(\sum_{i=1}^n \alpha_{\rho i}\frac{\partial H(N)}{\partial N_i}\right)\, ;  \\
&r^-_{\rho}=\varphi_{\rho}^- \exp\left(\sum_{i=1}^n \beta_{\rho i}\frac{\partial H(N)}{\partial N_i}\right)\, .
\end{split}
\end{equation}

The rate $r_{\rho}$ is defined as the difference $r_{\rho}=r_{\rho}^+-r_{\rho}^-$ and the kinetic equations have the same form (\ref{KinUrChem}). The detailed balance condition is:
\begin{equation}\label{detailed balance}
\varphi_{\rho}^+=\varphi_{\rho}^- .
\end{equation}

Under this condition, a symmetry relation holds: $\theta(\lambda)=\theta(1-\lambda)$. Therefore,  $\theta(1) = \theta(0)$ for every composition vector $N$ and according to Proposition~\ref{Prop1}, $\D H/ \D t \leq 0$. Direct calculation of $\D H/ \D t$ by virtue of the system of kinetic equations under the detailed balance condition gives  the classical result (compare to (\ref{ElementEntProd})):
\begin{equation}\label{EntropyProd}
\frac{\D H}{\D t}=
-V \sum_{\rho} (\ln r_{\rho}^+ - \ln r_{\rho}^-) (r_{\rho}^+ -r_{\rho}^-) \leq 0 \, .
\end{equation}

Because of this property, $H(N)$ is called the thermodynamic Lyapunov function.

The detailed analysis of entropy production in nonequilibrium systems was provided recently in~\cite{Grmela2019}. Grmela considered  the equilibrium and nonequilibrium thermodynamics as representations of the Dynamical Maximum Entropy Principle \cite{Grmela2013}.

\subsection{Conditionally Universal Lyapunov Functions and their Geometric Characterisation}

In this Subsection, we consider systems of kinetic equations (\ref{KinUrChem})
with the given thermodynamic Lyapunov function $H$, reaction rates presented by
GMAL (\ref{GMAL}), and detailed balance (\ref{detailed balance}) for a given reaction mechanism (\ref{reversibleMechanism}). According to inequality (\ref{EntropyProd}), $H$ is a Lyapunov function for such a system for any reaction mechanism. This means that  $H$ is a {\em universal Lyapunov function} for chemical kinetics. If the reaction mechanism is fixed then the {\em conditionally universal} Lyapunov functions are introduced.

\begin{Definition}[ \cite{Gorban2014}]
A convex function $F(N)$ in $U$ is a conditionally universal Lyapunov function for kinetic equations (\ref{KinUrChem}),  given $H$ and  reaction mechanism (\ref{reversibleMechanism}) if $$\frac{\D F}{\D t}\leq 0$$ for any values of kinetic factors, which satisfy the detailed balance conditions (\ref{detailed balance}).
\end{Definition}

For each elementary reaction $\sum_i\alpha_{\rho i}A_i \rightleftharpoons \sum_i
\beta_{\rho i} A_i$ from the reaction mechanism given by the stoichiometric Equations
(\ref{reversibleMechanism}) and any $X\in U$ we define an interval of a straight line
\begin{equation}
I_{X, \rho}= \{X+\lambda \gamma_{\rho} \, |
\, \lambda \in \mathbb{R}\} \cap U.
\end{equation}

\begin{Definition}[Partial equilibria criterion for GMAL]\label{def:partEquilGMAL}A convex function $F(N)$ on $U$ satisfies the partial
equilibria criterion with a given thermodynamic Lyapunov function $H$ and reversible
reaction mechanism given by stoichiometric Equations (\ref{reversibleMechanism}) if
\begin{equation}
\underset{{N\in I_{X, \rho}}}{\operatorname{argmin}} H(N) \subseteq \underset{{N\in I_{X, \rho}}}{\operatorname{argmin}} F(N)
\end{equation}
for all $X\in U$, $\rho=1,\ldots ,m$.
\end{Definition}

\begin{Theorem}\label{theorem:DetBalGenHthGMAL}[General $H$-theorem]A convex function $F(N)$ on $U$ is a conditionally universal Lyapunov function for kinetic Equations (\ref{KinUrChem}),  given $H$ and  reaction mechanism (\ref{reversibleMechanism}) if it satisfies the
partial equilibria criterion (Definition~\ref{def:partEquilGMAL}).
\end{Theorem}

\subsection{Complex Balance}

Let us return to the general form of the reaction mechanism without coupling direct and reverse reactions (\ref{stoichiometricequation}). The complex balance condition means that $\theta(1)\equiv \theta(0)$ for all
values of the gradient vectors from $\mathbb{R}^n$. More formally, it means that
\begin{equation}\label{thetaComplex}
\sum_{\rho}\varphi_{\rho}\exp\left[\sum_{i=1}^n \alpha_{\rho i} \mu_i\right] \equiv
\sum_{\rho}\varphi_{\rho}\exp\left[\sum_{i=1}^n \beta_{\rho i} \mu_i \right]
\end{equation}
for all vectors $\mu\in  \mathbb{R}^n$ with coordinates $\mu_i$. Functions $\exp(y,\mu)$ of vector $\mu \in  \mathbb{R}^n$ are linear independent for any finite set of $y \in  \mathbb{R}^n$. Therefore, the identity (\ref{thetaComplex}) can be split in the several linear conditions on the coefficients $\varphi_{\rho}$.

Assume that there are $q$ different  vectors $y_1, \ldots, y_q$ among $\{\alpha_{\rho},\beta_{\rho}\}$ ($\rho=1, \ldots ,m$). The identity~(\ref{thetaComplex}) is equivalent to $q$ conditions:
\begin{equation}\label{complexbalanceGENKIN}
\sum_{\rho, \, \alpha_{\rho}=y_j} \varphi_{\rho}= \sum_{\rho, \, \beta_{\rho}=y_j}\varphi_{\rho}\;\; (j=1, \ldots , q).
\end{equation}

Formal sums $\sum y_i A_i$ from stoichiometric equations are called complexes, so  conditions (\ref{complexbalanceGENKIN}) are called the complex balance conditions \cite{HornJackson1972}. In physics, the terms cyclic balance conditions or semidetailed balance conditions are also used. These conditions were derived from the Markov processes of microkinetics under two asymptotic assumptions: (i) the asymptotic intermediates are in fast equilibrium with the main components and (ii) the concentration of asymptotic intermediates is small (the Michaelis--Menten--Stueckelberg theorem \cite{GorbanShahzad2011}). If each complex $\sum y_i A_i$ is once and only once the left hand part of the stoichiometric equation from the reaction mechanism (\ref{stoichiometricequation}) and once the right hand part, for the reverse reaction equation, then the complex balance conditions literally coincide  with the detailed balance conditions.

\subsection{Cone Theorem and $H$-theorems for Complex Balancing Systems \label{SectionComplex}}

For analysis of conditionally universal Lyapunov functions, a notion of cone of possible velocities is useful \cite{G1,Gorban2014,Gorban2019}. This cone is defined for a cone of kinetic equations and a given composition vector $N$. It consists of all possible values of the velocity vector $\D N/\D t$ at this point for  equations from selected cone. For example, the systems with detailed balance for a given reaction mechanism and function $H$ form the convex cone in the space of vector fields on the composition space. The corresponding cone of possible velocities is
\begin{equation}
\mathbf{Q}_{\rm DB}(N)={\rm cone}\{\gamma_{\rho} {\rm sgn}(r_{\rho}(N)) | i=1, \ldots m\} ,
\end{equation}
where cone stands for the conical hull and the piecewise-constant functions ${\rm sgn}(r_{\rho}(N))$ do  not depend on (positive) values of kinetic factors  $\varphi_{\rho}$  under assumption of detailed balance. Indeed,
$${\rm sgn}(r_{\rho}(N))={\rm sgn}\left(\exp\left[\sum_{i=1}^n \alpha_{\rho i}\frac{\partial H(N)}{\partial N_i}\right]-\exp\left[\sum_{i=1}^n \beta_{\rho i}\frac{\partial H(N)}{\partial N_i}\right]\right)\, .$$

Consider the complex balance systems with a given reaction mechanism and function $H$. They are given by linear conditions (\ref{complexbalanceGENKIN}) and form a convex cone of vector fields in the composition space.  For a given composition vector $N$, the cone of all values of $\D N/ \D t$ is a convex cone in $\mathbb{R}^n$. We denote this cone by $\mathbf{Q}_{\rm CB}(N)$.

For a given function $H$, consider a reversible reaction mechanism (\ref{reversibleMechanism}) and kinetics with detailed balance. Calculate $\mathbf{Q}_{\rm DB}(N)$. Decouple the direct and reverse reactions, consider kinetics with complex balance (for the same reaction mechanism). Calculate $\mathbf{Q}_{\rm CB}(N)$. These cones coincide:
\begin{Theorem}[Cone Theorem \cite{Gorban2012arXiv,Gorban2014,Gorban2019}]\label{ConeTheorem} For the same set of elementary reactions,
 $$\mathbf{Q}_{\rm DB}(N)=\mathbf{Q}_{\rm CB}(N).$$
\end{Theorem}
This means that the possible directions of motion for the kinetic systems with detailed and for systems with complex balance at one point coincide. The difference between these two classes of systems appears if we consider several points or kinetic curves, not pointwise.

Time derivative of a function $F(N)$ by virtue of kinetic equations is computed pointwise, therefore, an obvious consequence of Theorem~\ref{ConeTheorem} is:
\begin{Corollary}
If a function $F(N)$ is a conditionally universal Lyapunov function for  systems with detailed balance, thermodynamic Lyapunov function $H$ and reaction mechanism (\ref{reversibleMechanism}) then it is a conditionally universal Lyapunov function for the systems with complex balance, the same $H$ and the list of elementary reactions.
\end{Corollary}

So, any construction of conditionally universal Lyapunov functions for systems with detailed balance can be easily generalised for systems with complex balance.

\section{Gorban's Lyapunov Functions $H_{\Gamma}$ \label{GorLya}}

Direct application of the general $H$-theorem (Theorem~\ref{theorem:DetBalGenHthGMAL}) gives the following construction of conditionally universal Lyapunov functions for GMAL kinetic equations with detailed and complex balance \cite{Gorban2012arXiv,Gorban2014,Gorban2019}. Consider a GMAL system with the reaction mechanism (\ref{reversibleMechanism}), the convex thermodynamic Lyapunov function $H$ and the detailed or complex balance. Let $\Gamma \subset \mathbb{R}^n$ be a finite set of non-zero vectors, which includes all the stoichiometric vectors $\gamma_{\rho}$. Assume, additionally, that the function $H$ is strictly convex on each non-empty interval $U \cap (N+\mathbb{R} \gamma)$ ($\gamma\in \Gamma$) and achieves its minimum on this interval in an internal point (this point of minimum is unique due to strict convexity of $H$ in direction $\gamma$). This property trivially holds for the $H$ function for perfect systems under isothermal isochoric conditions (\ref{MALLyapFreeEN}) as well as for perfect systems under all other classical conditions (for example, for isothermal isobaric systems or for isolated isochoric systems  if $\gamma$ has both positive and negative coordinates \cite{Yab1991}).

Two main operations in the construction of the conditionally universal Lapunov function $H_{\Gamma}(N)$ are \cite{Gorban2019}:
\begin{itemize}
\item For each  $\gamma \in \Gamma$ calculate
\begin{equation}\label{QEentropy}
H_{\gamma}(N)=\min_{N+\gamma x \in\mathbb{R}_{>0}^n}H(N+ \gamma  x).
\end{equation}
\item Find
\begin{equation}\label{GorbanFunction}
H_{\Gamma}(N)=\max_{\gamma \in \Gamma}H_{\gamma}(N).
\end{equation}
\end{itemize}

Thus, for calculation of $H_{\Gamma}(N)$ we have to solve several 1D convex minimization problems and select the maximum of these minima. These functions are indexed by finite set $\Gamma$. The construction of $H_{\Gamma}(N)$ does not depend on the length of the vectors $\gamma \in \Gamma$. Therefore, for theoretical purposes it makes sense to consider normalised vectors or, even better, the elements of the projective space (i.e., one-dimensional subspaces of $\mathbb{R}^n$). For calculations, such a normalisation is not necessary.

The quasi-equilibrium entropies (\ref{QEentropy}) and partial equilibria in direction $\gamma$
$$\underset{N+\gamma x \in\mathbb{R}_{>0}^n}{\operatorname{argmin}}H(N+ \gamma  x)$$
are standard and very old tools for description of fast equilibria and partial equilibrium approximations. For example, the classical work of Michaelis and Menten used assumption of fast equilibration of `compounds' with stable reagents \cite{Michaelis1913}. For detailed discussion of this approximation we refer to \cite{GorbanShahzad2011}, for application to thermodynamics of driven systems see \cite{Grmela1993},  more physical and chemical applications, from Boltzmann's equation to chemical kinetics, and general theory are presented in the book \cite{GorbanKarlin2005}.

Partial equilibria are used in the construction of Gorban's universal Lyapunov functions (\ref{GorbanFunction}) in a completely different way. They do not substitute the genuine kinetic trajectory as the partial equilibrium approximations, but rather follow the non-perturbed motion as the ensemble of its projections on the surfaces of partial equilibria (`partial equilibrium shadows'). For the calculation of Gorban's function, the closest shadow is selected. (It is the closest shadow in the entropic divergence~(\ref{GorbanFunction})). Which shadow is closest can be changed in the course of motion.

Such ensembles of quasi-equilibrium projections were used in 1979 \cite{Gorban1979} for construction of attainability regions for chemical kinetic equations with a given reaction mechanism (this problem is close to the problem of conditionally universal Lyapunov functions). Later on, this geometric approach was used in various applications \cite{G1,GorbKagan2006} and reappeared recently in the theory of toric differential inclusions of chemical kinetics \cite{Craciun2019}.

The `ensemble of equilibrium subsystems' has been intensively used during almost 40 years as an effective tool for mathematical analysis of complex catalytic reactions and has given rise to many useful methods reviewed in a recent book \cite{MarinYab2019}.

\section{Case Studies \label{SectionCase}}

``A picture is worth a thousand words.'' Examples are needed to  evaluate the difference between Gorban's entropies and the classical Boltzmann--Gibbs--Shannon entropy. The difference in their construction is obvious but we need to evaluate the difference between these functions values and between their changes in dynamics. In this section, we analyse the level sets of these functions and their  dynamic changes along kinetic trajectories. Several reaction mechanisms have been selected for~benchmarking:
\begin{itemize}
\item Linear isomerisation of three components (Section~\ref{SectioninIsom})

$$A_1\rightleftharpoons A_2 \rightleftharpoons A_3 \rightleftharpoons A_1;$$
\item Nonlinear isomerisation reaction (Section \ref{SectionlIsom})

$$A_1\rightleftharpoons A_2 \rightleftharpoons A_3,\;\; 2A_1\rightleftharpoons A_2+A_3;$$
\item Water Gas Shift (WGS) reaction (Section \ref{SectionWGS})

$$\mathrm{H}_2\mathrm{O}+\mathrm{red}\rightleftharpoons \mathrm{H}_2+\mathrm{Ox},\;\;
\mathrm{CO}+\mathrm{Ox}\rightleftharpoons \mathrm{CO}_2+\mathrm{red},$$
or in abstract notations

$$A_1+A_5\rightleftharpoons A_2+A_6,\;\;
A_3+A_6\rightleftharpoons A_4+A_5;$$
\item Hydrogen Chloride (HCl) reaction (Section \ref{SectionHCL})

$$\mathrm{H}_2 \rightleftharpoons 2\mathrm{H},\;\;
\mathrm{Cl}_2 \rightleftharpoons 2\mathrm{Cl},\;\;
\mathrm{H}+\mathrm{Cl}_2\rightleftharpoons \mathrm{HCl}+\mathrm{Cl},\;\;
\mathrm{Cl}+\mathrm{H}_2\rightleftharpoons \mathrm{HCl}+\mathrm{H},$$
or in abstract notations

$$A_1\rightleftharpoons 2A_2,\;\;
A_3\rightleftharpoons 2A_4,\;\;
A_2+A_3 \rightleftharpoons A_5+A_4,\;\;
A_4+A_1 \rightleftharpoons A_5+A_2;$$
\end{itemize}

For these reaction mechanisms, we selected various cort\'{e}ges of reaction rate constants: with detailed balance, with complex balance, more or less stiff, etc. The goal was to demonstrate various aspects of similarity and difference between the classical thermodynamic Lyapunov functions and Gorban's functions.

All the systems below were considered in perfect gases and under isothermal isochorich conditions. Therefore, the volume $V$ was constant and there was no need to use two sets of variables, amounts $N_i$ and concentrations $c_i$. We used the concentrations $c_i$ with the vector of concentrations $c$ and the classical thermodynamic Lyapunov function for these conditions $H(c)$ (\ref{LyapFreeEN}).

The first subsection below contains explicit formulae for points of partial equilibrium for five types of elementary reactions which are used in case studies. The following four subsections present four case studies for four different reaction systems.

\subsection{Partial Equilibria for Several Typical Reactions}

The calculation of Gorban's Lyapunov function $H_\Gamma$ requires finding the points
$$c^*_\gamma (c)=\argmin_{c+\gamma\chi\in \mathbb{R}^n_{>0}}H(c+\gamma\chi),$$
where $\gamma$ is a stoichiometric vector or any other vector with at least one positive and at least one negative element. There is no general formula for the explicit search for such points, but for some typical cases an explicit solution can be found analytically.

Since Boltzmann's $H$ is a strictly convex function in $\mathbb{R}^n_{>0}$ with $c_i\log c_i$ singularities at the borders, the minimizer in the direction $\gamma$ is a positive vector $c+\gamma\chi$, where $\D H(c+\gamma\chi)/ \D \chi=0$:

$$\frac{\mathrm{d}H(c+\gamma\chi)}{\mathrm{d}\chi}=\frac{\mathrm{d}}{\mathrm{d}\chi}\sum_{i=1}^{n}(c_i+\gamma_i\chi)\bigg(\ln \frac{c_i+\gamma_i\chi}{c^{\rm eq}_i}-1 \bigg)=\sum_{i=1}^{n}\gamma_i\ln \frac{c_i+\gamma_i\chi}{c^{\rm eq}_i}=0.$$

A partial equilibrium in direction $\gamma$ satisfies the following equation:
\begin{equation}\label{GenEq}
\prod_{\alpha_i>0}\left( \frac{c_i-\alpha_i\chi}{c^{\rm eq}_i}\right)^{\alpha_i}=\prod_{\beta_i>0}\bigg(\frac{c_i+\beta_i\chi}{c^{\rm eq}_i}\bigg)^{\beta_i}.
\end{equation}

Equation (\ref{GenEq}) is very similar to the usual condition of detailed balance but we have to emphasise that it does not include any reaction rate constant, does not assume the reversibility of any reaction or microreversibility and just describes the minimisers of $H$ in the given direction. It can be considered as the thermodynamic equilibrium condition for the elementary reaction with the stoichiometric vector $\gamma$ and can differ from the kinetic equilibrium condition if the detailed balance is not assumed. Possibility of such a difference in general kinetics is sometimes called the `Wegscheider paradox' \cite{MarinYab2019} to celebrate the work of Wegscheider \cite{Wegscheider1902}.

Let us consider the isomerisation reaction $A_1\rightleftharpoons A_2$. The corresponding stoichiometric vector is $\gamma=(-1, 1)$. For this vector, there is one stoichiometric conservation law $c_1+c_2=b$, where $b$ is a positive constant. Equation (\ref{GenEq}) for this vector has the form:

$$ \frac{c_1^{}-\chi}{c^{\rm eq}_1}=\frac{c_2^{}+\chi}{c^{\rm eq}_2}.$$

The root of this polynomial is

$$\chi=\frac{c_1^{}c^{\rm eq}_2-c_2^{}c^{\rm eq}_1}{c^{\rm eq}_1+c^{\rm eq}_2}$$
and the point of partial equilibrium is

\begin{equation}\label{GenIsom}
\begin{split}
c^*_1=\frac{bc^{\rm eq}_1}{c^{\rm eq}_1+c^{\rm eq}_2},\\
c^*_2=\frac{bc^{\rm eq}_2}{c^{\rm eq}_1+c^{\rm eq}_2}.
\end{split}
\end{equation}

Let us consider the reaction of dissociation $A_1\rightleftharpoons 2A_2$. The corresponding stoichiometric vector is $\gamma=(-1, 2)$. For this vector, there is one stoichiometric conservation law $2c_1+c_2=b$, where $b$ is a positive constant. Equation (\ref{GenEq}) for this vector has the form

$$ \frac{c_1^{}-\chi}{c^{\rm eq}_1}=\bigg(\frac{c_2^{}+\chi}{c^{\rm eq}_2}\bigg)^2.$$

The roots of this polynomial are

$$\chi=\frac{-4c_2^{}-k\pm\sqrt{8kb+k^2}}{8},$$
where

$$ k = \frac{\big(c_2^{\rm eq})^2}{c_1^{\rm eq}}.$$

The sign of the root can be determined from the condition of non-negativity of concentrations. The point of partial equilibrium is

\begin{equation}\label{GenDiss}
\begin{split}
c^*_1&=\frac{4b+k-\sqrt{8kb+k^2}}{8},\\
c^*_2&=\frac{-k+\sqrt{8kb+k^2}}{4}.
\end{split}
\end{equation}

Let us consider the reaction $A_1+A_2\rightleftharpoons A_3$. The corresponding stoichiometric vector is $\gamma=(-1, -1, 1)$. For this vector, there are two stoichiometric conservation laws, $c_2-c_1=b_1$ and $c_1+c_3=b_2$, where $b_1$ and $b_2$ are positive constants. Equation (\ref{GenEq}) for this vector has the form

$$ \frac{c_1^{}-\chi}{c^{\rm eq}_1} \frac{c_2^{}-\chi}{c^{\rm eq}_2}= \frac{c_3^{}+\chi}{c^{\rm eq}_3}.$$

The point of partial equilibrium is

\begin{equation}\label{Gen111}
\begin{split}
c^*_1&=\frac{-b_1-k+\sqrt{(k+b_1)^2+kb_2}}{2},\\
c^*_2&=\frac{b_1-k+\sqrt{(k+b_1)^2+kb_2}}{2},\\
c^*_3&=\frac{b_1+2b_2+k-\sqrt{(k+b_1)^2+kb_2}}{2},
\end{split}
\end{equation}
where

$$ k = \frac{c_1^{\rm eq}c_2^{\rm eq}}{c_3^{\rm eq}}.$$

For the reaction $A_1+A_2\rightleftharpoons 2A_3$ we have a stoichiometric vector $\gamma=(-1, -1, 2)$. For this vector, there are two stoichiometric conservation laws $c_2-c_1=b_1$ and $c_1+c_2+c_3=b_2$, where $b_1$ and $b_2$ are positive constants. Equation (\ref{GenEq}) for this vector has the form

$$ \frac{c_1^{}-\chi}{c^{\rm eq}_1} \frac{c_2^{}-\chi}{c^{\rm eq}_2}= \bigg(\frac{c_3^{}+2\chi}{c^{\rm eq}_3}\bigg)^2.$$

The point of partial equilibrium is

\begin{equation}\label{Gen112}
\begin{split}
c^*_1&=\frac{k(b_2-b_1)+b_2-\sqrt{(k+1)b_2^2-kb_1^2}}{2k},\\
c^*_2&=\frac{k(b_2+b_1)+b_2-\sqrt{(k+1)b_2^2-kb_1^2}}{2k},\\
c^*_3&=\frac{-b_2+\sqrt{(k+1)b_2^2-kb_1^2}}{k},
\end{split}
\end{equation}
where

$$ k = 4\frac{c_1^{\rm eq}c_2^{\rm eq}}{\big(c_3^{\rm eq}\big)^2}-1.$$

For the reaction $A_1+A_2\rightleftharpoons A_3+A_4$ we have a stoichiometric vector $\gamma=(-1, -1, 1, 1)$. For this vector, there are three stoichiometric conservation laws $c_2-c_1=b_1$, $c_4-c_3=b_2$ and $c_1+c_2+c_3+c_4=b_3$, where $b_1, b_2$ and $b_3$ are positive constants. Equation (\ref{GenEq}) for this vector has the form

$$ \frac{c_1^{}-\chi}{c^{\rm eq}_1} \frac{c_2^{}-\chi}{c^{\rm eq}_2}= \frac{c_3^{}+\chi}{c^{\rm eq}_3} \frac{c_4^{}+\chi}{c^{\rm eq}_4}.$$

The point of partial equilibrium is

\begin{equation}\label{Gen1111}
\begin{split}
c^*_1&=\frac{b_3+k(b_3-b_1)-\sqrt{(k+1)b_3^2+2k^2b_2^2-kb_1^2}}{2k},\\
c^*_2&=\frac{b_3+k(b_3+b_1)-\sqrt{(k+1)b_3^2+2k^2b_2^2-kb_1^2}}{2k},\\
c^*_3&=\frac{-b_3-kb_2+\sqrt{(k+1)b_3^2+2k^2b_2^2-kb_1^2}}{2k},\\
c^*_4&=\frac{-b_3+kb_2+\sqrt{(k+1)b_3^2+2k^2b_2^2-kb_1^2}}{2k},
\end{split}
\end{equation}
where

$$ k = \frac{c_1^{\rm eq}c_2^{\rm eq}}{c_3^{\rm eq}c_4^{\rm eq}}-1.$$

An analytical representation of partial equilibria can also be found for many other reactions. In this subsection, we have presented only all the reactions that are used in case studies.

\subsection{Linear Kinetics \label{SectioninIsom}}

Let us consider the isomerisation cycle

\begin{equation}\label{SystIsom}
A_1\rightleftharpoons A_2 \rightleftharpoons A_3 \rightleftharpoons A_1.
\end{equation}

There is one conservation law for this system: $c_1+c_2+c_3=b$. The line of partial equilibrium for each of the three stoichiometric vectors is defined by (\ref{GenIsom}). For example, for the first reaction, this partial equilibrium line is

\begin{equation*}
\begin{split}
c^*_1&=\frac{(b-c_3^{})c^{\rm eq}_1}{c^{\rm eq}_1+c^{\rm eq}_2},\\
c^*_2&=\frac{(b-c_3^{})c^{\rm eq}_2}{c^{\rm eq}_1+c^{\rm eq}_2},\\
c^*_3&=c^{}_3.
\end{split}
\end{equation*}

\begin{figure}[htb]
\centering
(a)\includegraphics[width=0.3\textwidth]{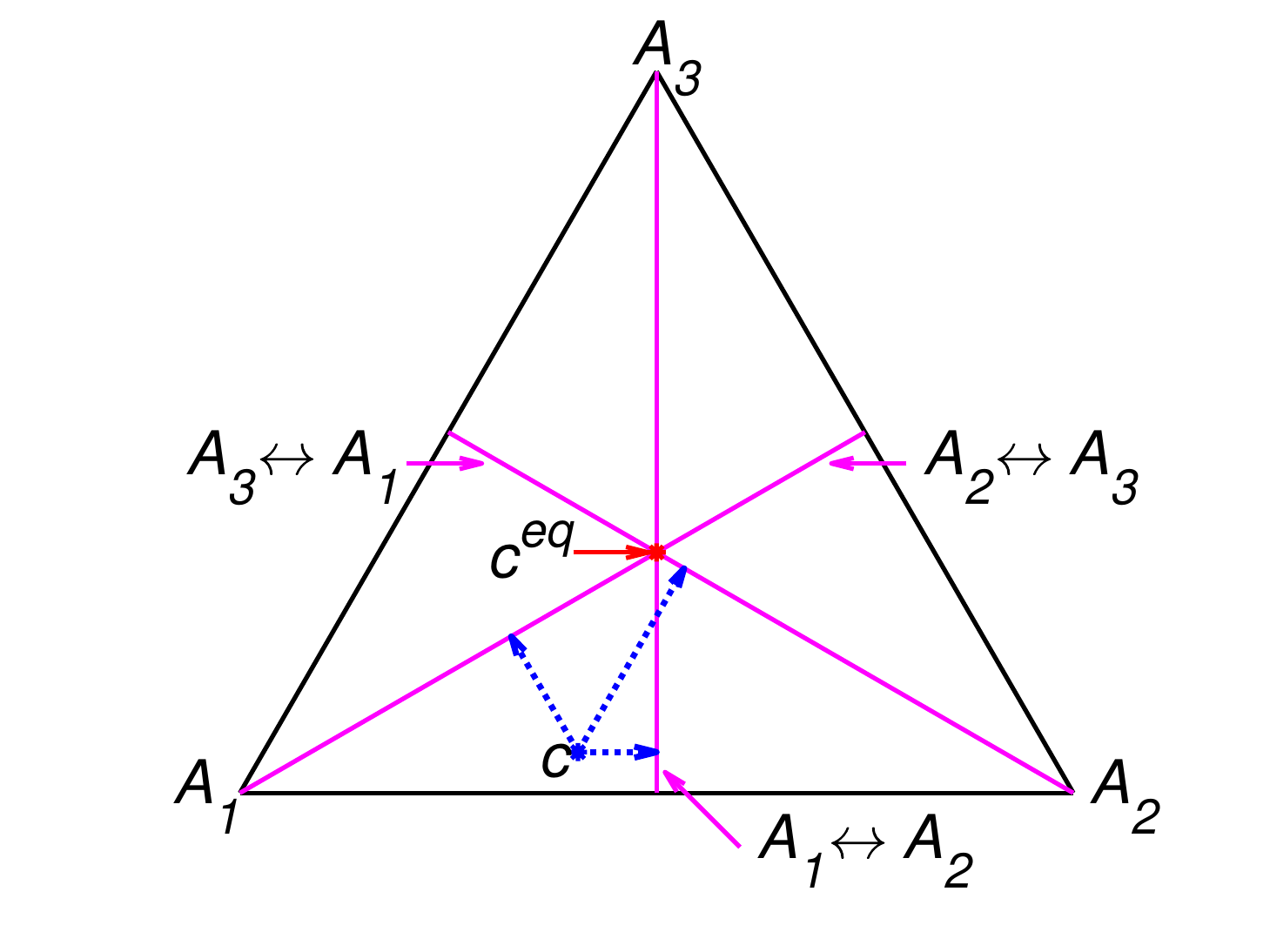}
(b)\includegraphics[width=0.3\textwidth]{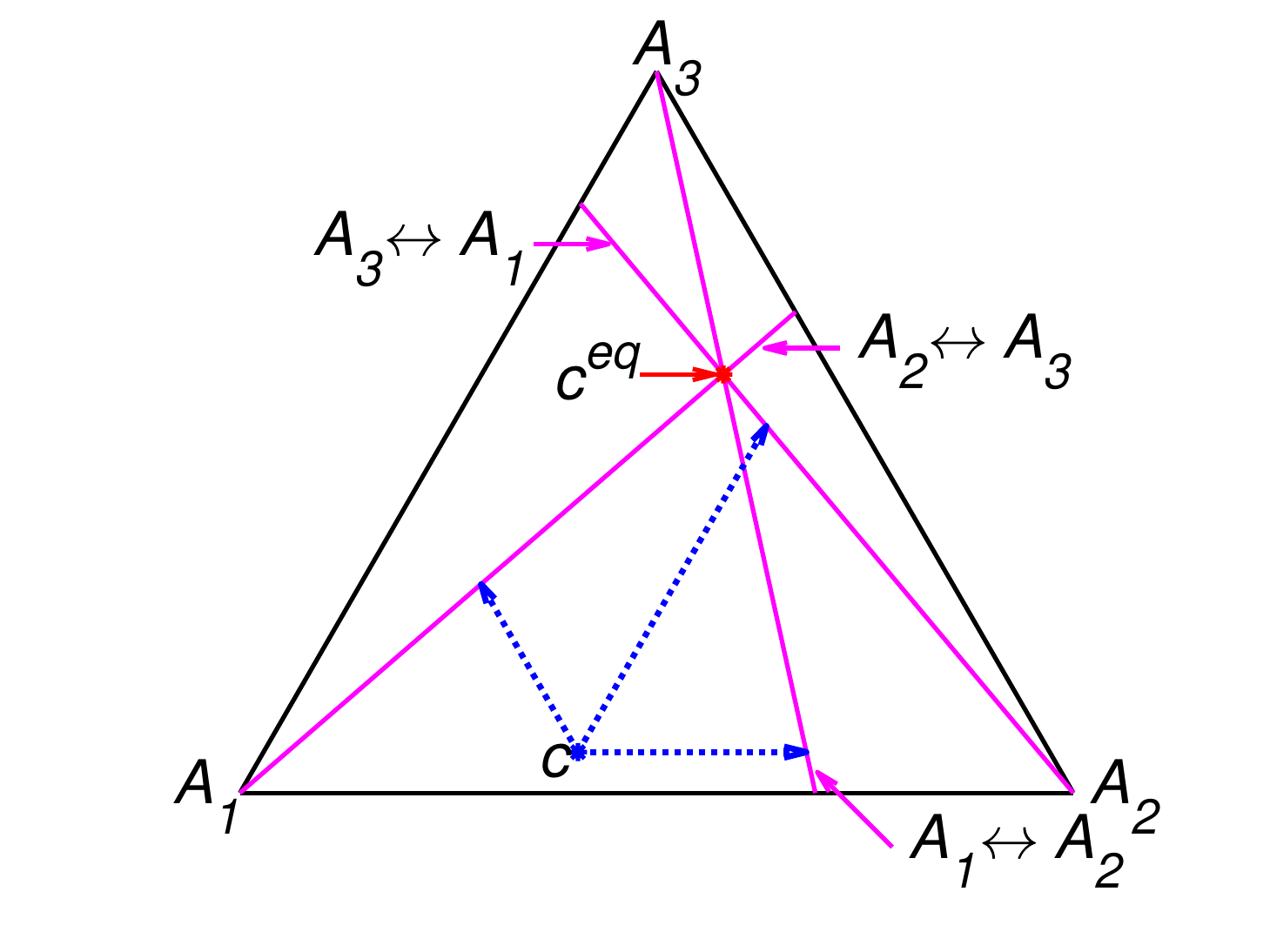}
(c)\includegraphics[width=0.3\textwidth]{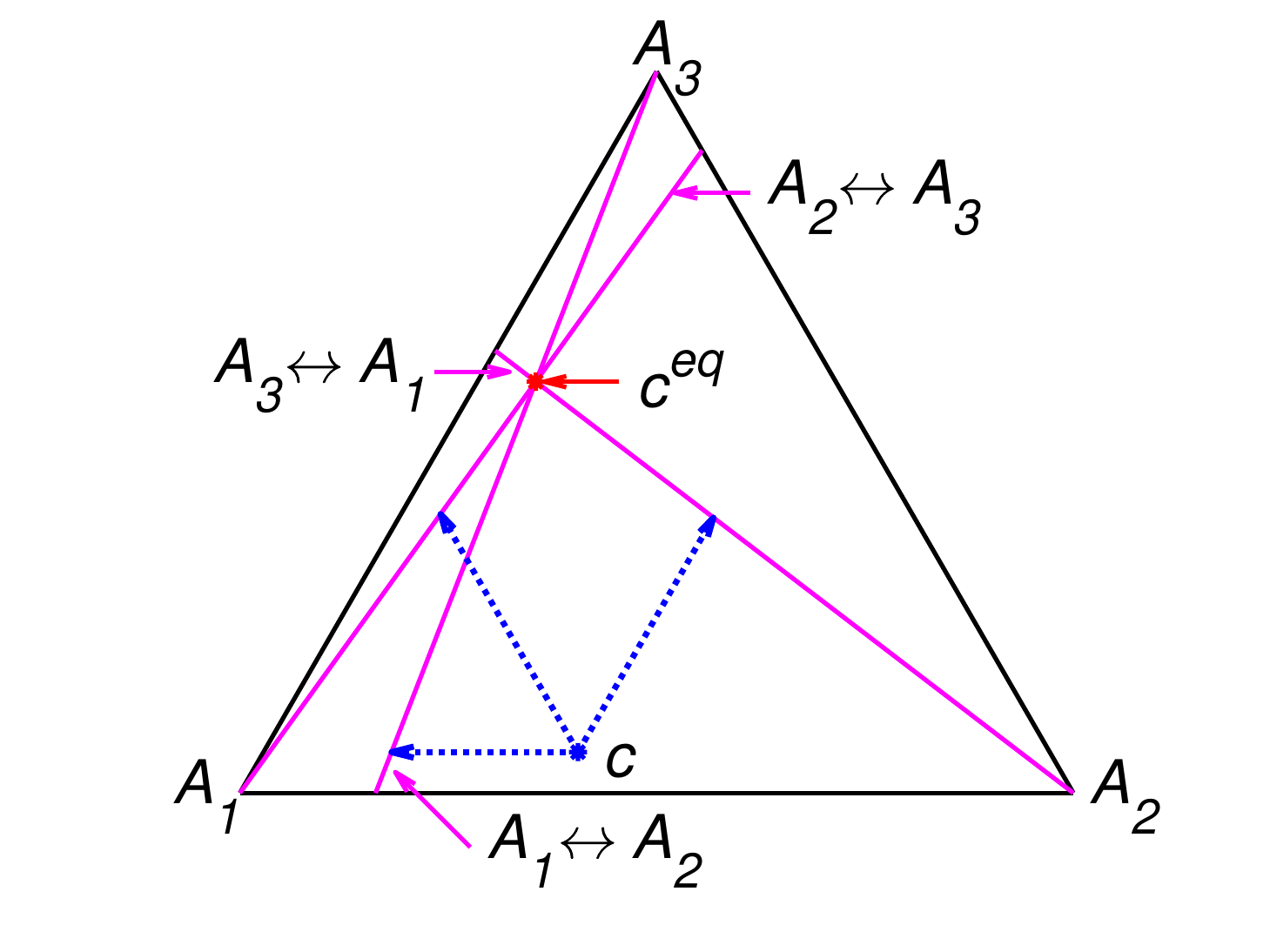}

\caption{Partial equilibrium lines (solid magenta lines) and points of partial equilibrium for point c (dotted arrows) for the reaction system $A_1\rightleftharpoons A_2 \rightleftharpoons A_3 \rightleftharpoons A_1$ with several equilibria: (\textbf{a}) $c^{\rm eq}=(1/3,1/3,1/3)$, (\textbf{b}) $c^{\rm eq}=(0.13,0.29,0.58)$, and (\textbf{c}) $c^{\rm eq}=(0.36,0.07,0.57)$.}
\label{FigLinEquil}
\end{figure}

\begin{figure}[htb]
\centering
(a)\includegraphics[width=0.3\textwidth]{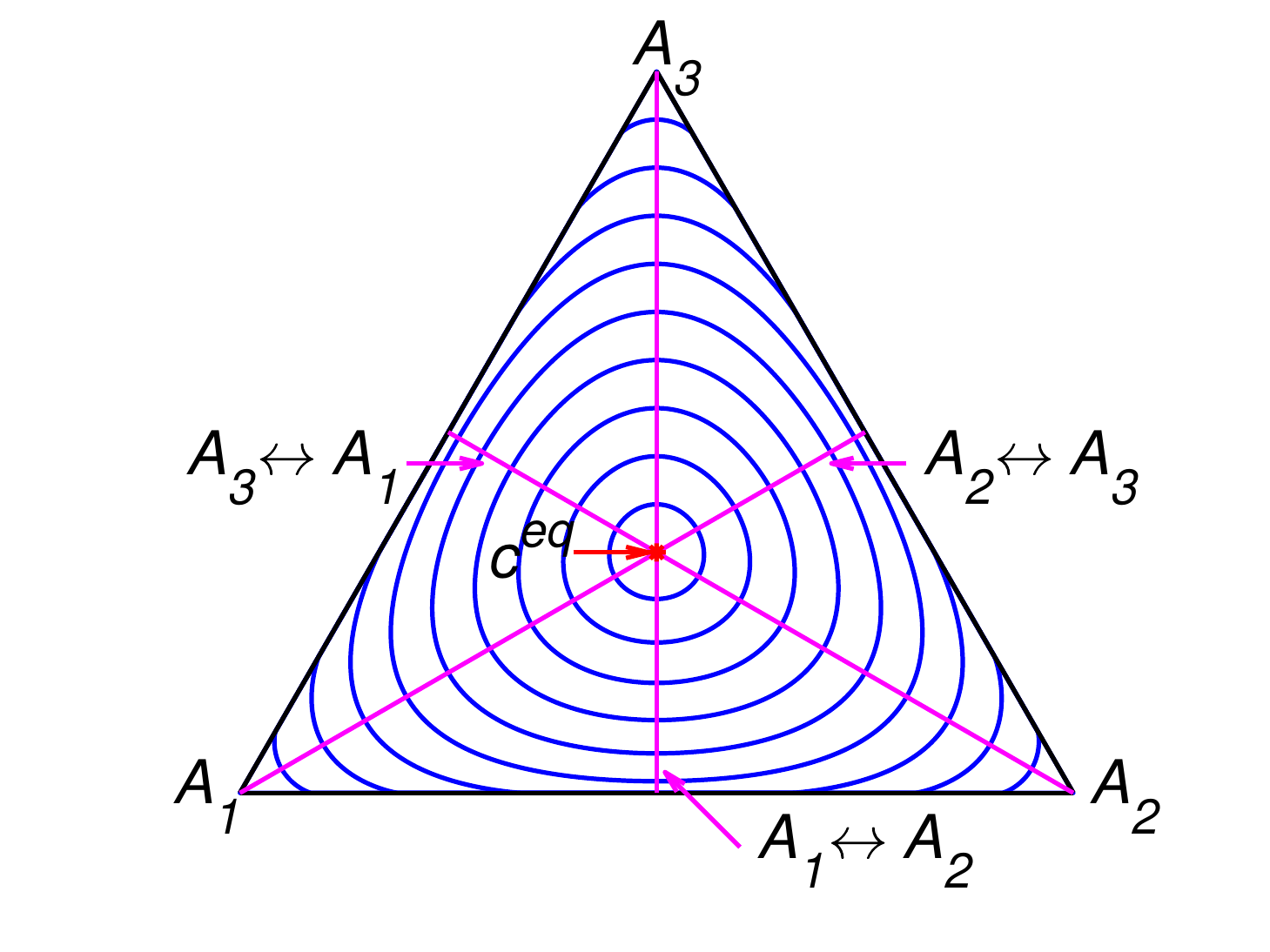}
(b)\includegraphics[width=0.3\textwidth]{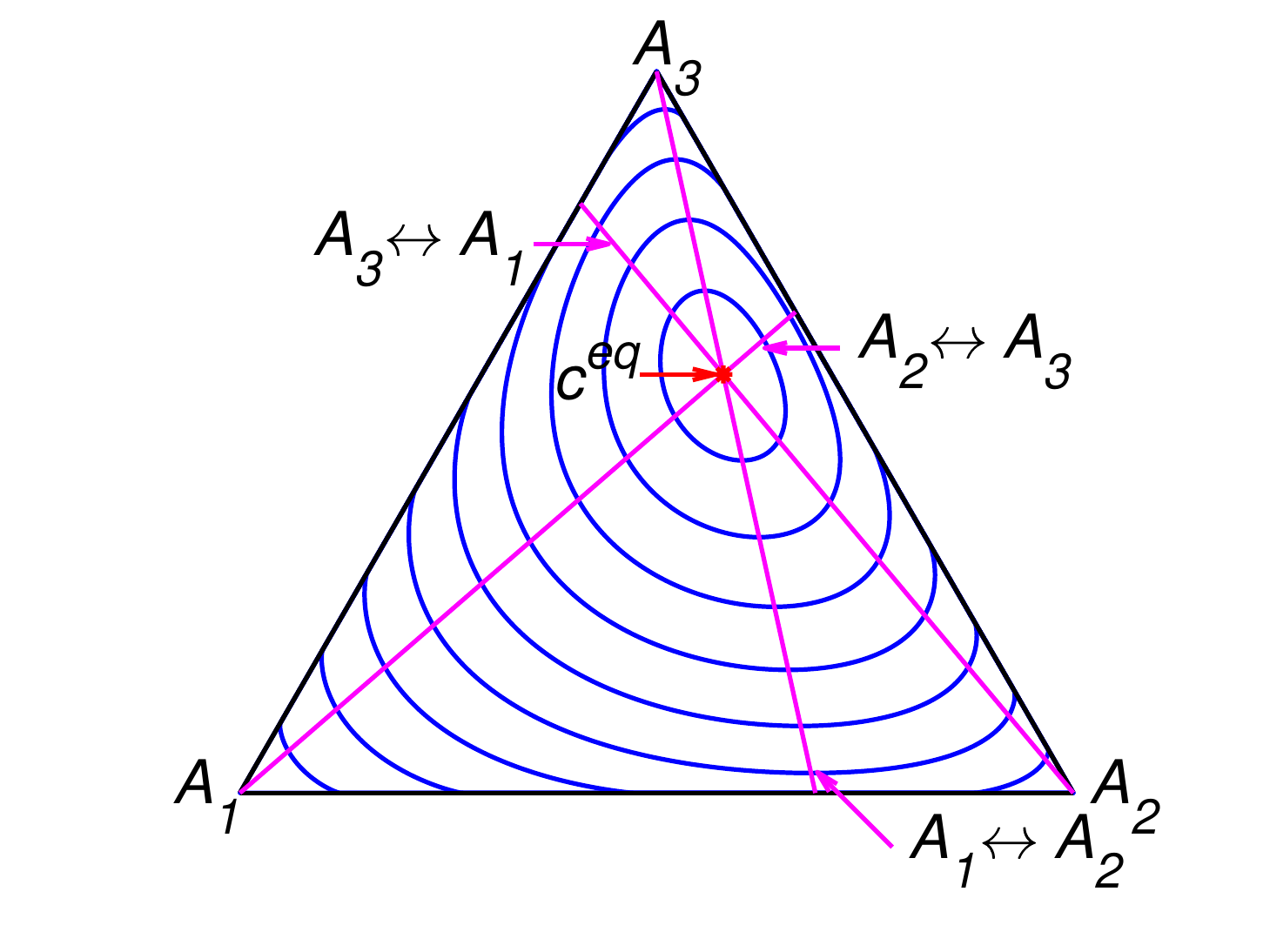}
(c)\includegraphics[width=0.3\textwidth]{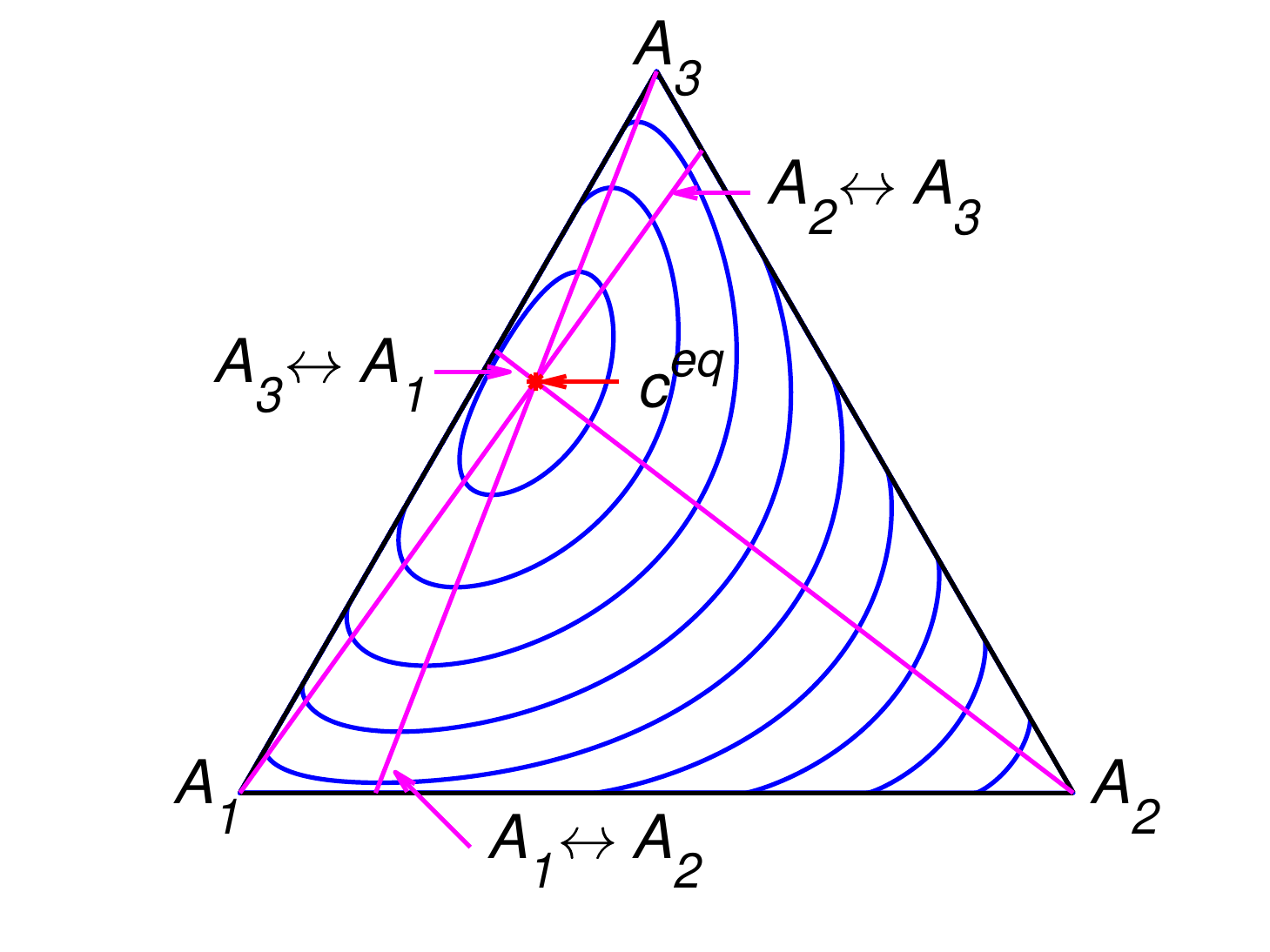}\\
(d)\includegraphics[width=0.3\textwidth]{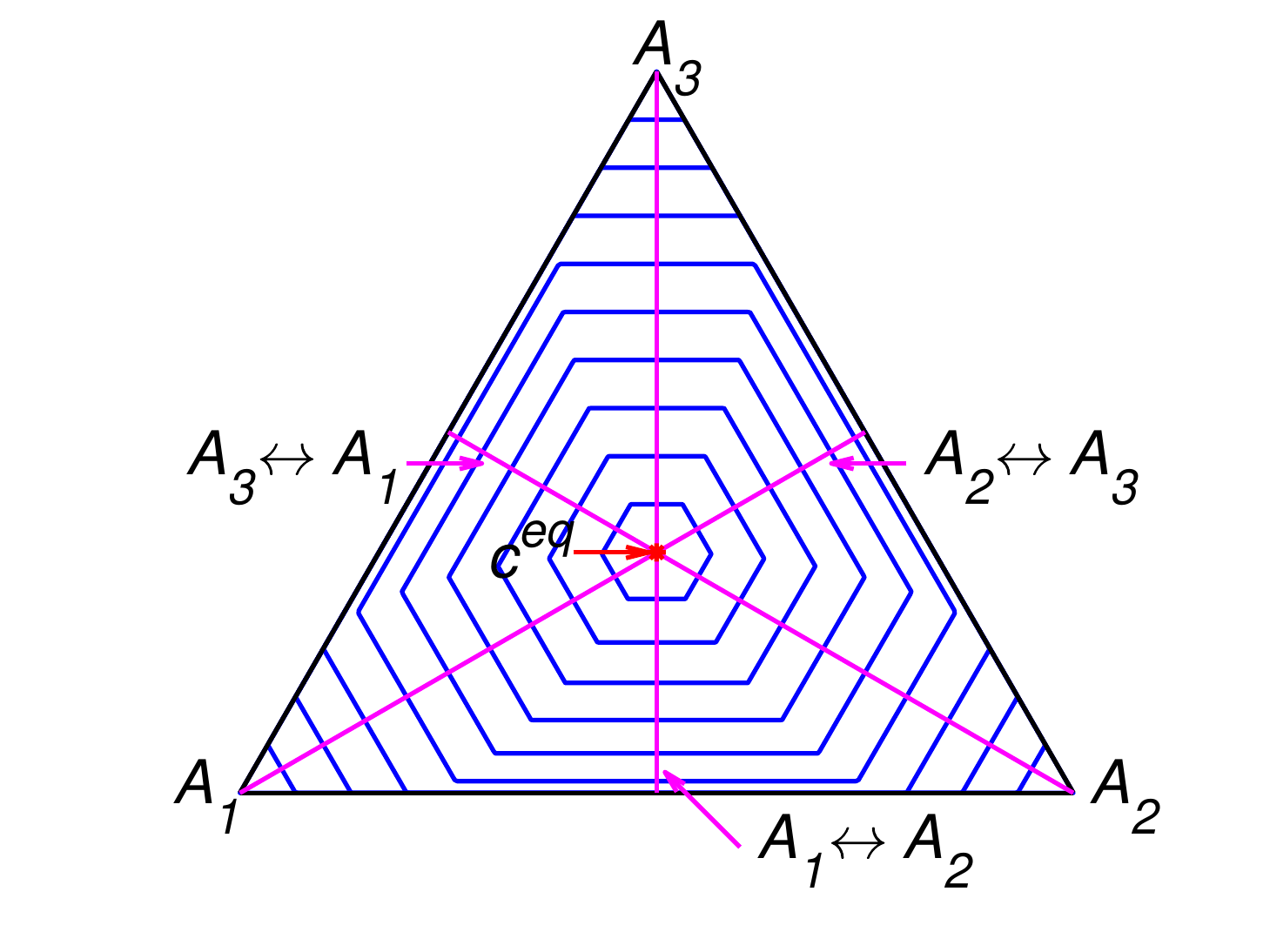}
(e)\includegraphics[width=0.3\textwidth]{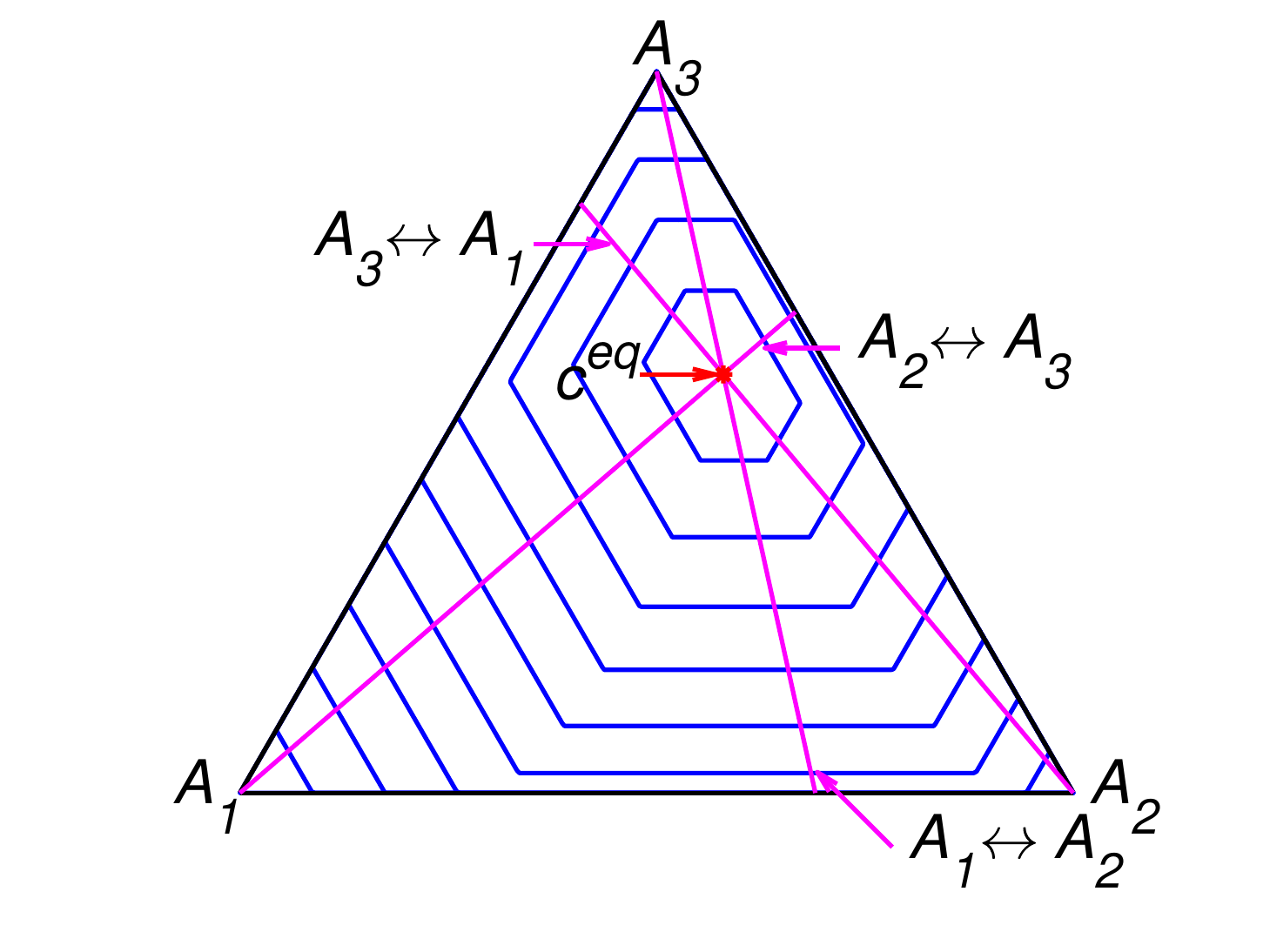}
(f)\includegraphics[width=0.3\textwidth]{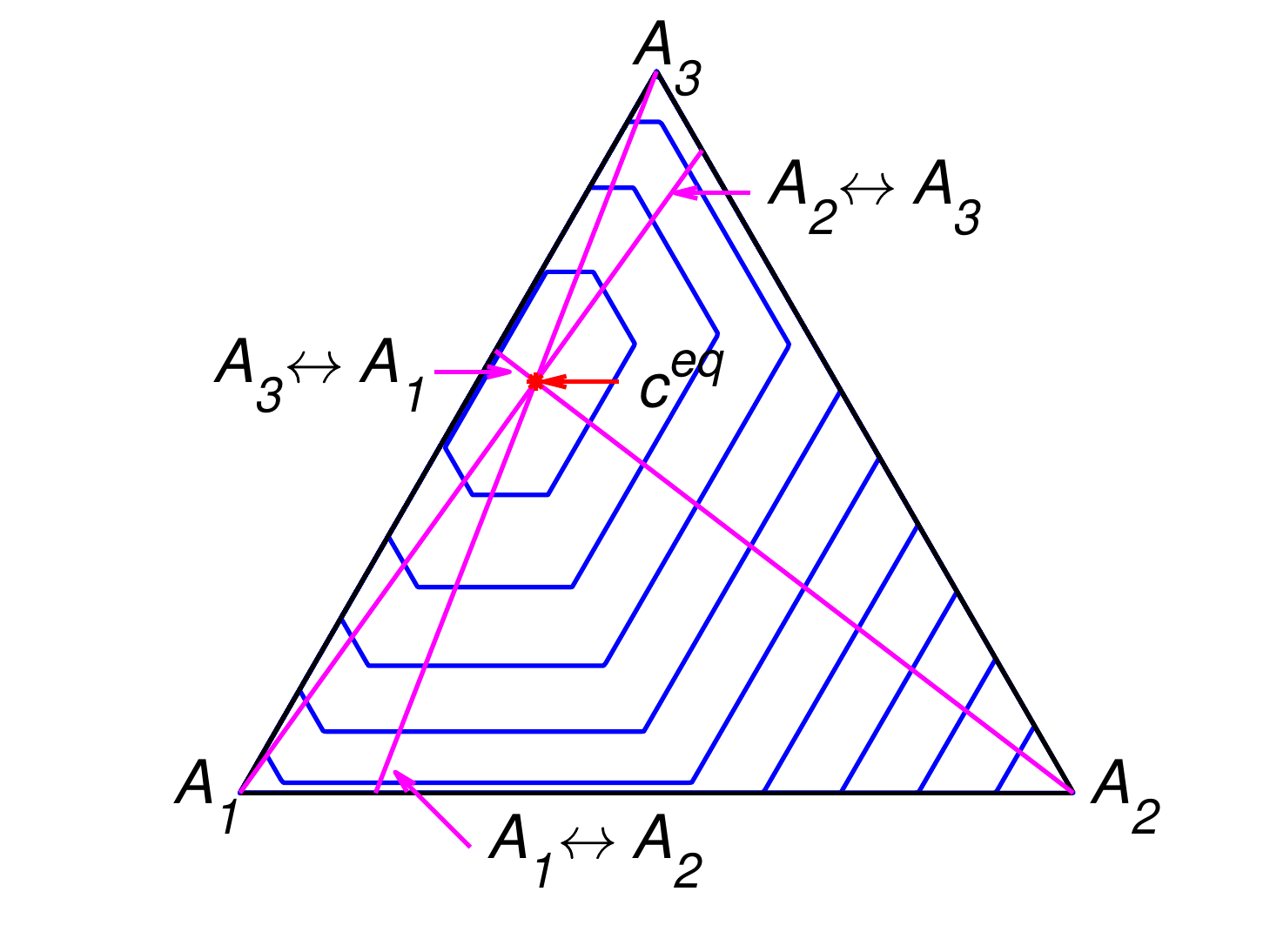}

\caption{The level sets for Boltzmann's $H$ function in top row and the corresponding level sets for Gorban's $H_\Gamma$ function in bottom row for several equilibria: (\textbf{a, d}) $c^{\rm eq}=(1/3,1/3,1/3)$, (\textbf{b, e}) $c^{\rm eq}=(0.13,0.29,0.58)$, and (\textbf{c, f}) $c^{\rm eq}=(0.36,0.07,0.57)$.}
\label{FigLinHBHG}
\end{figure}

The lines of partial equilibrium and partial equilibrium points for a given point $c$ are presented in Figure~\ref{FigLinEquil}.
The level sets for Boltzmann's $H$ function and Gorban's $H_\Gamma$ function are presented in Figure~\ref{FigLinHBHG}. It is important to emphasise that these level sets are independent of kinetic constants and are completely determined by the equilibrium for Boltzmann's $H$ function and by the equilibrium and set of stoichiometric vectors $\Gamma$ for Gorban's $H_\Gamma$ function.

The kinetic equations for the system (\ref{SystIsom}) are:

\begin{equation}\label{EqLinKin}
\begin{split}
\frac{\mathrm{d}c_1}{\mathrm{d}t}&=-k^+_1c_1+k^-_1c_2+k^+_3c_3-k^-_3c_1,\\
\frac{\mathrm{d}c_2}{\mathrm{d}t}&=k^+_1c_1-k^-_1c_2-k^+_2c_2+k^-_2c_3,\\
c_3&=b-c_1-c_2.
\end{split}
\end{equation}

For system (\ref{SystIsom}) with detailed balance the conditions for the reaction rate constants are

$$k^+_1c_1^{\rm eq}=k^-_1c_2^{\rm eq},\;\;  k^+_2c_2^{\rm eq}=k^-_2c_3^{\rm eq},\;\;  k^+_3c_3^{\rm eq}=k^-_3c_1^{\rm eq}.$$

The system can be completely parametrised by three equilibrium concentrations $c^{\rm eq}_i$ and three reaction rate constants, for example, by the constants $k^+_1, k^+_2, k^+_3$.
To obtain the complex balance condition it is necessary to list all the different stoichiometric vectors $\alpha_\rho$ and $\beta_\rho$:

\begin{equation*}
\begin{split}
\alpha_{-3}=\alpha_{1}=\beta_{3}=\beta_{-1}=(1,0,0),\\
\alpha_{-1}=\alpha_{2}=\beta_{1}=\beta_{-2}=(0,1,0),\\
\alpha_{-2}=\alpha_{3}=\beta_{2}=\beta_{-3}=(0,0,1).
\end{split}
\end{equation*}

The conditions of complex balance are

\begin{equation}\label{EqLinComplex}
\begin{split}
k^-_3c_1^{\rm eq}+k^+_1c_1^{\rm eq}=k^+_3c_3^{\rm eq}+k^-_1c_2^{\rm eq},\\
k^-_1c_2^{\rm eq}+k^+_2c_2^{\rm eq}=k^+_1c_1^{\rm eq}+k^-_2c_3^{\rm eq},\\
k^-_2c_3^{\rm eq}+k^+_3c_3^{\rm eq}=k^+_2c_2^{\rm eq}+k^-_3c_1^{\rm eq}.
\end{split}
\end{equation}

We can see that the complex balance conditions for this system are equivalent to the condition of stationarity of the point $c^{\rm eq}$ and are not equivalent to the detailed balance conditions. The first two equations in (\ref{EqLinComplex}) are linearly independent, but the third equation is linearly dependent on the first two equations because the sum of these three equations is equivalent to the trivial equality $0=0$. As a result, this system can be parametrised by three equilibrium concentrations $c^{\rm eq}_i$ and four reaction rate constants, for example, by the constants $k^+_1, k^+_2, k^+_3, k^+_{-3}$. This means that system with complex balance has one additional degree of freedom. Since system (\ref{SystIsom}) can have a complex balance equilibrium, which is not a point of detailed balance, a set of parameters with a stable focus in equilibrium instead of a stable node is not a priori forbidden. To illustrate the possible behaviour of system (\ref{SystIsom}), we selected the parameters presented in Table~\ref{TabLinMod}.

\begin{table}[htb]
\caption{The set of parameters used in the simulations and the corresponding type of equilibrium.}
\label{TabLinMod}
\centering
\begin{tabular}{cccccccccc}
\toprule
\textbf{Set Name}	& \textbf{$c^{\rm eq}_1$}	& \textbf{$c^{\rm eq}_2$} & \textbf{$c^{\rm eq}_3$} & \textbf{$k^+_1$} & \textbf{$k^+_2$} & \textbf{$k^+_3$} & \textbf{$k^+_{-3}$} & \textbf{Equilibrium Type}\\
\midrule
S1.1 & 1/3  & 1/3  & 1/3  & 0.1       & 0.2       & 0.3       & 0.6  & Stable node\\
S1.2 & 1/3  & 1/3  & 1/3  & 1/3-0.001 & 1/3-0.001 & 1/3-0.001 & 0.001 & Stable focus\\
S2.1 & 0.13 & 0.29 & 0.58 & 0.5       & 0.6       & 0.1       & 1.1   & Stable node\\
S2.2 & 0.13 & 0.29 & 0.58 & 0.5       & 0.6       & 0.1       & 10    & Stable focus\\
S3.1 & 0.36 & 0.07 & 0.57 & 0.2       & 0.5       & 0.1       & 0.1   & Stable node\\
S3.2 & 0.36 & 0.07 & 0.57 & 0.0005    & 0.001     & 0.00853   & 0.02  & Stable focus\\
\bottomrule
\end{tabular}
\end{table}

\begin{figure}[htbp]
\centering
(a)
\includegraphics[width=0.31\textwidth]{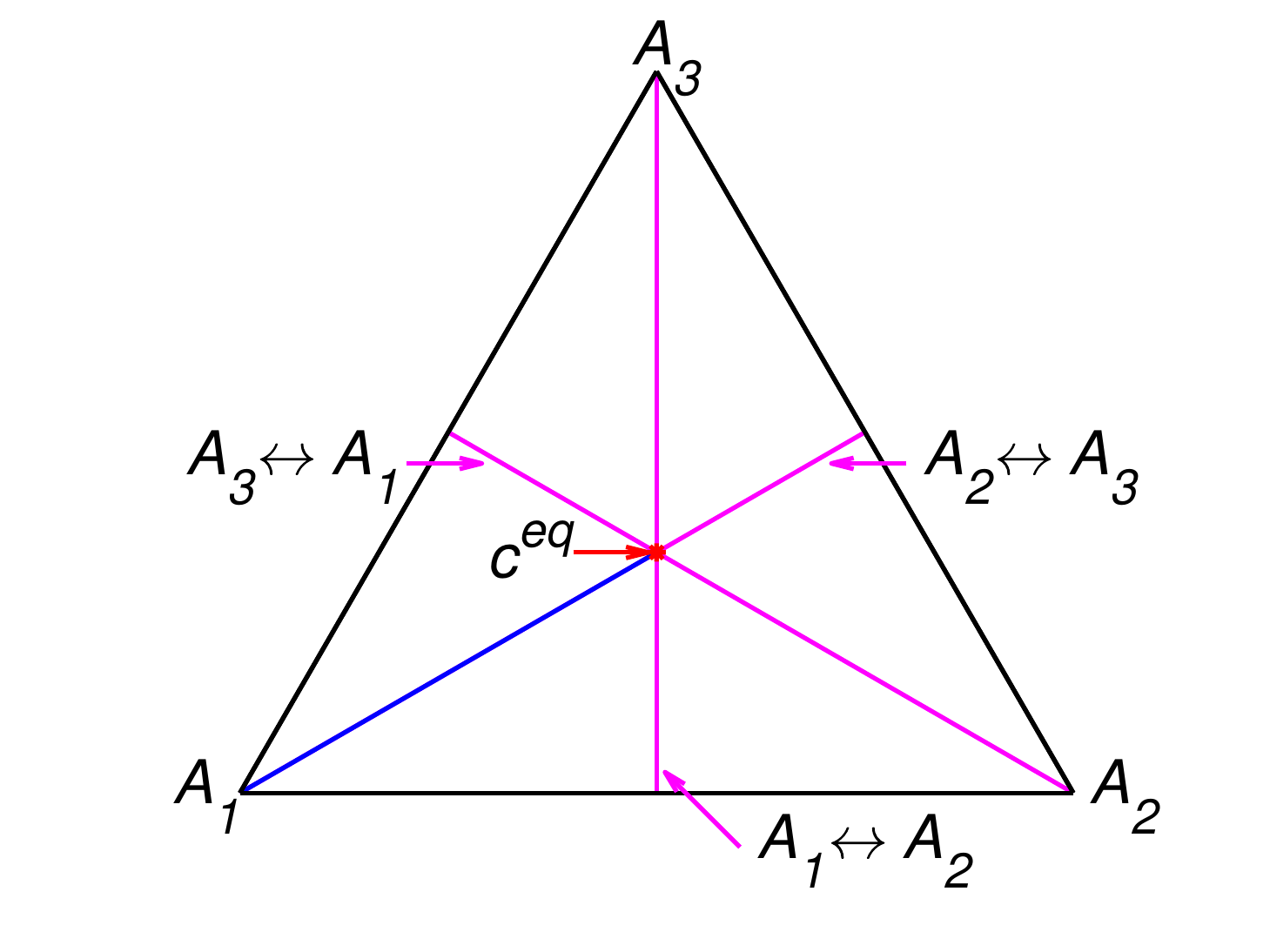}
\includegraphics[width=0.31\textwidth]{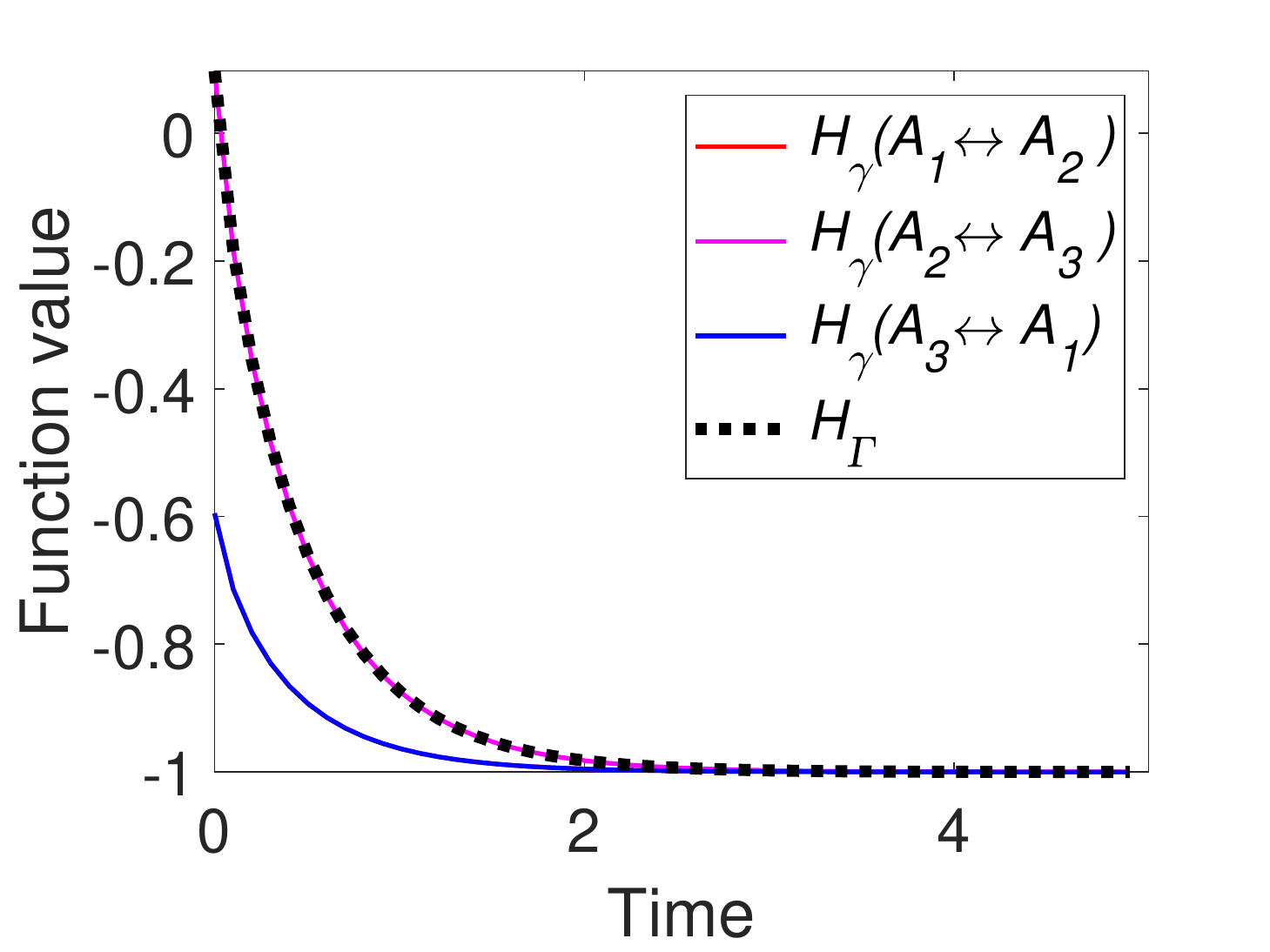}
\includegraphics[width=0.31\textwidth]{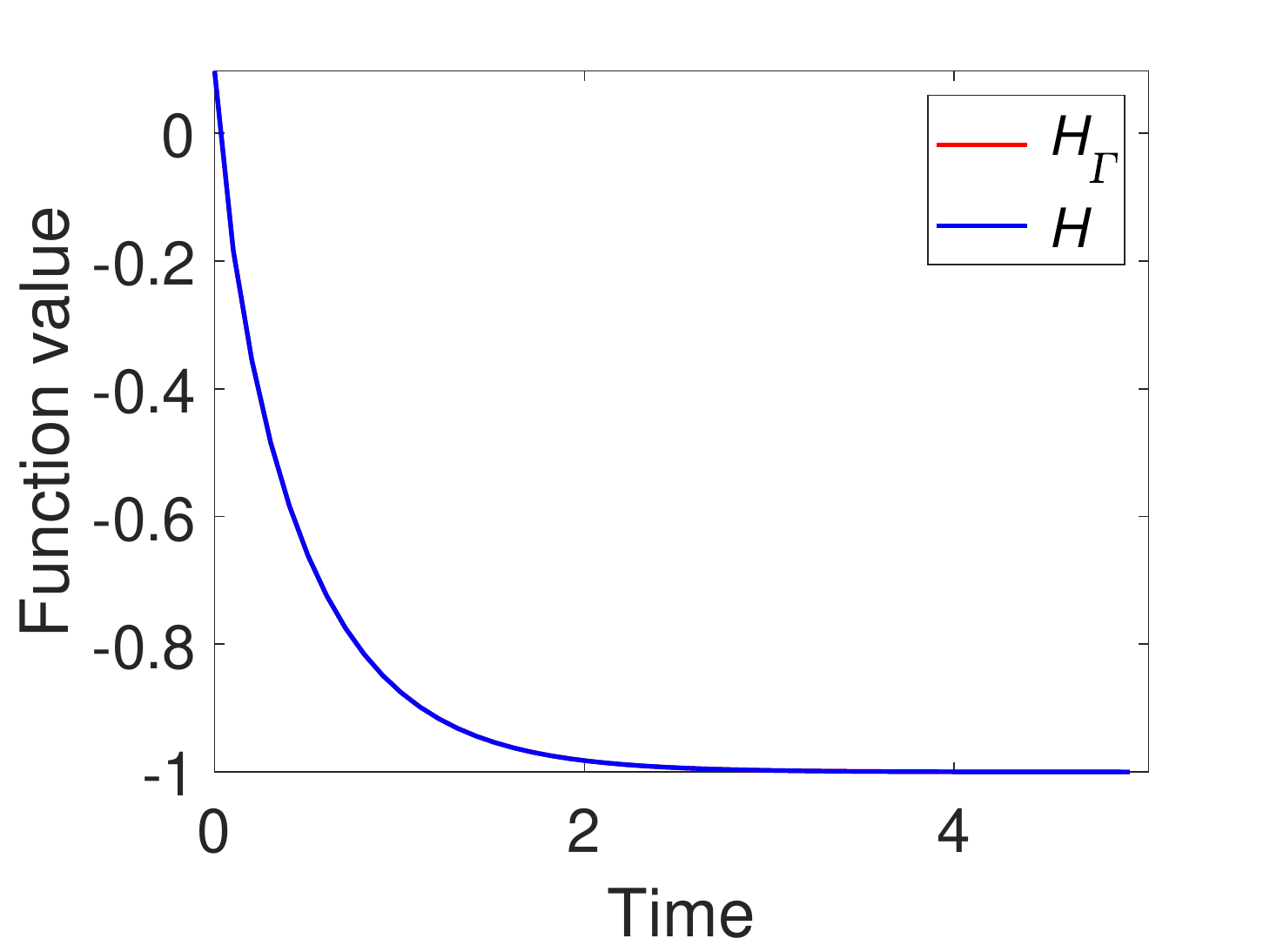}

(b)
\includegraphics[width=0.31\textwidth]{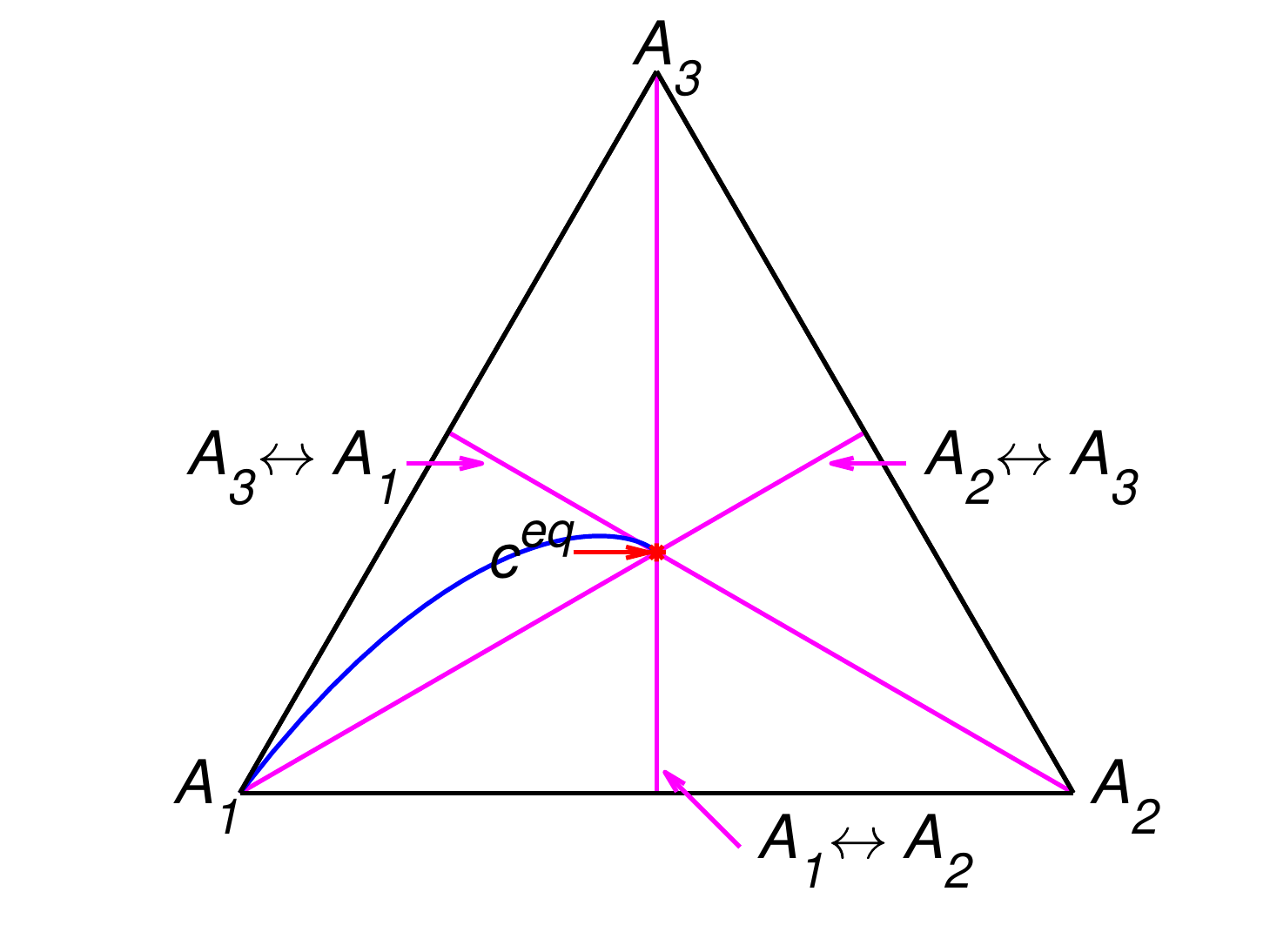}
\includegraphics[width=0.31\textwidth]{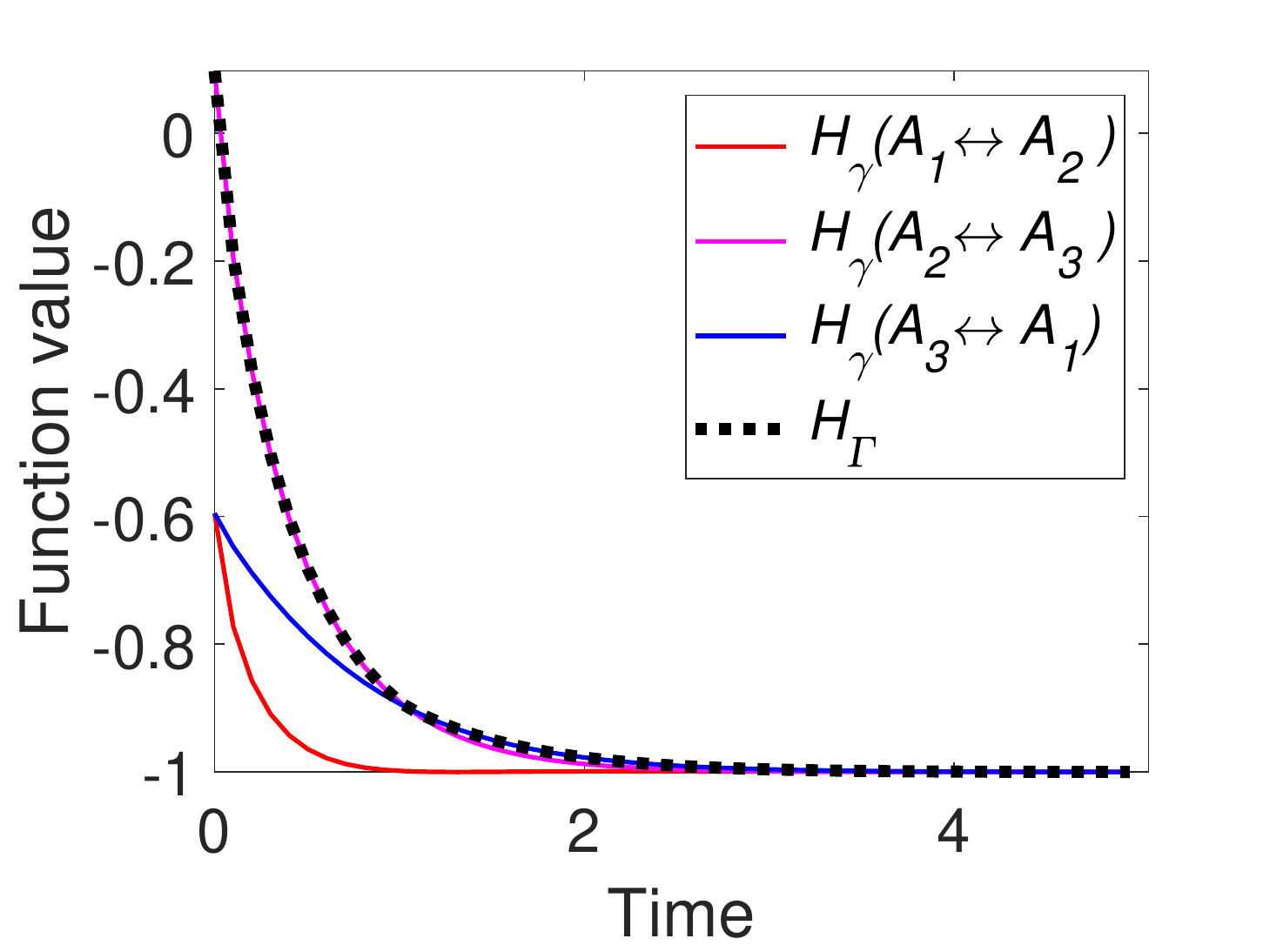}
\includegraphics[width=0.31\textwidth]{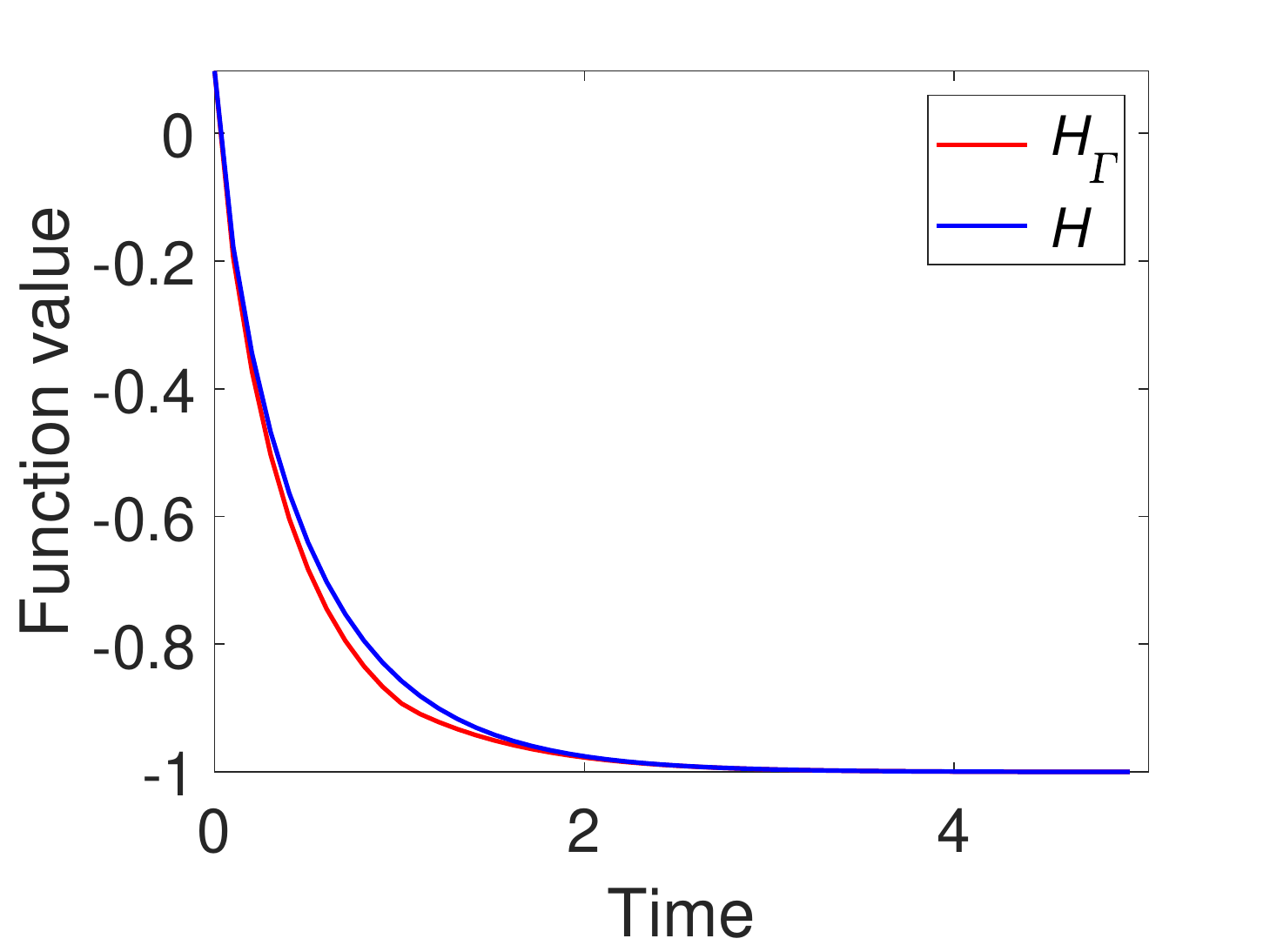}

(c)
\includegraphics[width=0.31\textwidth]{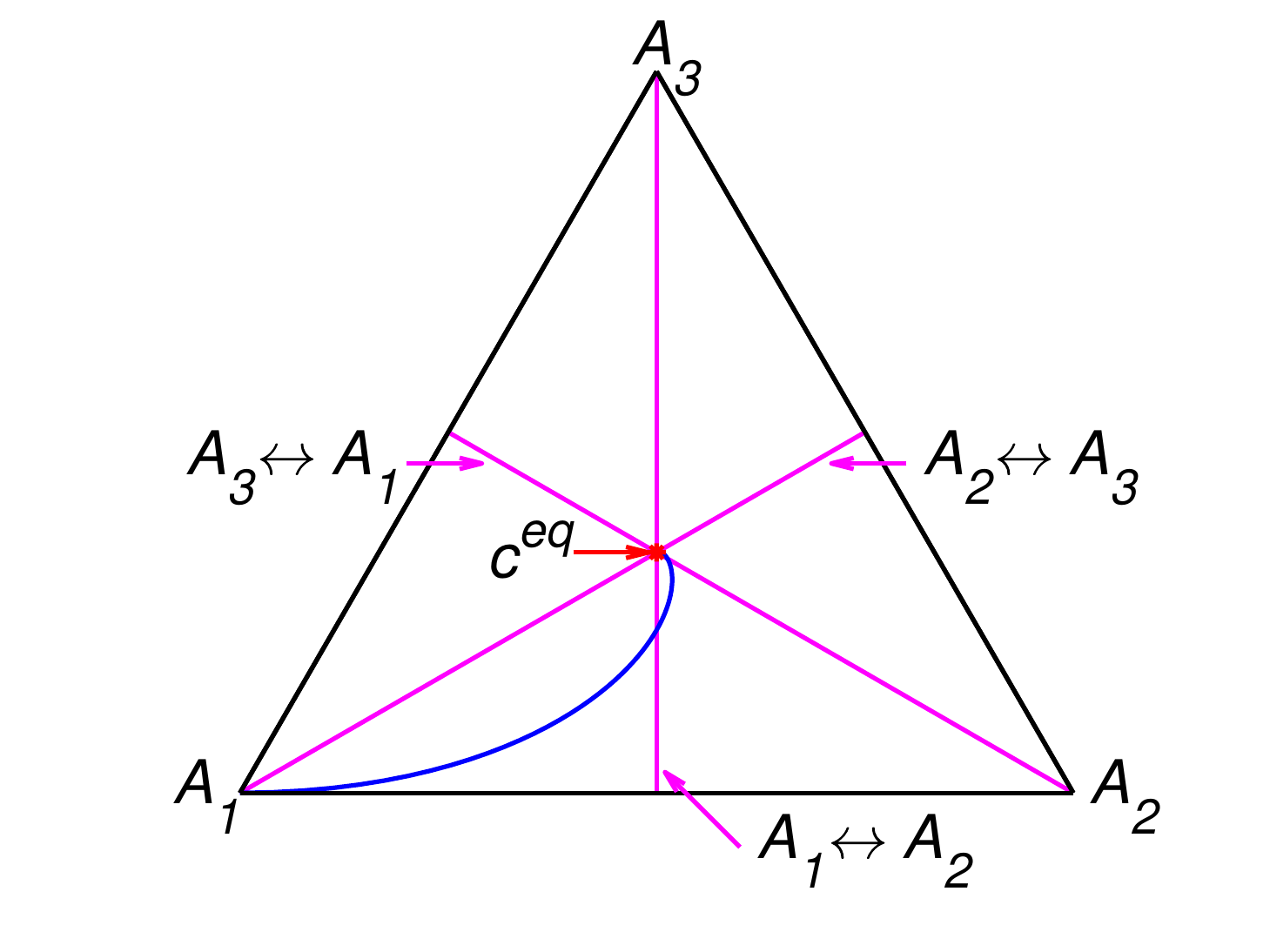}
\includegraphics[width=0.31\textwidth]{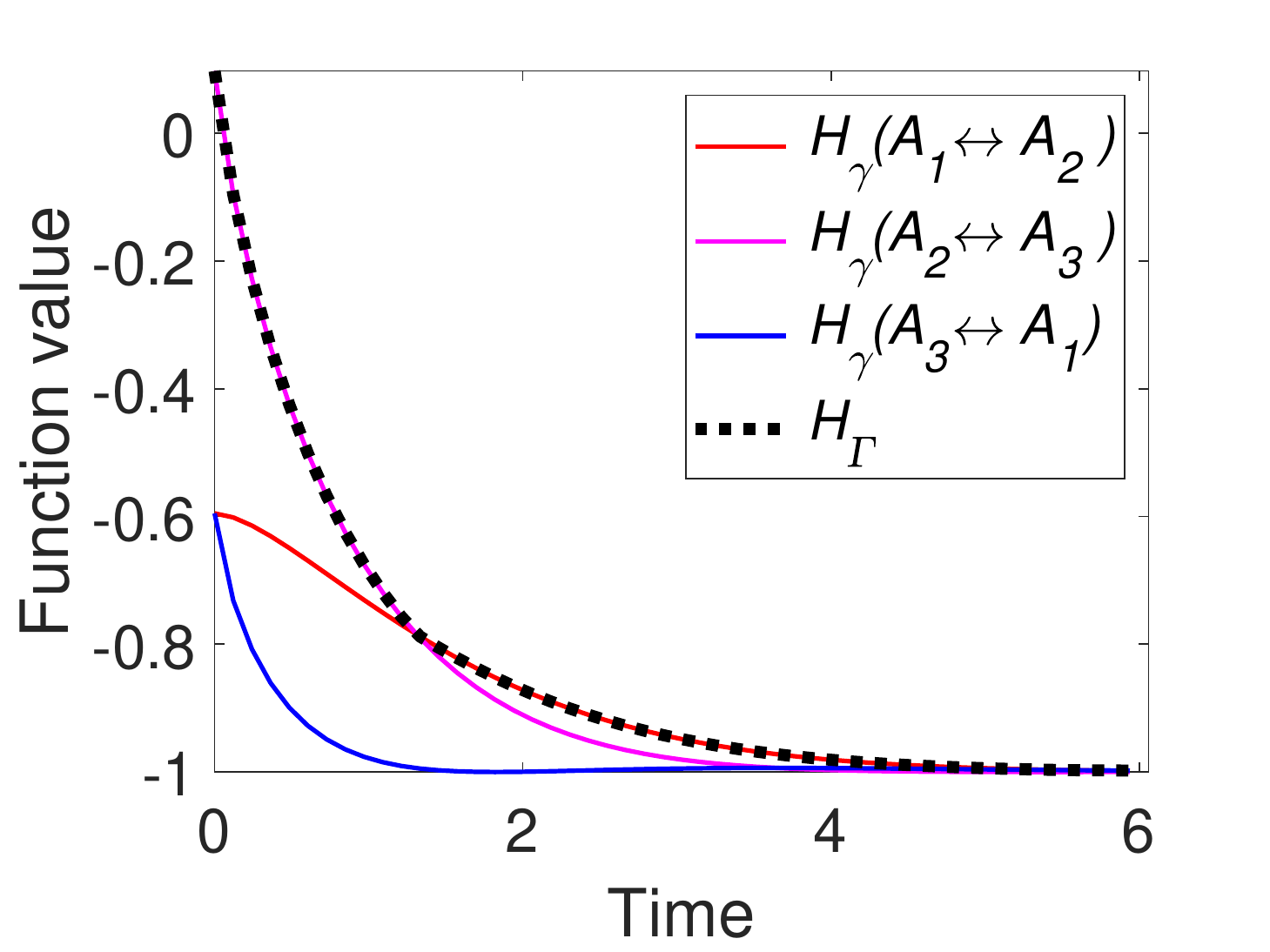}
\includegraphics[width=0.31\textwidth]{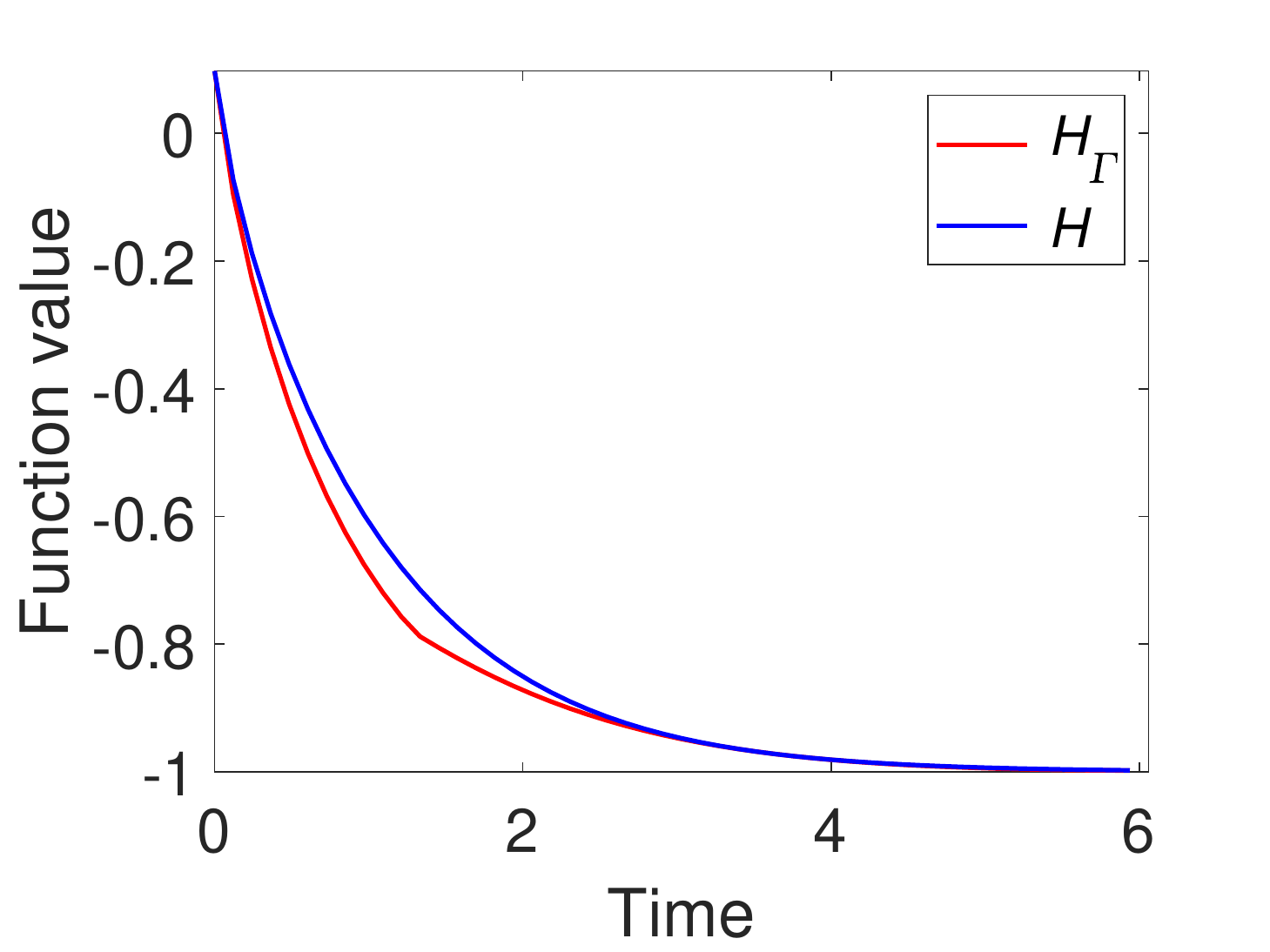}

(d)
\includegraphics[width=0.31\textwidth]{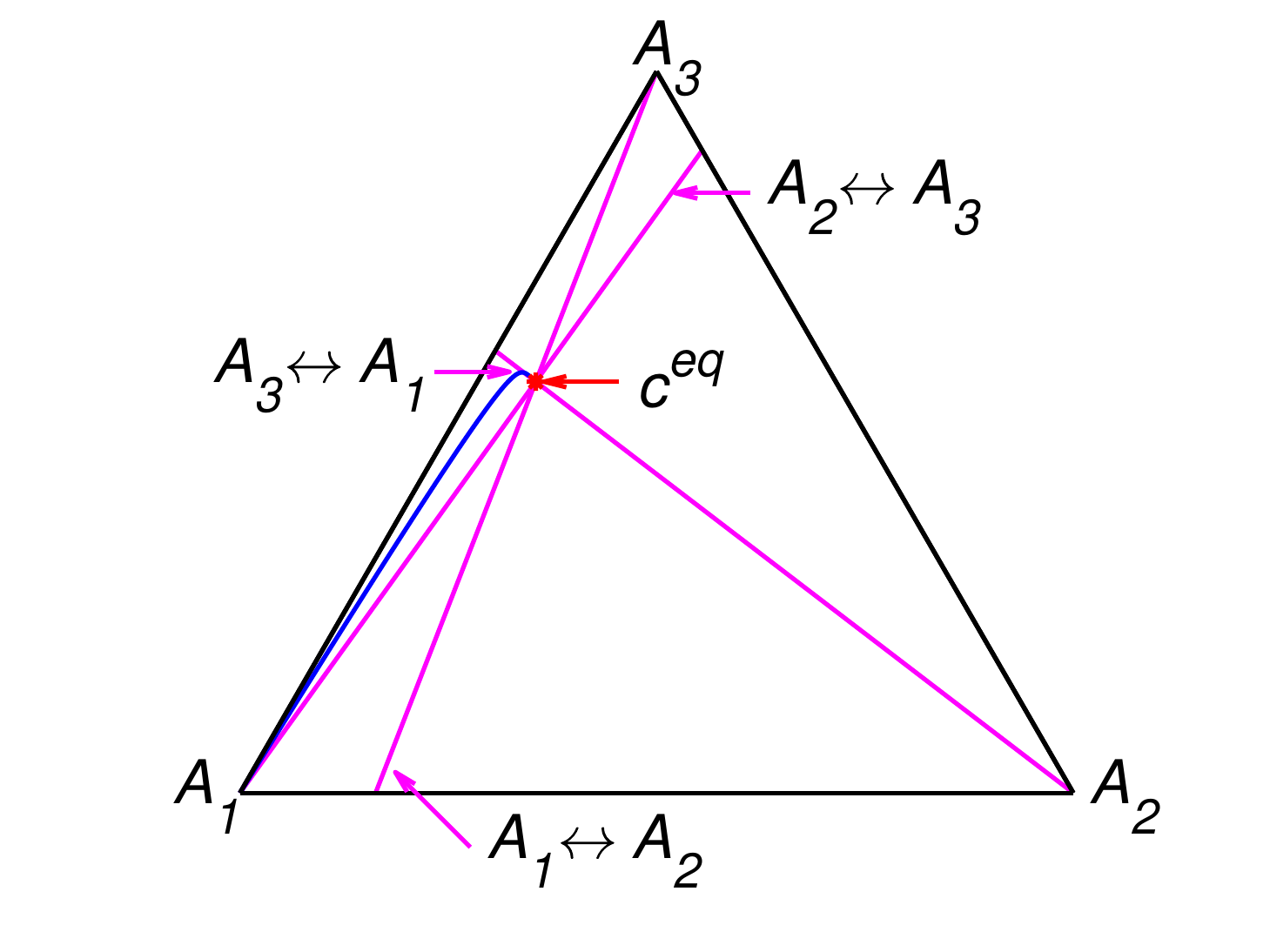}
\includegraphics[width=0.31\textwidth]{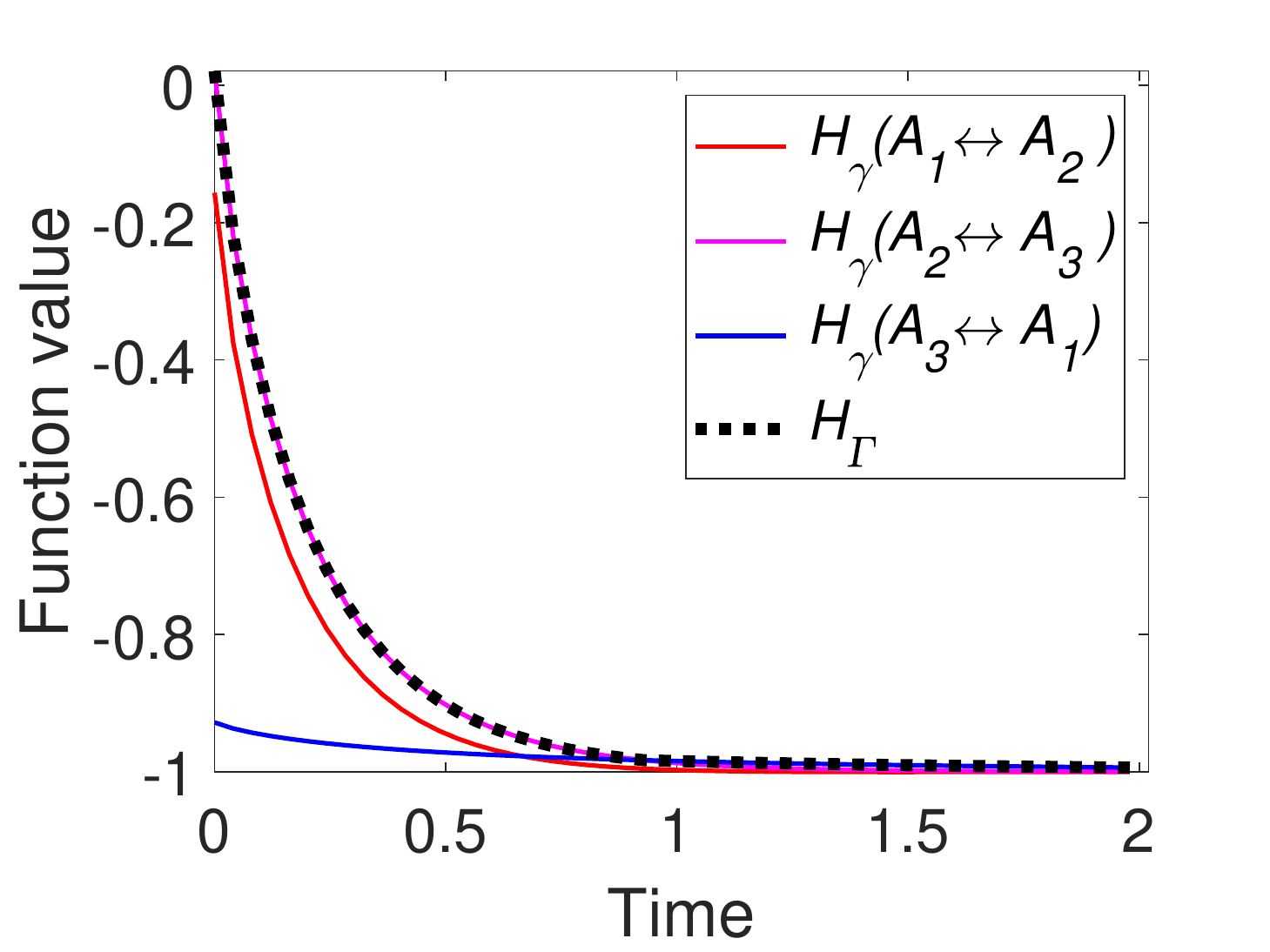}
\includegraphics[width=0.31\textwidth]{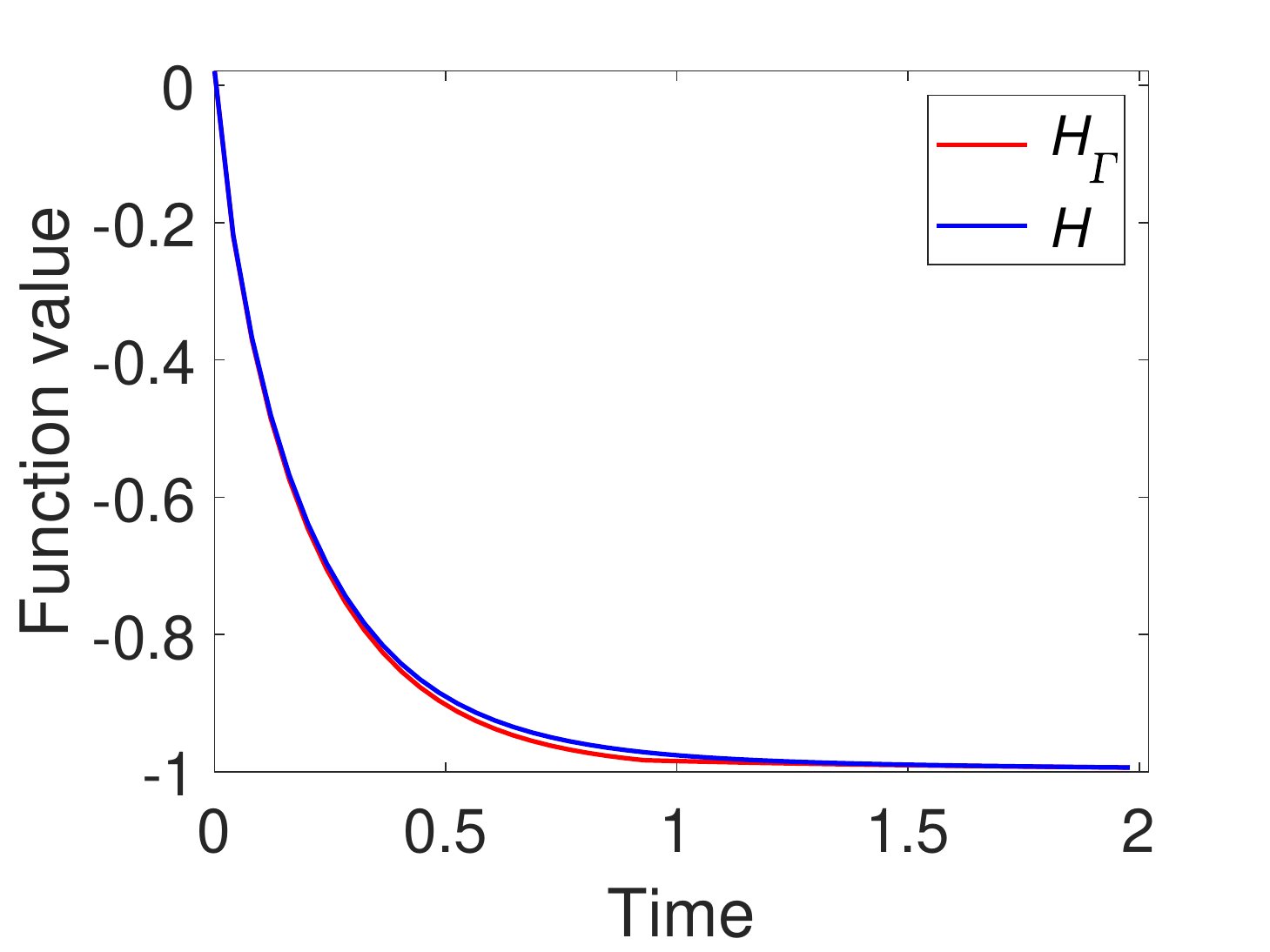}

\caption{The left column presents the trajectories of the system (\ref{EqLinKin}) in the phase plane, the middle column contains graphs of $H_{\gamma_i}$ and $H_\Gamma$ versus time, and the right column depicts graphs of Boltzmann's $H$ and Gorban's $H_\Gamma$ versus time. Each row present system with one set of parameters: (\textbf{a}) Set S1.2 with detailed balance, (\textbf{b}) Set S1.1 without detailed balance, (\textbf{c}) Set S1.2 without detailed balance, and (\textbf{d}) Set S3.2 with detailed balance.}
\label{FigLin}
\end{figure}

The results of simulation of system (\ref{EqLinKin}) with the parameters listed in Table~\ref{TabLinMod} are partially presented in Figure~\ref{FigLin}. All other figures can be found online in \cite{MirkesGit}. For a system with detailed balance, the reaction rate constants $ k_1 ^ +, k_2 ^ +, k_3 ^ + $ presented in the Table~\ref{TabLinMod} were used. We can see the different behaviour of the two $H$ functions. For the system with detailed balance, equal equilibrium concentrations and equal reaction rate constants of direct reactions (Set S1.2, Figure~\ref{FigLin}a) there is no apparent difference between $H$ and $H_\Gamma$ and $H_\Gamma=H_{\gamma_1}=H_{\gamma_3}$ all the time. A system with a set of parameters S1.1 demonstrates the difference between $H$ and $H_\Gamma$: there is the switch from $H_{\gamma_2}$ to $H_{\gamma_3}$ (see Figure~\ref{FigLin}b). Figure~\ref{FigLin}c also demonstrates the difference between $H$ and $H_\Gamma$ and the switch from $H_{\gamma_1}$ to $H_{\gamma_2}$. Figure~\ref{FigLin}c demonstrates weak nonmonotonicity of $H_{\gamma_3}$ near the time of 4 seconds.  Figure~\ref{FigLin}d presents a system with detailed balance and a set of parameters S3.2 and demonstrates fast movement from the initial point to the patrial equilibrium of the third reaction (approximately 0.9 seconds, defined by switch from $H_{\gamma_2}$ to $H_{\gamma_3}$) and then slowly tends to equilibrium.

It can be concluded that the simplest linear isomerisation cycle demonstrates the coincidence of behaviour of $H$ and $H_\Gamma$ for certain set of parameters (see Figure~\ref{FigLin}a) and the differences between these two Lyapunov functions for other parameters. In the case when the equilibrium is a stable focus (see Figure~\ref{FigLin}c) there are an infinite number of switches between $H_{\gamma_i}$, but the high rate of convergence does not allow this effect to be graphically illustrated. In the case of a see a stable node as equilibrium (see Figure~\ref{FigLin}b,d) we can observe only a finite number (usually one or two) of switches.

\subsection{Nonlinear Isomerisation Reaction \label{SectionlIsom}}

Let us consider isomerisation reaction

\begin{equation}\label{SystIsom2}
A_1\rightleftharpoons A_2 \rightleftharpoons A_3,\;\; 2A_1\rightleftharpoons A_2+A_3.
\end{equation}

There is one conservation law for this system: $c_1+c_2+c_3=b$. The lines of partial equilibrium for the first two stoichiometric vectors are defined by (\ref{GenIsom}). For example, for the first reaction, this partial equilibrium line is

\begin{equation*}
\begin{split}
c^*_1&=\frac{(b-c_3^{})c^{\rm eq}_1}{c^{\rm eq}_1+c^{\rm eq}_2},\\
c^*_2&=\frac{(b-c_3^{})c^{\rm eq}_2}{c^{\rm eq}_1+c^{\rm eq}_2},\\
c^*_3&=c^{}_3.
\end{split}
\end{equation*}

For the last reaction the  partial equilibrium is defined by (\ref{Gen112}):

\begin{equation*}
\begin{split}
c^*_1&=\frac{-b_2+\sqrt{(k+1)b^2-k(c_3-c_2)^2}}{k},\\
c^*_2&=\frac{k(b+c_2-c_3)+b-\sqrt{(k+1)b^2-k(c_3-c_2)^2}}{2k},\\
c^*_3&=\frac{k(b+c_3-c_2)+b-\sqrt{(k+1)b^2-k(c_3-c_2)^2}}{2k},
\end{split}
\end{equation*}
where

$$ k = 4\frac{c_2^{\rm eq}c_3^{\rm eq}}{\big(c_1^{\rm eq}\big)^2}-1.$$

\begin{figure}[htb]
\centering
(a)\includegraphics[width=0.3\textwidth]{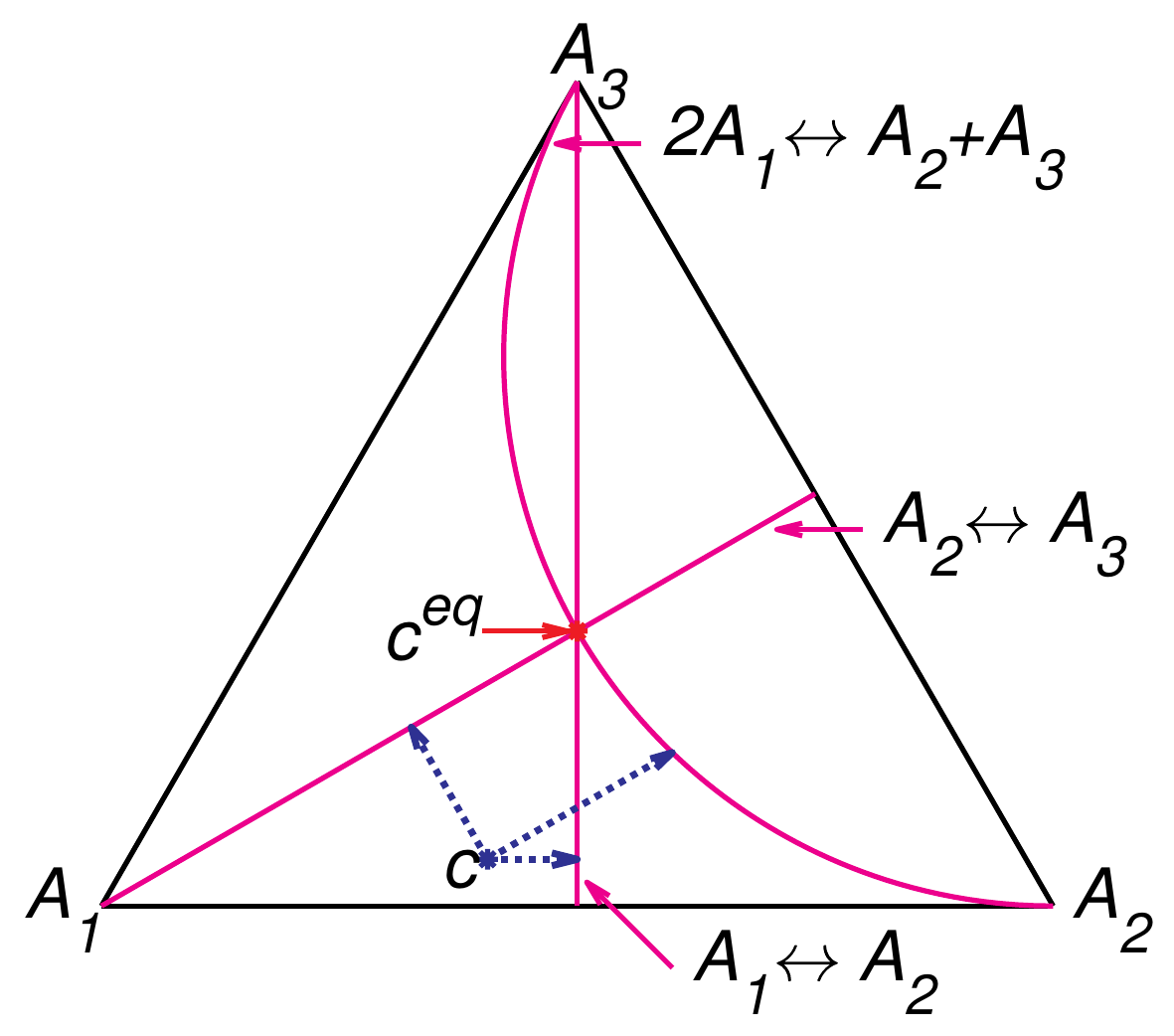}
(b)\includegraphics[width=0.3\textwidth]{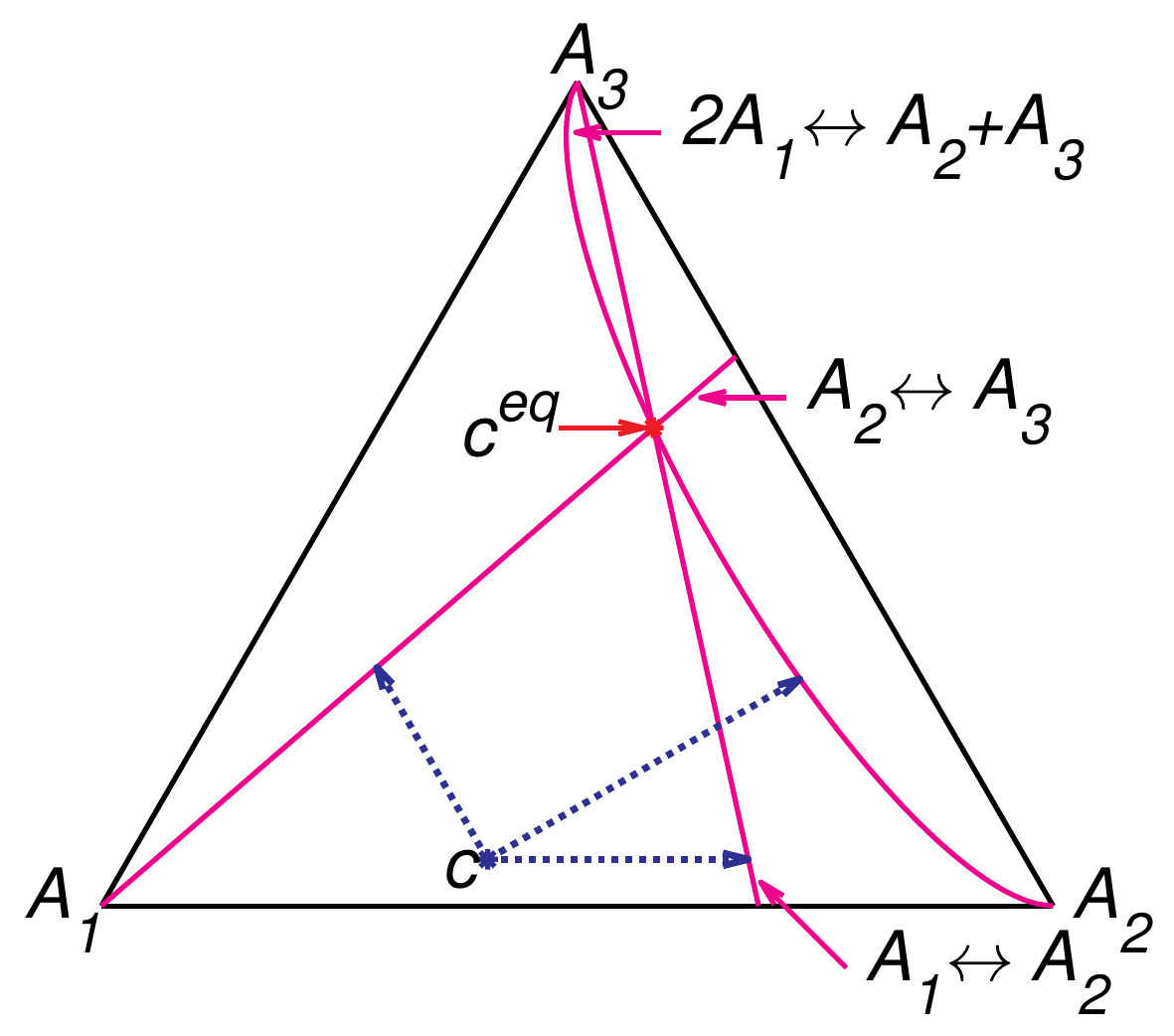}
(c)\includegraphics[width=0.3\textwidth]{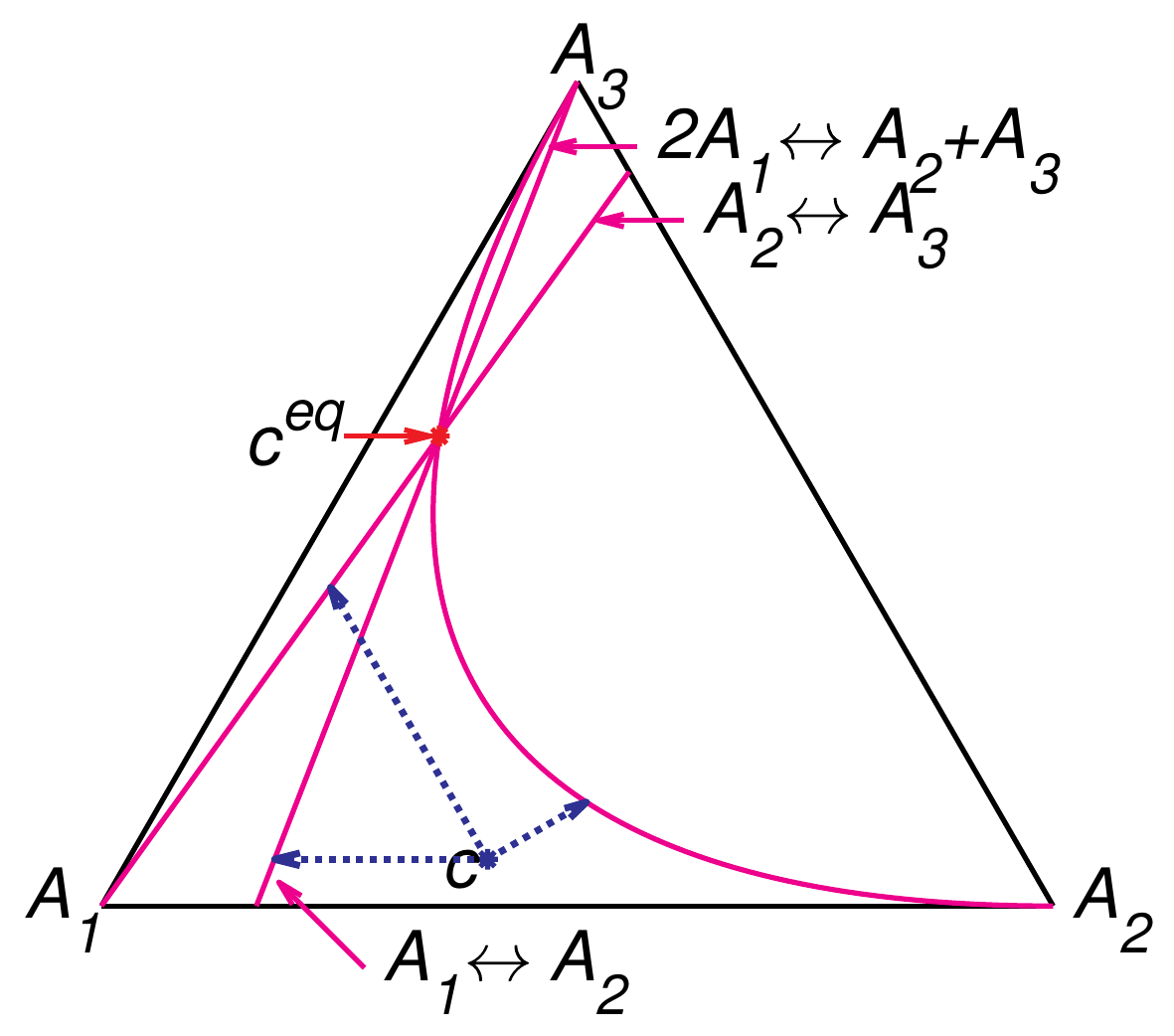}

\caption{Partial equilibrium lines (solid magenta lines) and points of partial equilibrium for point c (dotted arrows) for the reaction system $A_1\rightleftharpoons A_2 \rightleftharpoons A_3 \rightleftharpoons A_1$ with several equilibria: (\textbf{a}) $c^{\rm eq}=(1/3,1/3,1/3)$, (\textbf{b}) $c^{\rm eq}=(0.13,0.29,0.58)$, and (\textbf{c}) $c^{\rm eq}=(0.36,0.07,0.57)$.}
\label{FigNLinEquil}
\end{figure}

\begin{figure}[htb]
\centering
(a)\includegraphics[width=0.3\textwidth]{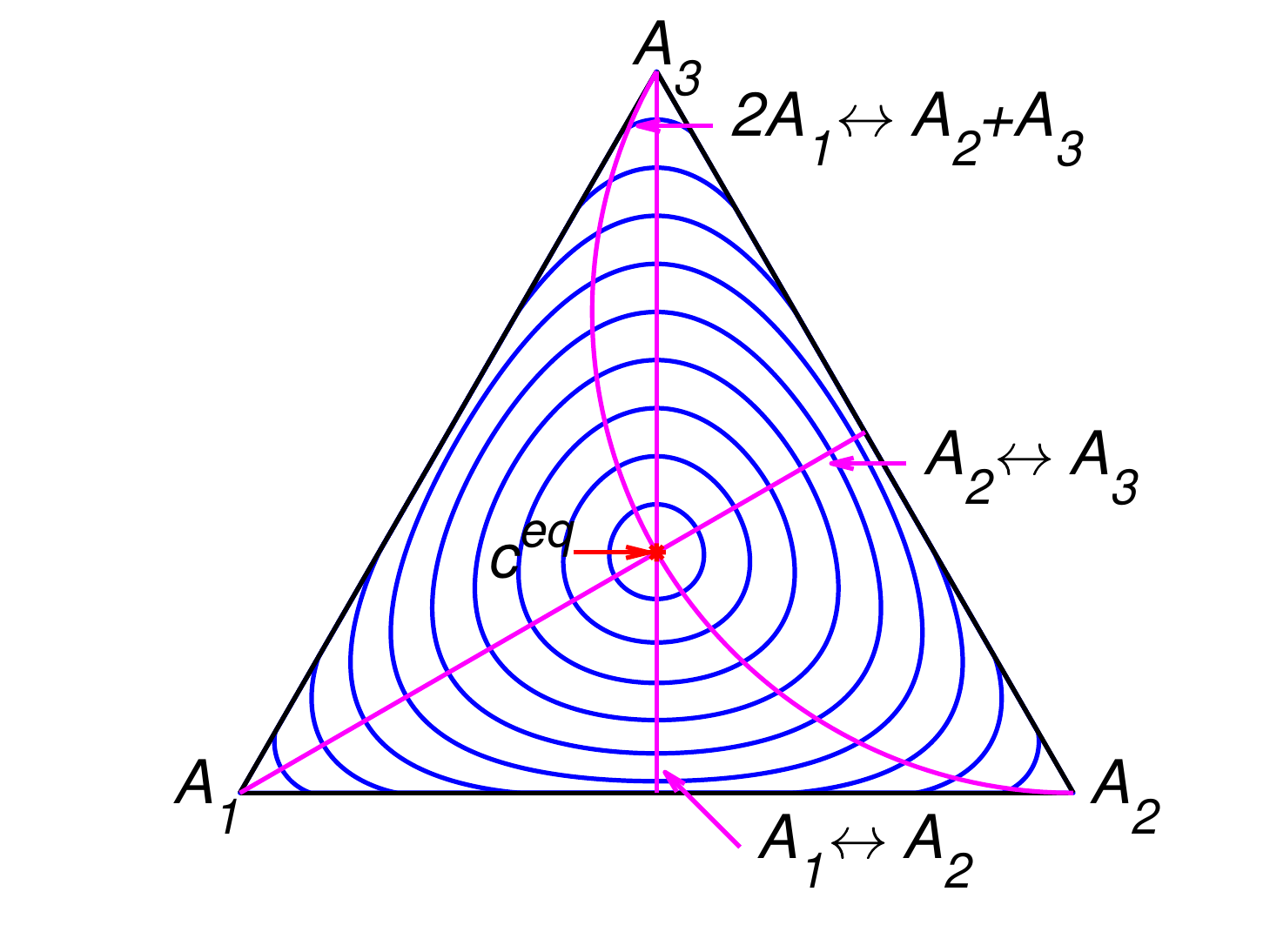}
(b)\includegraphics[width=0.3\textwidth]{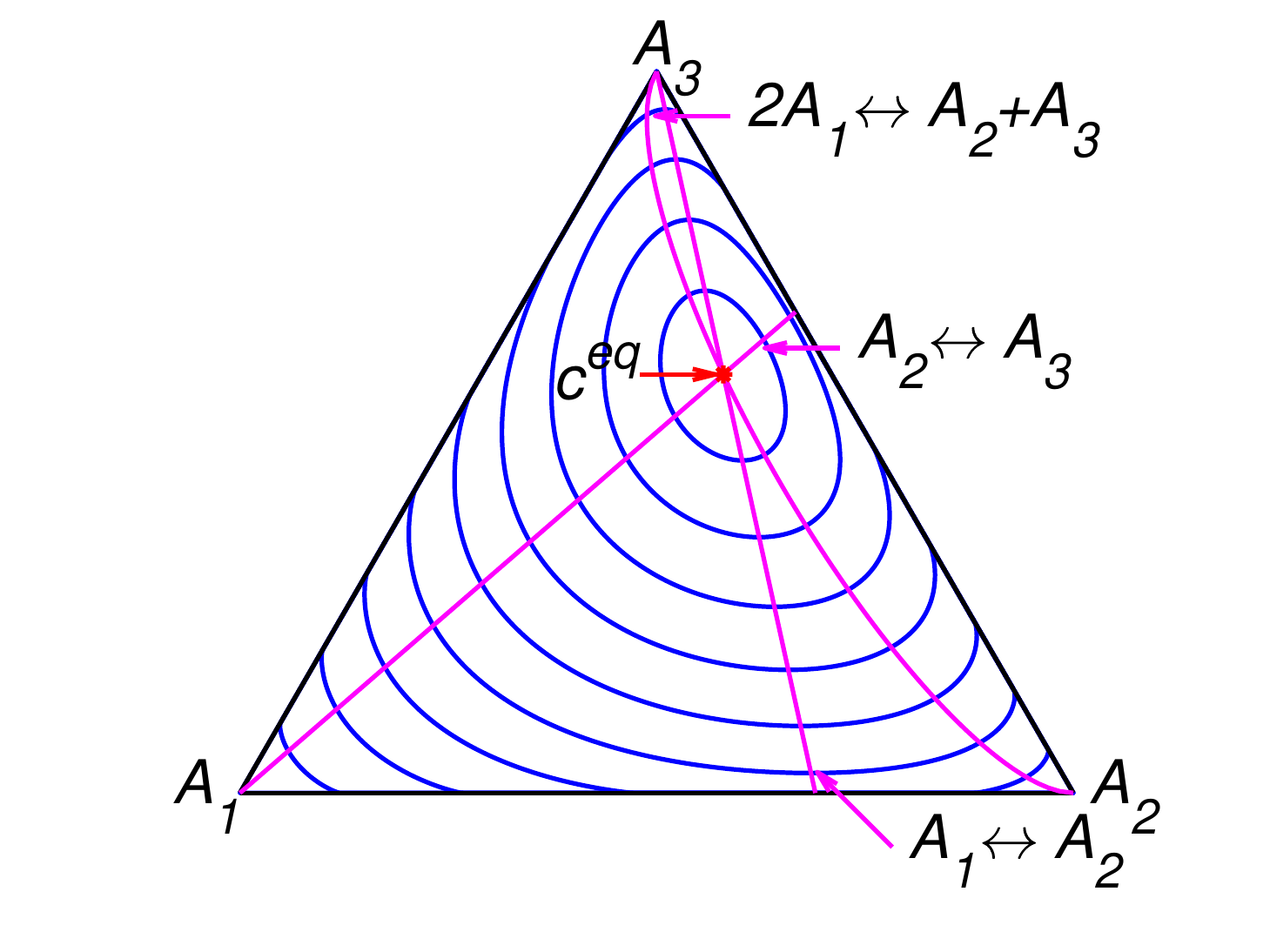}
(c)\includegraphics[width=0.3\textwidth]{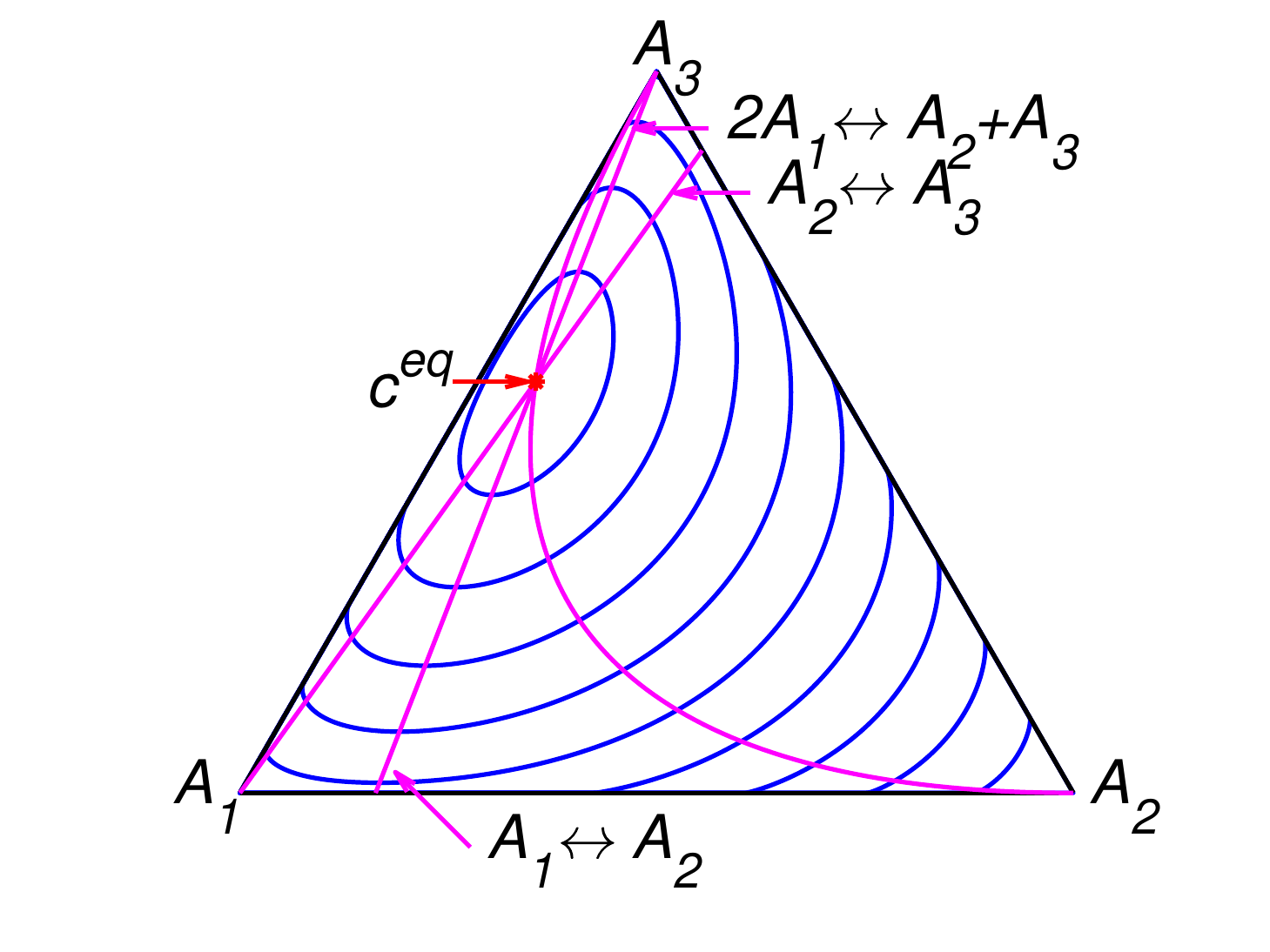}\\
(d)\includegraphics[width=0.3\textwidth]{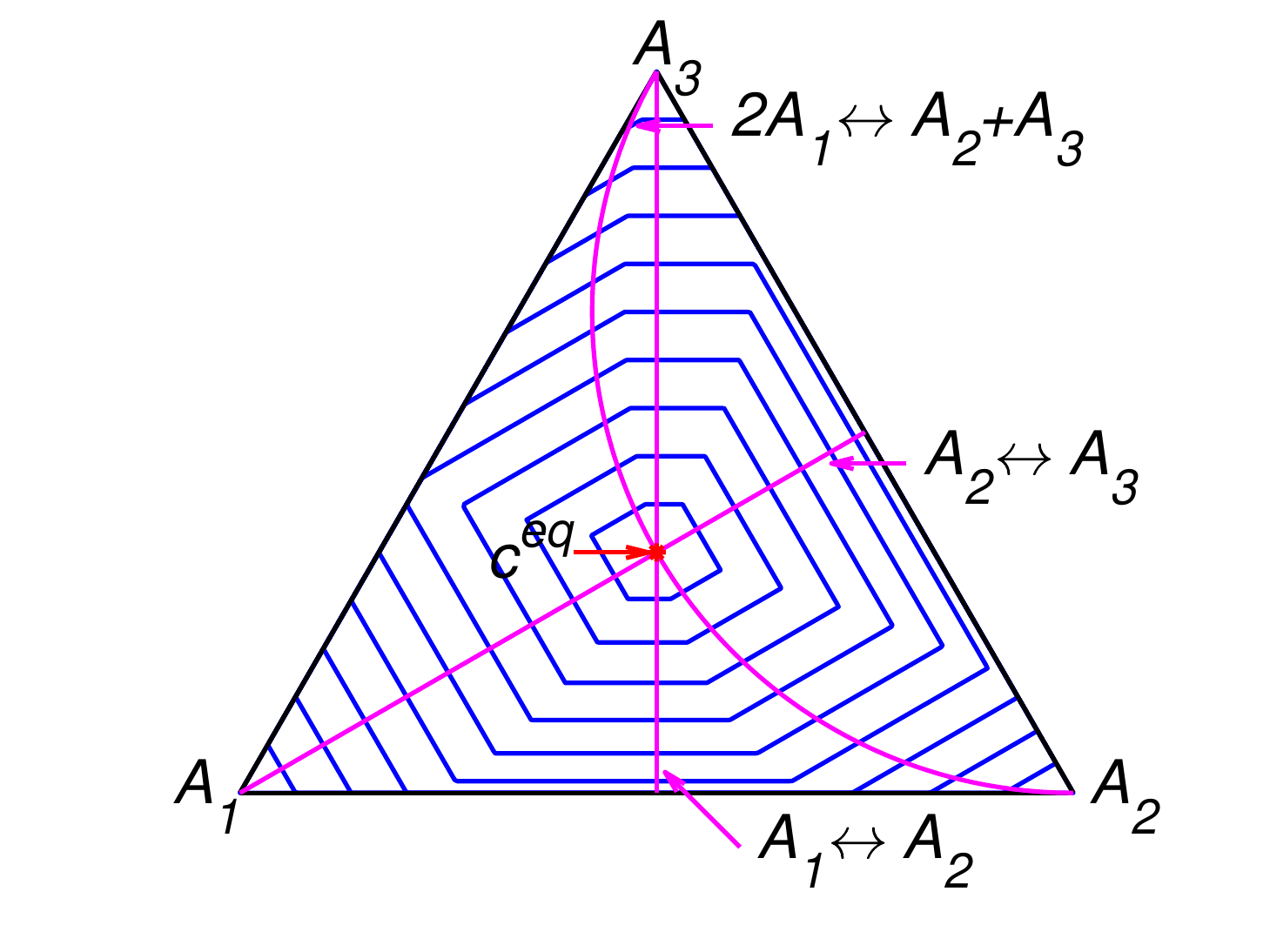}
(e)\includegraphics[width=0.3\textwidth]{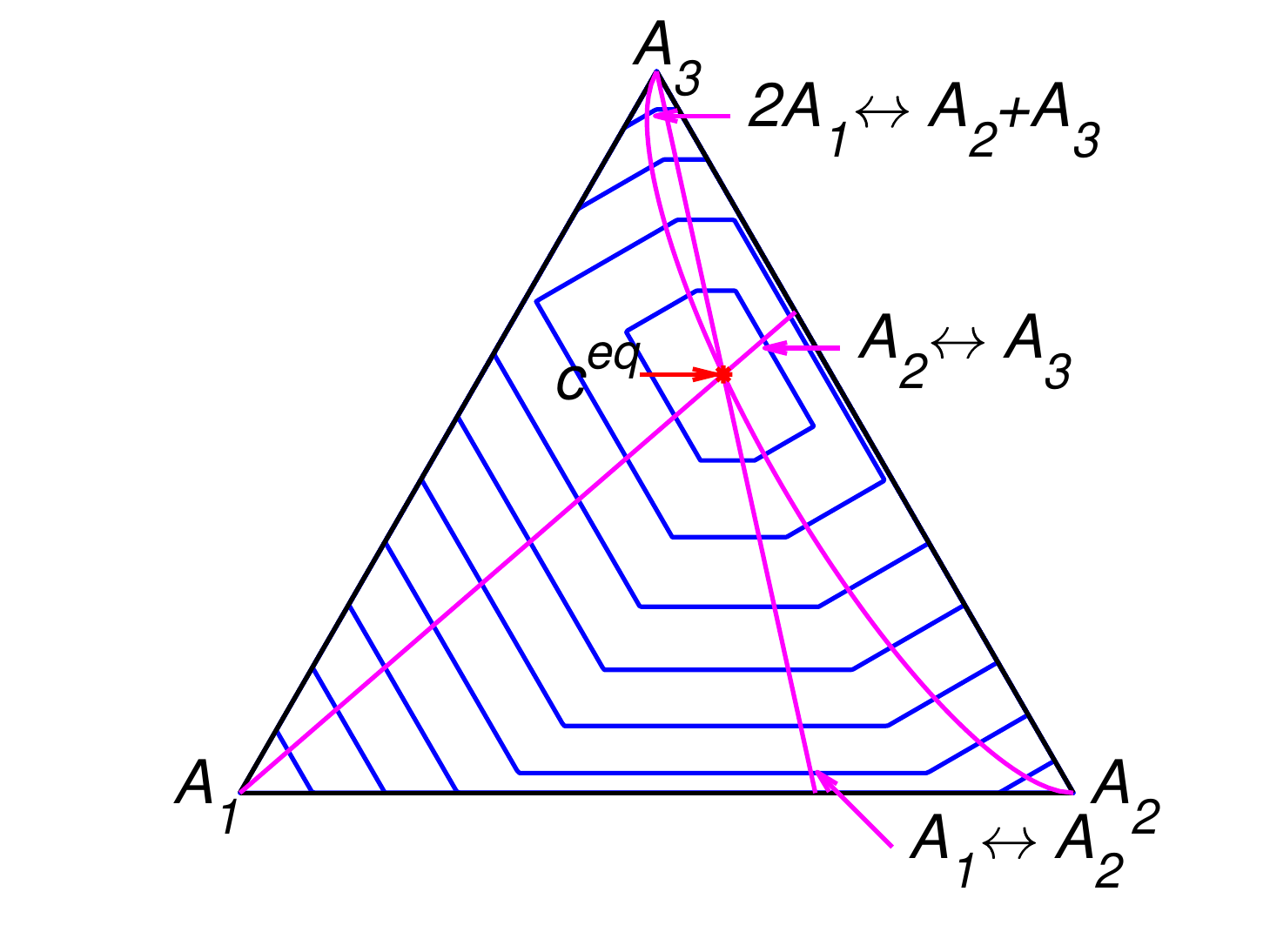}
(f)\includegraphics[width=0.3\textwidth]{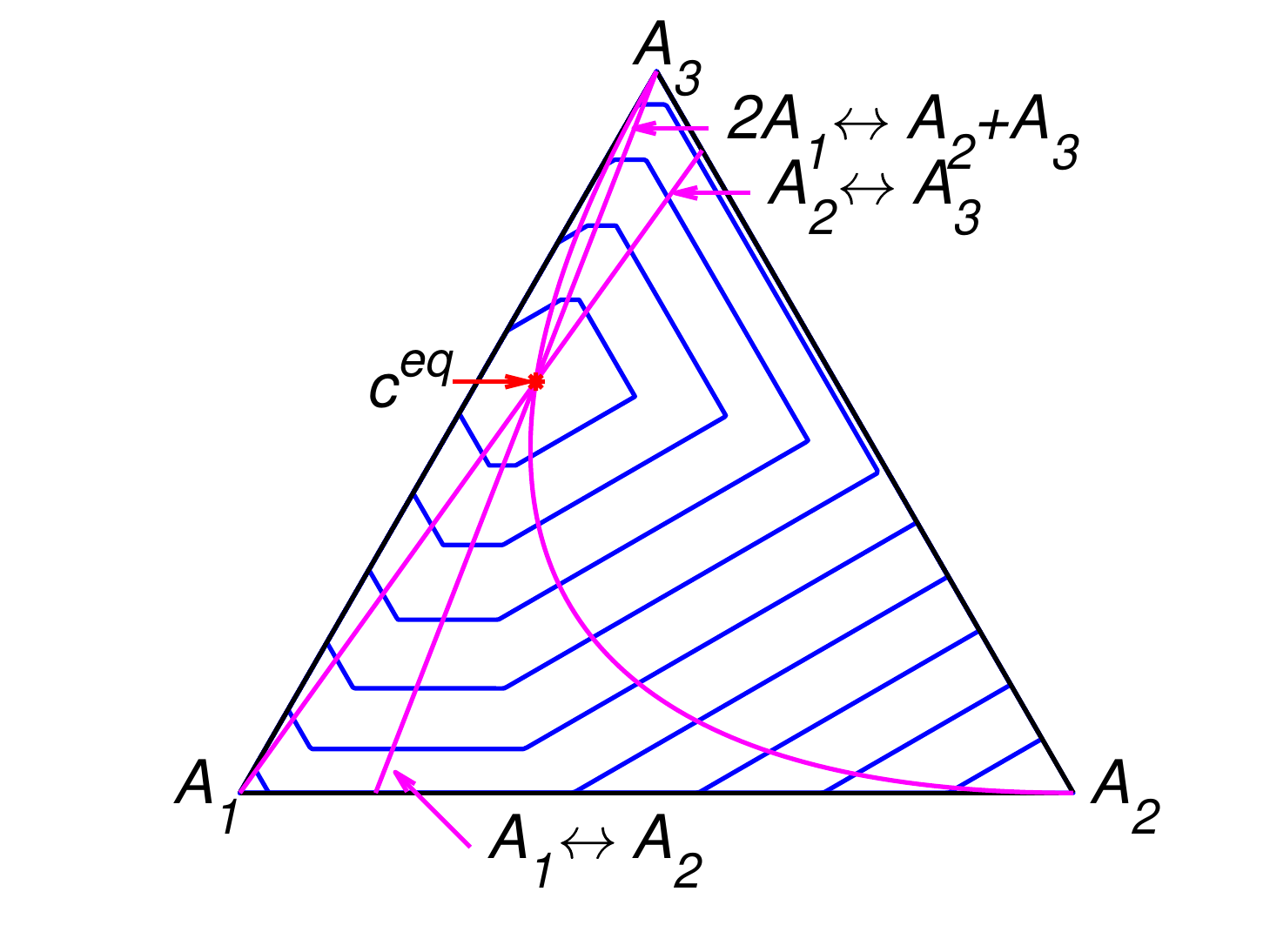}

\caption{The level sets for Boltzmann's $H$ function in top row and the corresponding level sets for Gorban's $H_\Gamma$ function in bottom row for several equilibria: (\textbf{a, d}) $c^{\rm eq}=(1/3,1/3,1/3)$, (\textbf{b, e}) $c^{\rm eq}=(0.13,0.29,0.58)$, and (\textbf{c, f}) $c^{\rm eq}=(0.36,0.07,0.57)$.}
\label{FigNLinHBHG}
\end{figure}

The lines of partial equilibrium and partial equilibrium points for a given point $c$ are presented in Figure~\ref{FigNLinEquil}.
The level sets for Boltzmann's $H$ function and Gorban's $H_\Gamma$ function are presented in Figure~\ref{FigNLinHBHG}.  It is important to emphasise that these level sets are independent of kinetic constants and are completely determined by the equilibrium for Boltzmann's $H$ function (the same level sets in Figure~\ref{FigLinHBHG}a--c and Figure~\ref{FigNLinHBHG}a--c) and by the equilibrium and set of stoichiometric vectors $\Gamma$ for Gorban's $H_\Gamma$ function (different level sets in Figure~\ref{FigLinHBHG}d--f and Figure~\ref{FigNLinHBHG}d--f).

The kinetic equations for the system (\ref{SystIsom2}) are:

\begin{equation}\label{EqNLinKin}
\begin{split}
\frac{\mathrm{d}c_1}{\mathrm{d}t}&=-k^+_1c_1+k^-_1c_2-2k^+_3c_1^2+2k^-_3c_2c_3,\\
\frac{\mathrm{d}c_3}{\mathrm{d}t}&=k^+_2c_2-k^-_2c_3+k^+_3c_1^2-k^-_3c_2c_3,\\
c_2&=b-c_1-c_3.
\end{split}
\end{equation}

For system (\ref{SystIsom2}) with detailed balance the conditions for the reaction rate constants are:

$$k^+_1c_1^{\rm eq}=k^-_1c_2^{\rm eq},\;\;  k^+_2c_2^{\rm eq}=k^-_2c_3^{\rm eq},\;\;  k^+_3(c_1^{\rm eq})^2=k^-_3c_2^{\rm eq}c_3^{\rm eq}.$$

The system can be completely parametrised by three equilibrium concentrations $c^{\rm eq}_i$ and three reaction rate constants, for example, by the constants $k^+_1, k^+_2, k^+_3$.
To obtain the complex balance condition it is necessary to list all the different stoichiometric vectors $\alpha_\rho$ and $\beta_\rho$:

\begin{equation*}
\begin{split}
\alpha_{1}=\beta_{-1}&=(1,0,0),\\
\alpha_{-1}=\alpha_{2}=\beta_{1}=\beta_{-2}&=(0,1,0),\\
\alpha_{-2}=\beta_{2}&=(0,0,1),\\
\alpha_{3}=\beta_{-3}&=(2,0,0),\\
\alpha_{-3}=\beta_{3}&=(0,1,1).
\end{split}
\end{equation*}

The conditions of  complex balance are

\begin{equation}\label{EqnLinComplex}
\begin{split}
k^+_1c_1^{\rm eq}&=k^-_1c_2^{\rm eq},\\
k^-_1c_2^{\rm eq}+k^+_2c_2^{\rm eq}&=k^+_1c_1^{\rm eq}+k^-_2c_3^{\rm eq},\\
k^-_1c_3^{\rm eq}&=k^+_2c_2^{\rm eq},\\
k^+_3(c_1^{\rm eq})^2&=k^-_3c_2^{\rm eq}c_3^{\rm eq},\\
k^-_3c_2^{\rm eq}c_3^{\rm eq}&=k^+_3(c_1^{\rm eq})^2.
\end{split}
\end{equation}

We can see that the first, third and fourth complex balance conditions for this system are equivalent to detailed balance conditions. This means that for system (\ref{EqNLinKin}) the detailed and complex balances are the same. For simulation of system (\ref{EqNLinKin}), we selected three equilibria and four sets of reaction rate constants, presented in Table~\ref{TabNLinMod}.

\begin{table}[htb]
\caption{Set of equilibrium concentrations and set of reaction rate constants for simulation of system (\ref{EqNLinKin}).}
\label{TabNLinMod}
\centering
\begin{tabular}{ccc}
\toprule
\textbf{$c^{\rm eq}_1$}	& \textbf{$c^{\rm eq}_2$} & \textbf{$c^{\rm eq}_3$}\\
\midrule
1/3  & 1/3  & 1/3\\
0.13 & 0.29 & 0.58\\
0.36 & 0.07 & 0.57\\
\bottomrule
\end{tabular}
\quad\quad
\begin{tabular}{ccc}
\toprule
\textbf{$k^+_1+k^-_1$} & \textbf{$k^+_2+k^-_2$} & \textbf{$k^+_3+k^-_3$}\\
\midrule
1  & 1 & 1\\
10 & 1 & 1\\
10 & 5 & 1\\
10 & 1 & 5\\
\bottomrule
\end{tabular}

\end{table}

\begin{figure}[H]
\centering
(a)
\includegraphics[width=0.31\textwidth]{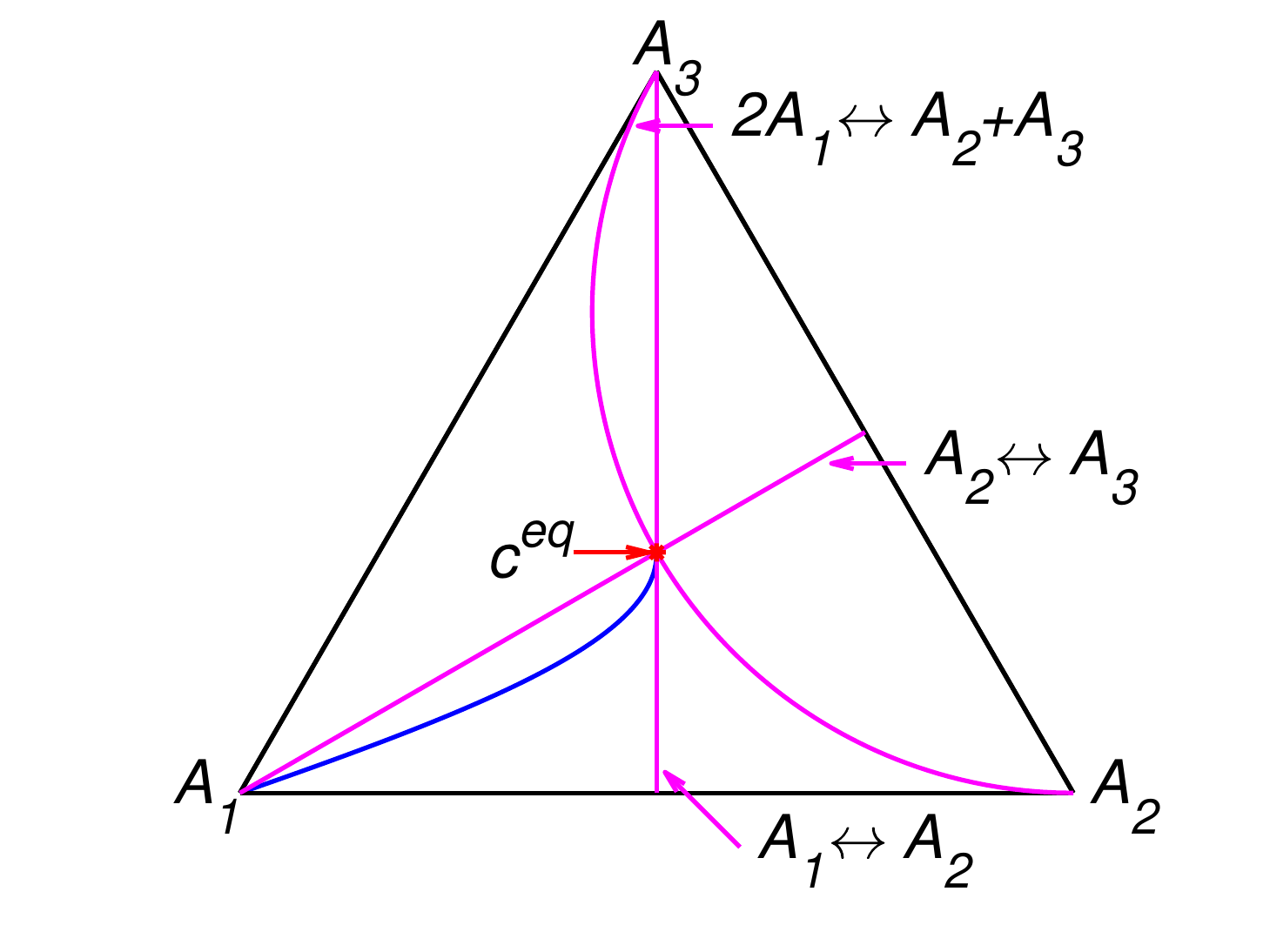}
\includegraphics[width=0.31\textwidth]{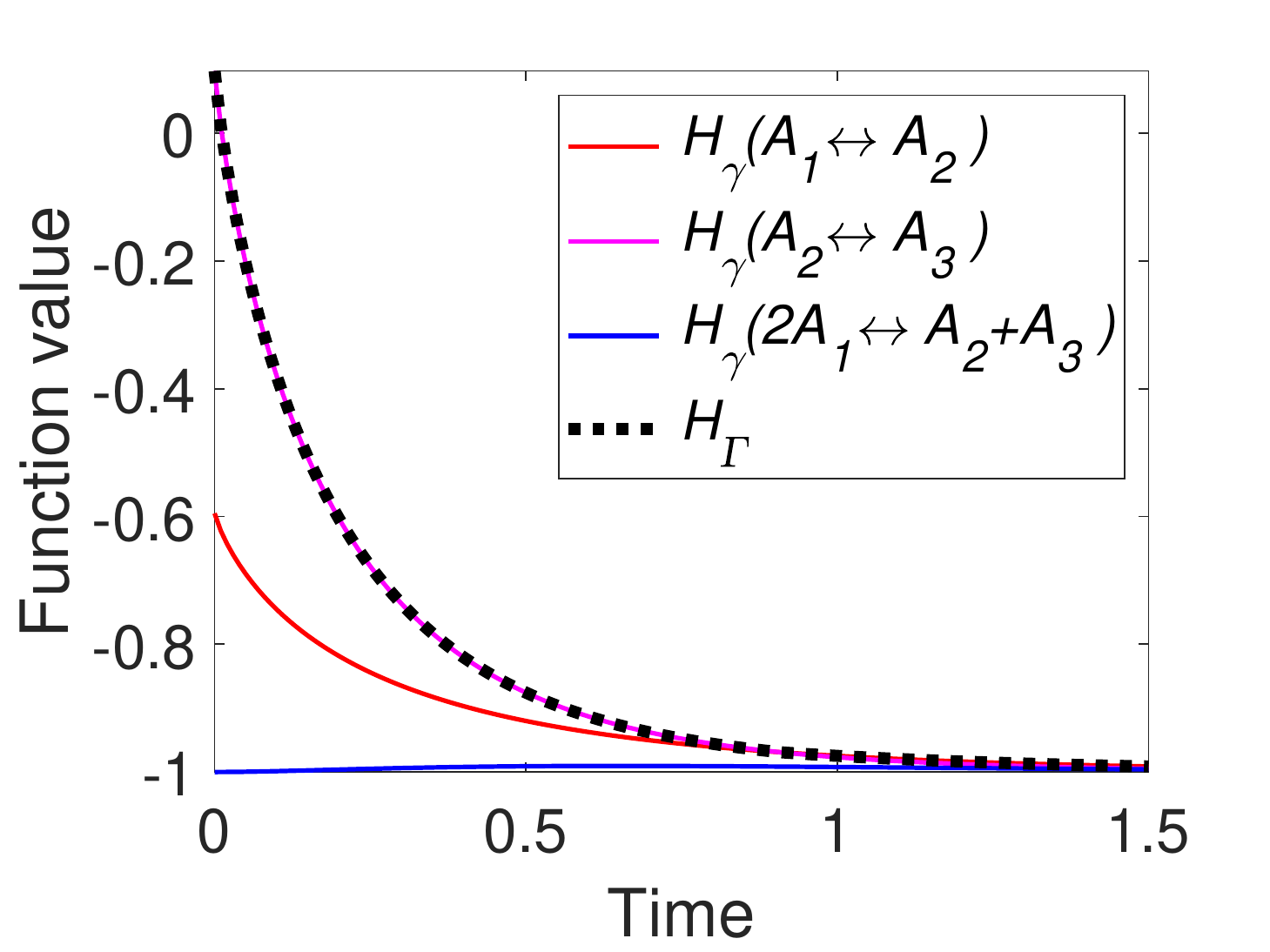}
\includegraphics[width=0.31\textwidth]{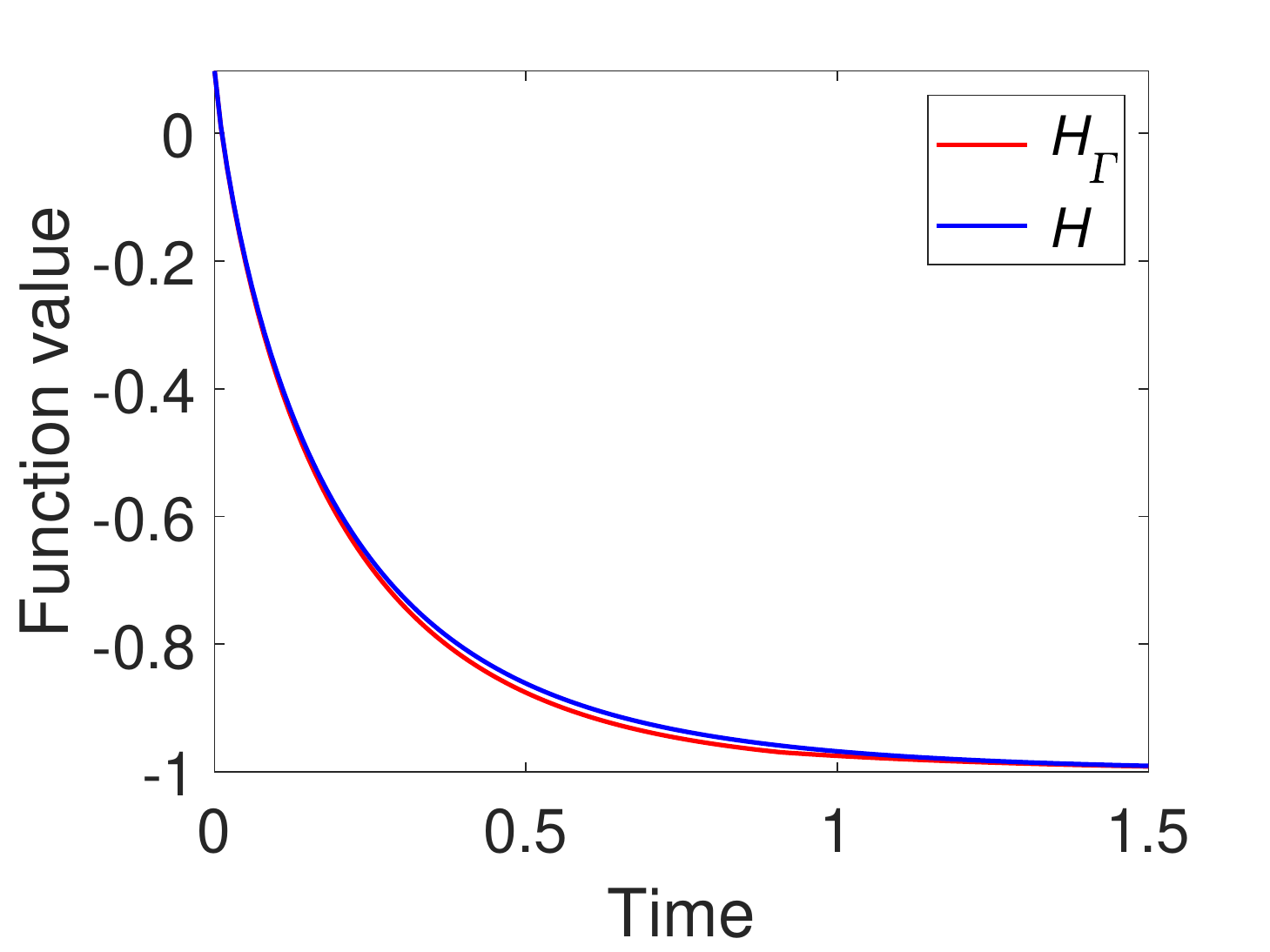}

(b)
\includegraphics[width=0.31\textwidth]{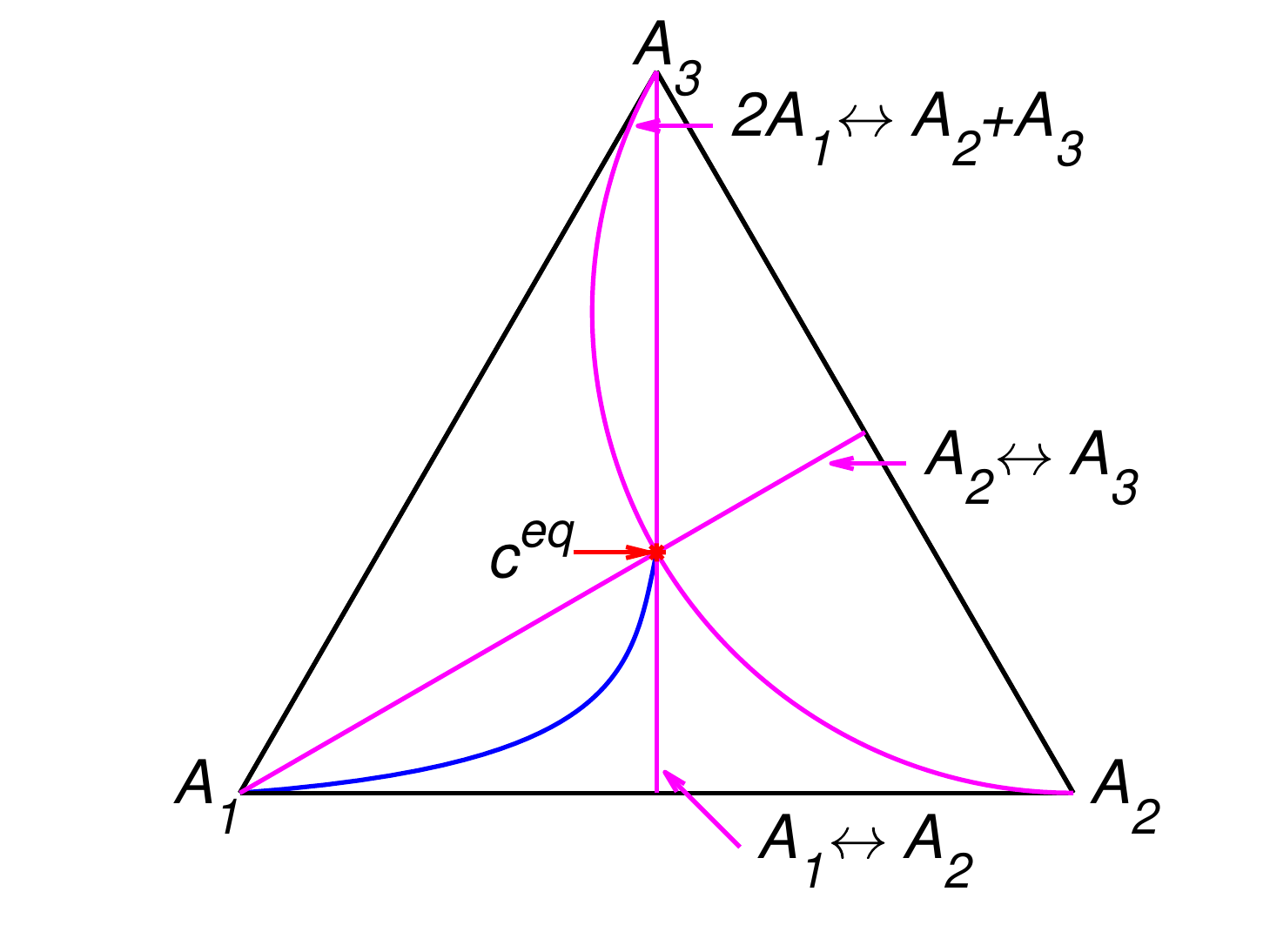}
\includegraphics[width=0.31\textwidth]{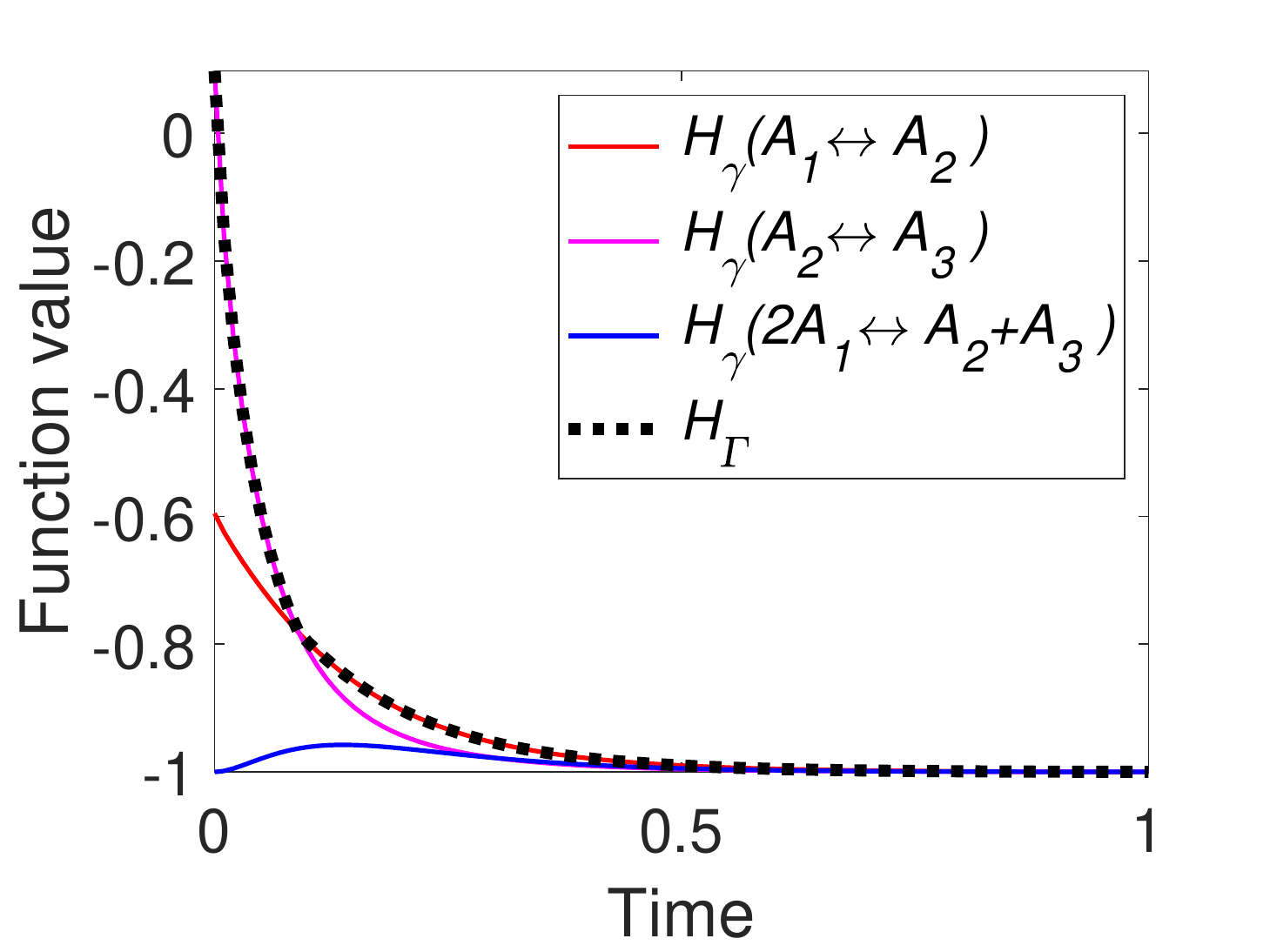}
\includegraphics[width=0.31\textwidth]{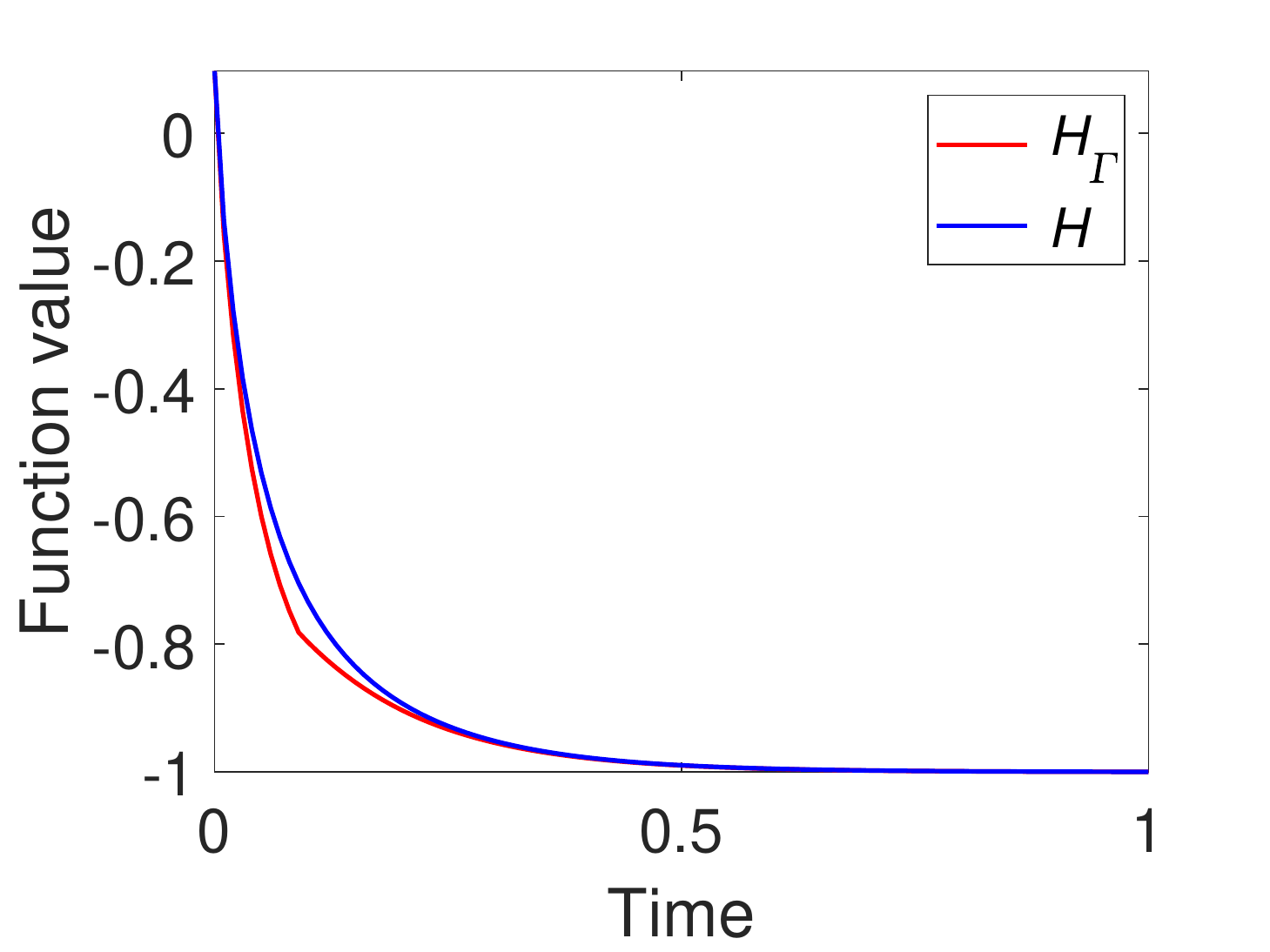}

(c)
\includegraphics[width=0.31\textwidth]{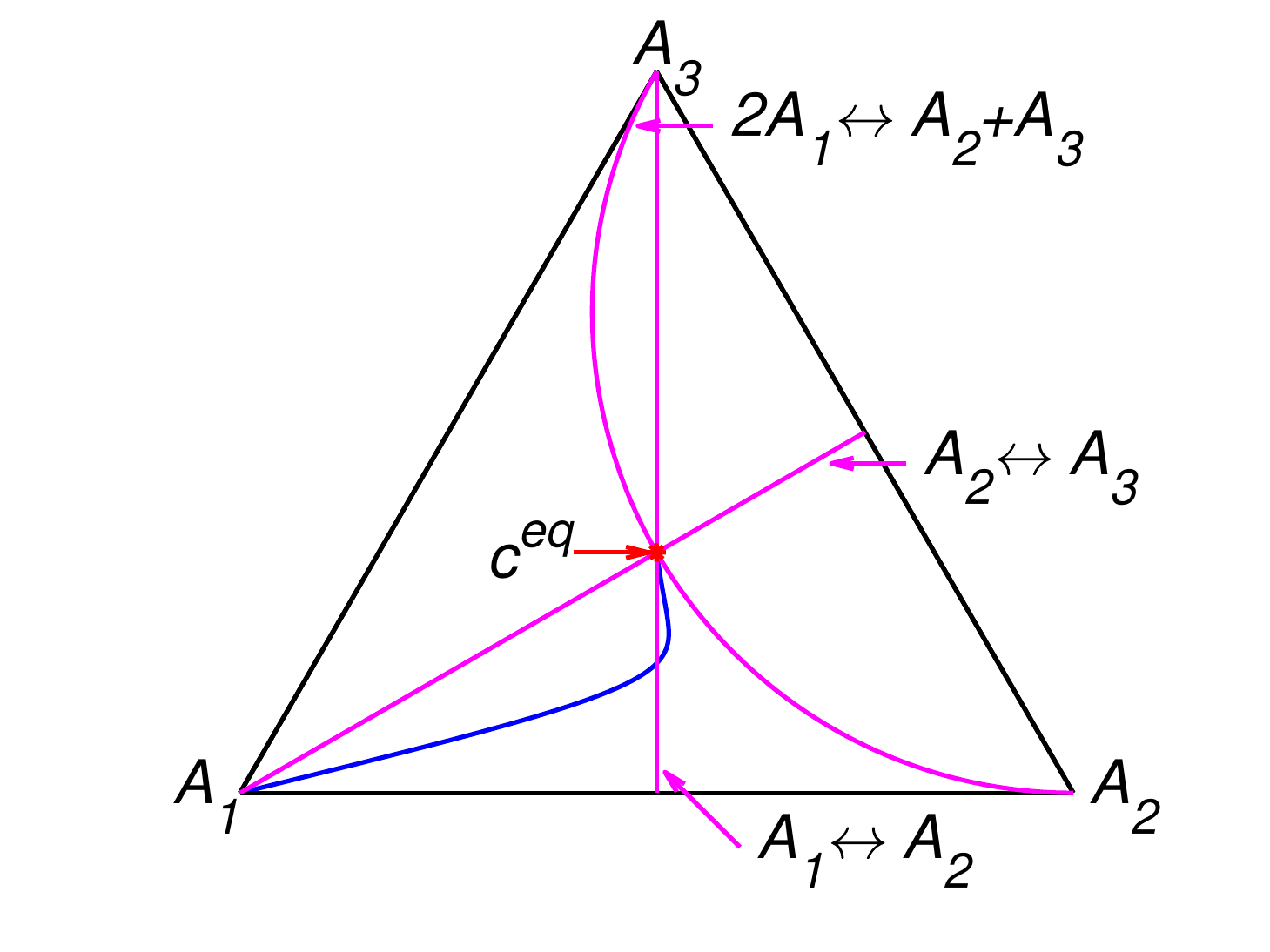}
\includegraphics[width=0.31\textwidth]{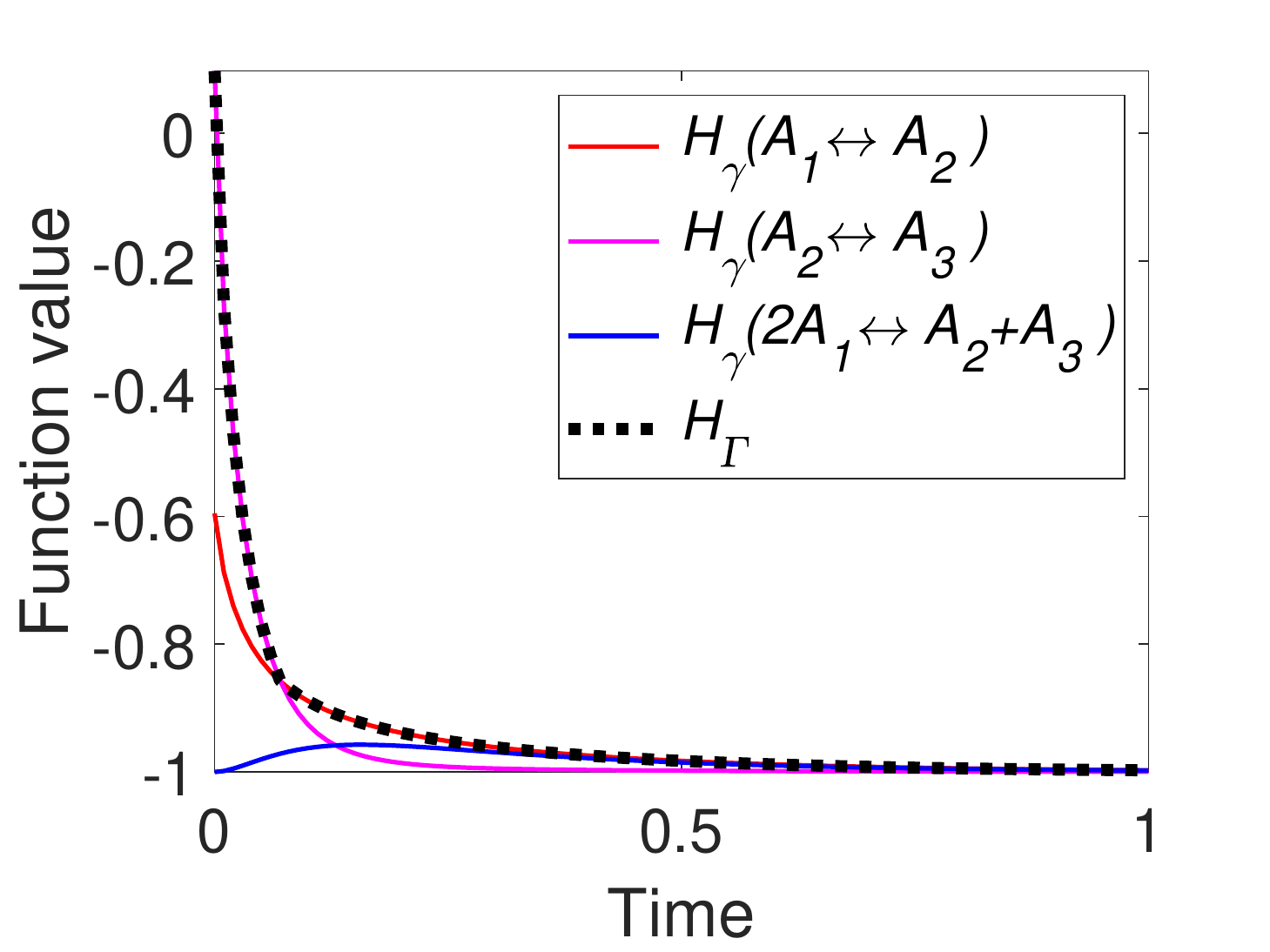}
\includegraphics[width=0.31\textwidth]{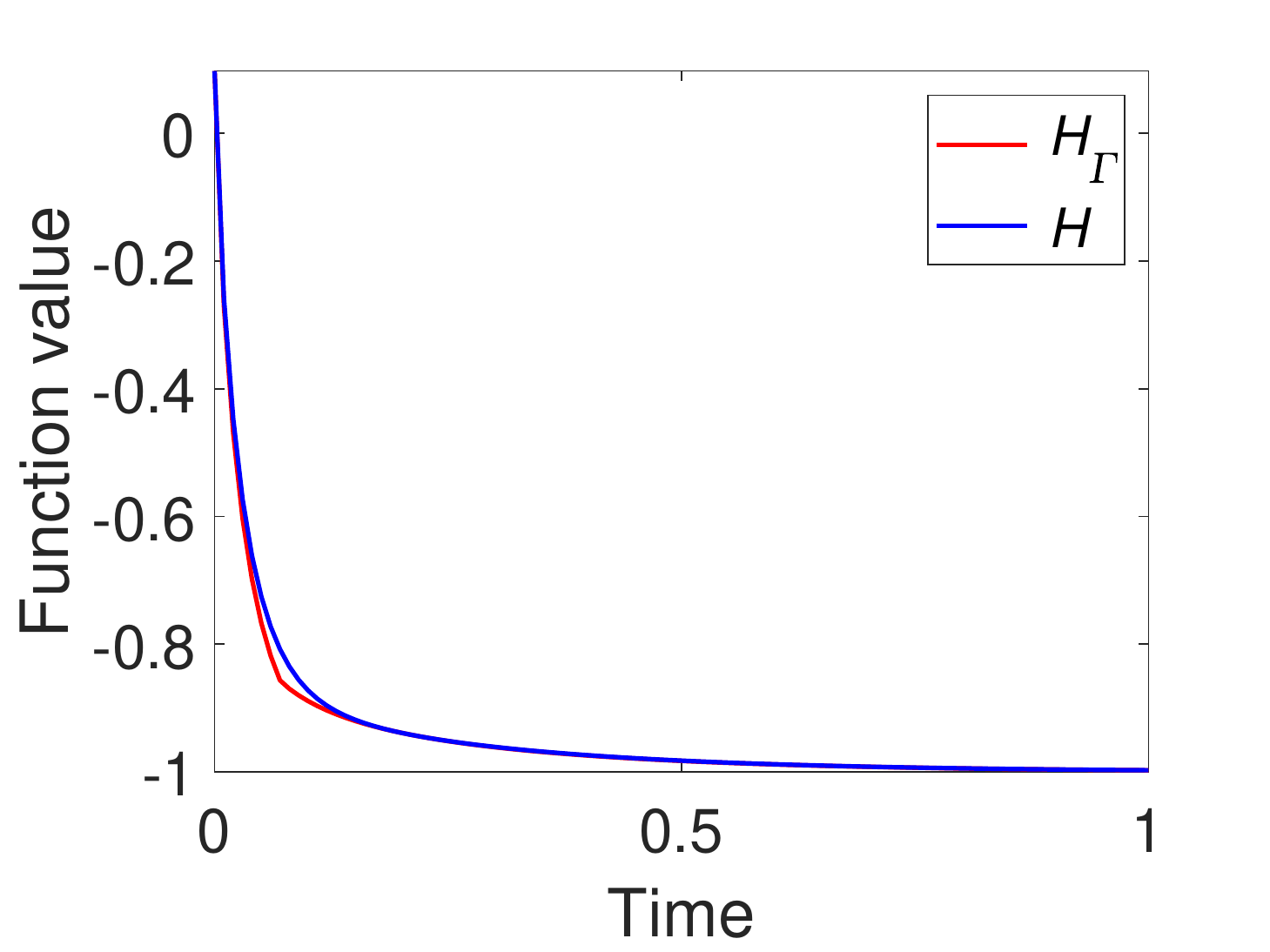}

\caption{The left column presents trajectories of system (\ref{EqLinKin}) in the phase plane, the middle column contains graphs of $H_{\gamma_i}$ and $H_\Gamma$ versus time, and the right column depicts graphs of Boltzmann's $H$ and Gorban's $H_\Gamma$ versus time. Each row present system with equal equilibrium concentrations and different sum of reaction rate constants of direct and inverse reactions: (\textbf{a}) $k^++k^-=(1,1,1)$, (\textbf{b}) $k^++k^-=(10,5,1)$, (\textbf{c}) $k^++k^-=(10,1,5)$.}
\label{FigNLin}
\end{figure}

Part of the simulation results of system (\ref{EqNLinKin}) with the parameters listed in Table~\ref{TabNLinMod} are presented in Figure~\ref{FigNLin}. All other figures can be found in \cite{MirkesGit}.

We can see the different behaviour of the two $H$ functions. Figure~\ref{FigNLin}a presents the results for a system with equal equilibrium concentrations and equal reaction rate constants of direct reactions. In contrast to the behavior of the linear system (Figure~\ref{FigLin}a), there is a difference between $H$ and $H_\Gamma$ and $H_\Gamma$ switches from $H_{\gamma_2}$ to $H_{\gamma_1}$. All three models in Figure~\ref{FigNLin} demonstrate the non-monotonicity of $H_{\gamma_3}$ and the difference between $H$ and $H_{\Gamma}$. The system in Figure~\ref{FigNLin}b demonstrates the switch $H_\Gamma$ from $H_{\gamma_2}$ to $H_{\gamma_1}$. The system in Figure~\ref{FigNLin}c demonstrates the switch $H_\Gamma$ from $H_{\gamma_2}$ to $H_{\gamma_1}$ and then to $H_{\gamma_3}$. We also see that in this case the trajectory intersects the partial equilibrium of the first reaction and then is attracted back to this partial equilibrium. Opposite to system in Figure~\ref{FigLin}c the equilibrium of this system is a stable node but not a stable focus.

We can conclude that the nonlinear isomerisation reaction demonstrates the difference in the behaviour of $H$ and $H_\Gamma$ for almost all set of parameters. Since the complex balance condition for this system is always is equivalent to the detailed balance condition, the equilibrium point always is a stable node and the number of switches between $H_{\gamma_i}$ is finite and usually equal to one.

\subsection{Water Gas Shift Reaction \label{SectionWGS}}

In this subsection, we consider the famous Water Gas Shift reaction (WGS) \cite{WGS1}. More precisely, we consider the redox mechanism proposed by \cite{WGS2} and described in details in \cite{WGS3, WGS4, WGS5}. In the first part of this subsection, we consider the WGS reaction with arbitrary chosen kinetic parameters. To avoid confusion, we call this reaction `abstract WGS'.
In the last part of this subsection we consider the real WGS reaction with all the parameters defined for this reaction.
The redox mechanism includes six substances: $\mathrm{H}_2\mathrm{O}, \mathrm{H}_2, \mathrm{CO}, \mathrm{CO}_2, \mathrm{red}, \mathrm{Ox}$. For the abstract WGS model, we use the following substances:  $A_1, A_2, A_3, A_4, A_5, A_6$. There are two reactions in the WGS mechanism:

\begin{equation}\label{SystWGS}
\begin{split}
\mathrm{H}_2\mathrm{O}+\mathrm{red}&\rightleftharpoons \mathrm{H}_2+\mathrm{Ox},\\
\mathrm{CO}+\mathrm{Ox}&\rightleftharpoons \mathrm{CO}_2+\mathrm{red}.
\end{split}
\end{equation}

The abstract WGS mechanism include the following reactions:

\begin{equation}\label{SystWGSAb}
\begin{split}
A_1+A_5\rightleftharpoons A_2+A_6,\\
A_3+A_6\rightleftharpoons A_4+A_5.
\end{split}
\end{equation}

Systems (\ref{SystWGS}) and (\ref{SystWGSAb}) have four stoichiometric conservation laws:

\begin{equation}\label{SystWGSAbCons}
\begin{split}
c_1+c_2&=b_\mathrm{H},\\
c_3+c_4&=b_\mathrm{C},\\
c_1+c_3+2c_4+c_6&=b_\mathrm{O},\\
c_5+c_6&=b_\mathrm{A}.
\end{split}
\end{equation}

For the WGS reaction these conservations laws mean the conservation of hydrogen, carbon, oxygen and catalyst (accelerator). For the abstract WGS reaction we use the same names of conservation laws.
For the simulation we choose the following balance values: hydrogen balance $b_\mathrm{H}=1$, carbon balance $b_\mathrm{C}=1$, oxygen balance $b_\mathrm{O}=b_\mathrm{H}+b_\mathrm{C}=2$, and catalyst balance $b_\mathrm{A}=0.5$. These values of balances correspond to one of the standard modes of WGS reaction \cite{WGS3}: "1:1 molar feed ratio $[\mathrm{H}_2\mathrm{O}/\mathrm{CO}]$" without hydrogen and carbon dioxide in the initial composition.
The line of partial equilibrium for both stoichiometric vectors is defined by (\ref{Gen1111}). For example, for the first reaction, the line of partial equilibrium is

\begin{equation*}
\begin{split}
c^*_1&=\frac{b_3+k(b_3-b_1)-\sqrt{(k+1)b_3^2+2k^2b_2^2-kb_1^2}}{2k},\\
c^*_2&=\frac{-b_3-kb_2+\sqrt{(k+1)b_3^2+2k^2b_2^2-kb_1^2}}{2k},\\
c^*_3&=c^{}_3,\\
c^*_4&=c^{}_4,\\
c^*_5&=\frac{b_3+k(b_3+b_1)-\sqrt{(k+1)b_3^2+2k^2b_2^2-kb_1^2}}{2k},\\
c^*_6&=\frac{-b_3+kb_2+\sqrt{(k+1)b_3^2+2k^2b_2^2-kb_1^2}}{2k},
\end{split}
\end{equation*}
where

$$ k = \frac{c_1^{\rm eq}c_5^{\rm eq}}{c_2^{\rm eq}c_6^{\rm eq}}-1,\;\;b_1 = b_\mathrm{A}-b_\mathrm{O}+c_3+2c_4,\;\;b_2 = b_\mathrm{O}-c_3-2c_4-b_\mathrm{H},\;\; b_3 = b_\mathrm{H}+b_\mathrm{A}.$$

\begin{figure}[htb]
\centering
(a)\includegraphics[width=0.45\textwidth]{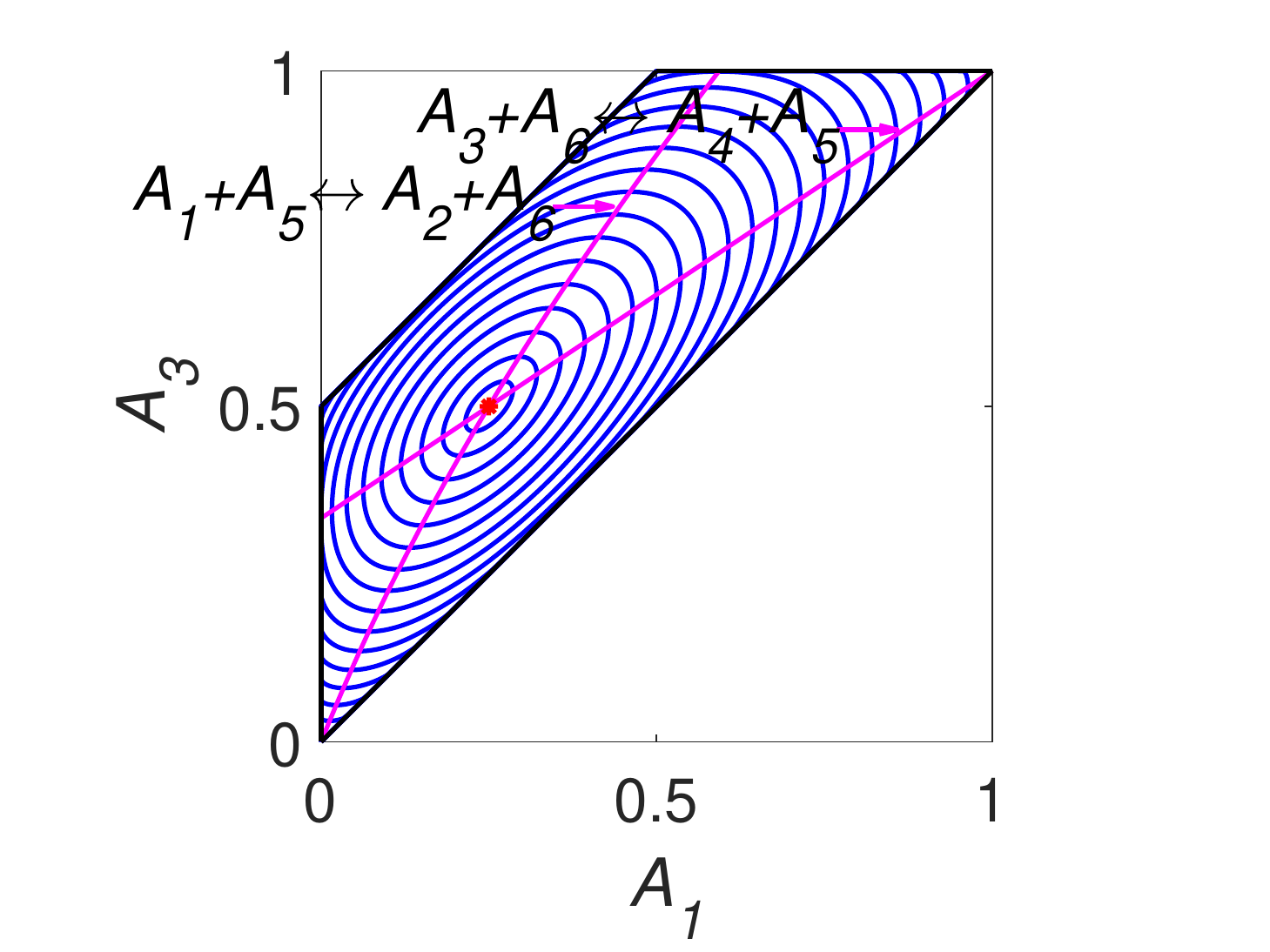}
(b)\includegraphics[width=0.45\textwidth]{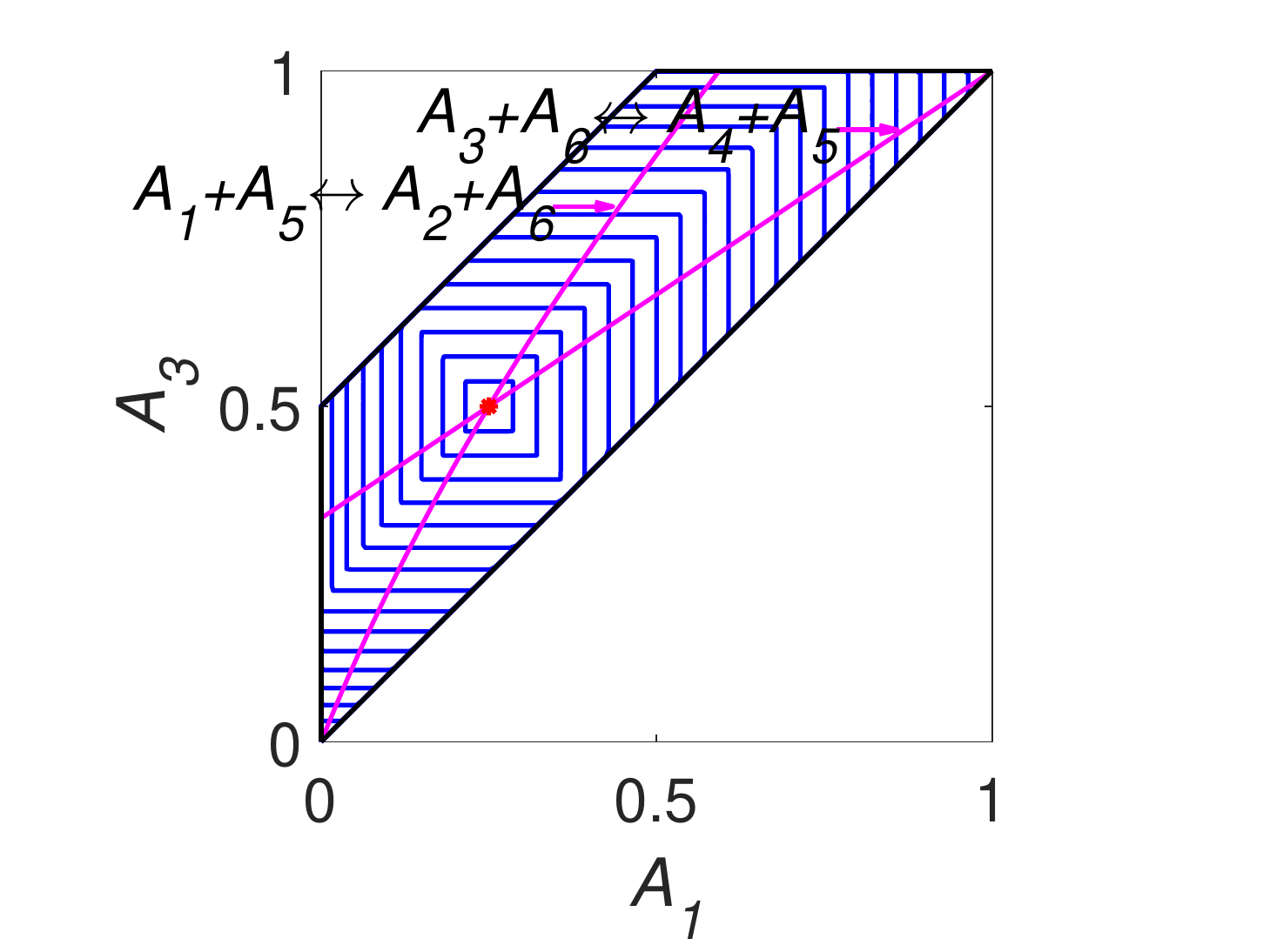}

\caption{Partial equilibrium lines (solid magenta lines) and level sets for: (\textbf{a}) Boltzmann's $H$ function and (\textbf{b}) Gorban's $H_\Gamma$ function.}
\label{FigWGSAbLev}
\end{figure}

The lines of partial equilibrium and the level sets for Boltzmann's $H$ function and Gorban's $H_\Gamma$ function are shown in Figure~\ref{FigWGSAbLev}. It is important to emphasise that these level sets are independent of kinetic constants and are completely determined by the equilibrium for Boltzmann's $H$ function and by the equilibrium and set of stoichiometric vectors $\Gamma$ for Gorban's $H_\Gamma$ function.

The kinetic equations for the system (\ref{SystWGSAb}) are:

\begin{equation}\label{EqWGSKin}
\begin{split}
\frac{\mathrm{d}c_1}{\mathrm{d}t}&=-k^+_1c_1c_5+k^-_1c_2c_6,\;\;c_2=b_\mathrm{H}-c_1,\\
\frac{\mathrm{d}c_3}{\mathrm{d}t}&=-k^+_2c_3c_6+k^-_2c_4c_5,\;\;c_4=b_\mathrm{C}-c_3,\\
c_5&=b_\mathrm{O}-c_1-c_3-2c_4,\;\;c_6=c_\mathrm{A}-c_5.
\end{split}
\end{equation}

For system (\ref{SystWGSAb}) with detailed balance, the conditions for the reaction rate constants are:

$$k^+_1c_1^{\rm eq}c_5^{\rm eq}=k^-_1c_2^{\rm eq}c_6^{\rm eq},\;\;k^+_2c_3^{\rm eq}c_6^{\rm eq}=k^-_2c_4^{\rm eq}c_5^{\rm eq}.$$

The system can be completely parametrised by six equilibrium concentrations $c^{\rm eq}_i$ and two reaction rate constants, for example, by the constants $k^+_1, k^+_2$.
To obtain the complex balance condition it is necessary to list all the different stoichiometric vectors $\alpha_\rho$ and $\beta_\rho$:

\begin{equation*}
\begin{split}
\alpha_{1}=\beta_{-1}&=(1,0,0,0,1,0),\\
\alpha_{-1}=\beta_{1}&=(0,1,0,0,0,1),\\
\alpha_{2}=\beta_{-2}&=(0,0,1,0,0,1),\\
\alpha_{-2}=\beta_{2}&=(0,0,0,1,1,0).
\end{split}
\end{equation*}

The conditions of complex balance are

\begin{equation}\label{EqnWGSComplex}
\begin{split}
k^+_1c_1^{\rm eq}c_5^{\rm eq}&=k^-_1c_2^{\rm eq}c_6^{\rm eq},\\
k^-_1c_2^{\rm eq}c_6^{\rm eq}&=k^+_1c_1^{\rm eq}c_5^{\rm eq},\\
k^+_2c_3^{\rm eq}c_6^{\rm eq}&=k^-_2c_4^{\rm eq}c_5^{\rm eq},\\
k^-_2c_4^{\rm eq}c_5^{\rm eq}&=k^+_2c_3^{\rm eq}c_6^{\rm eq}.
\end{split}
\end{equation}

\begin{figure}[htb]
\centering
(a)\includegraphics[width=0.30\textwidth]{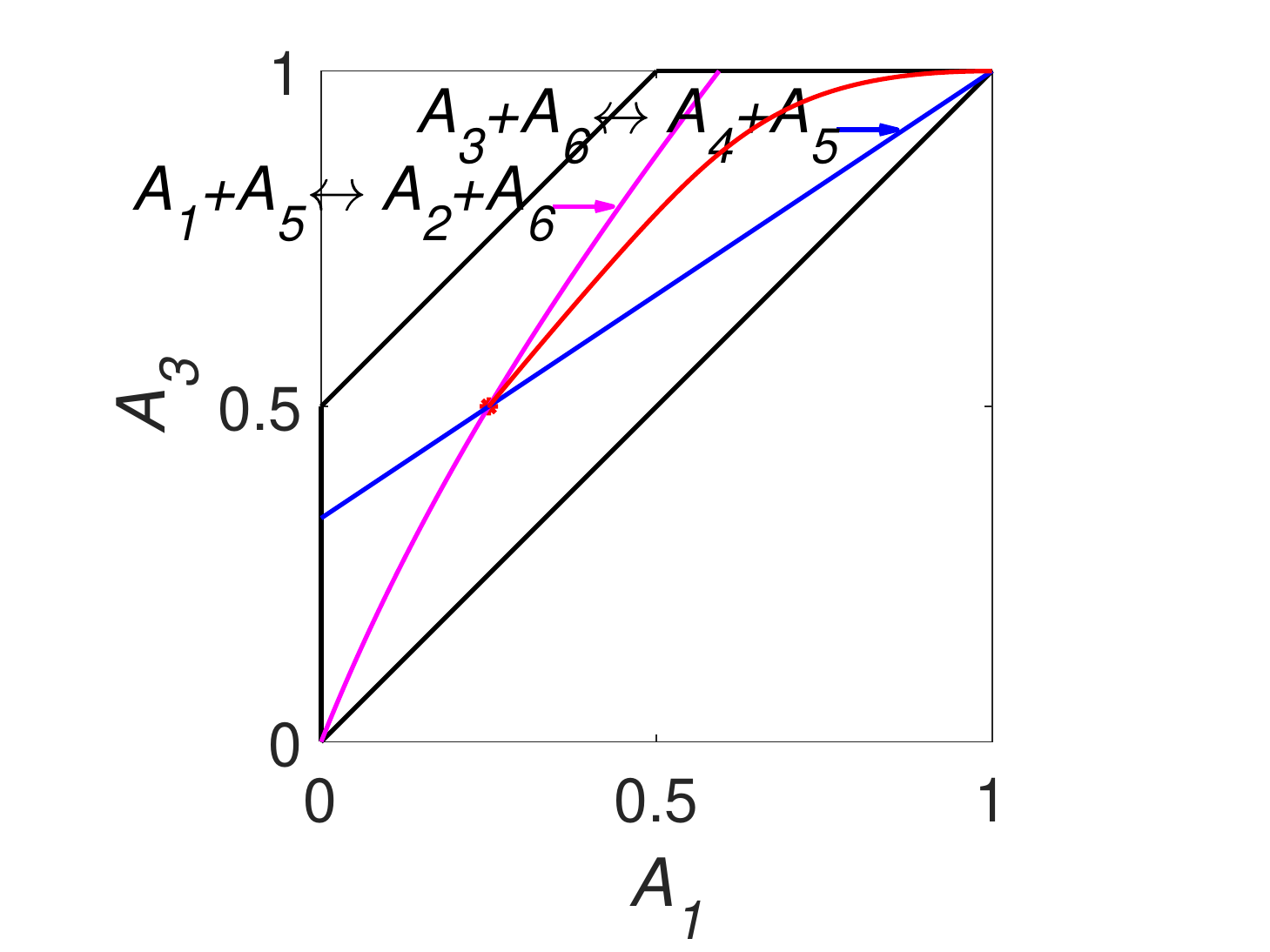}
(b)\includegraphics[width=0.30\textwidth]{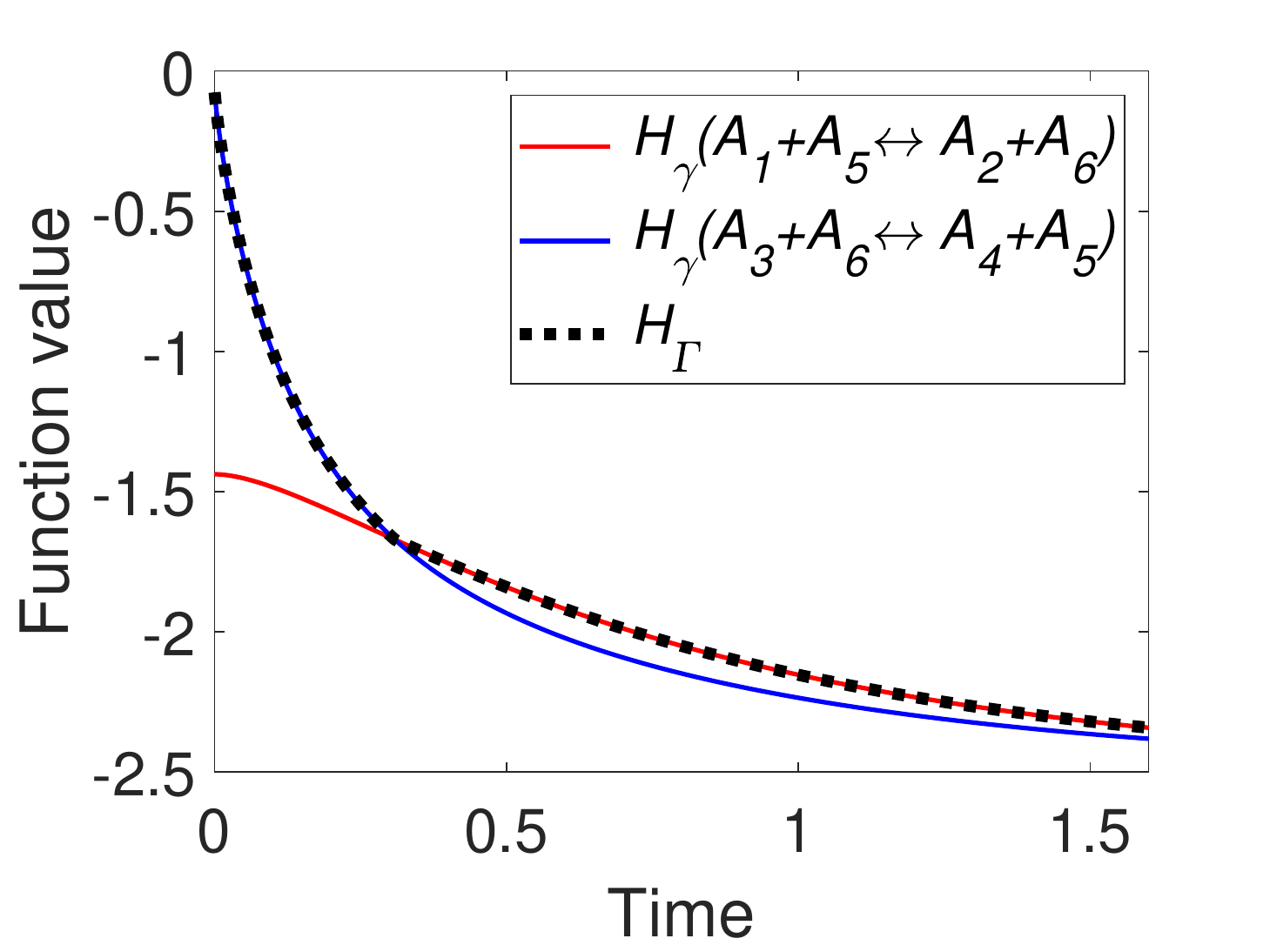}
(c)\includegraphics[width=0.30\textwidth]{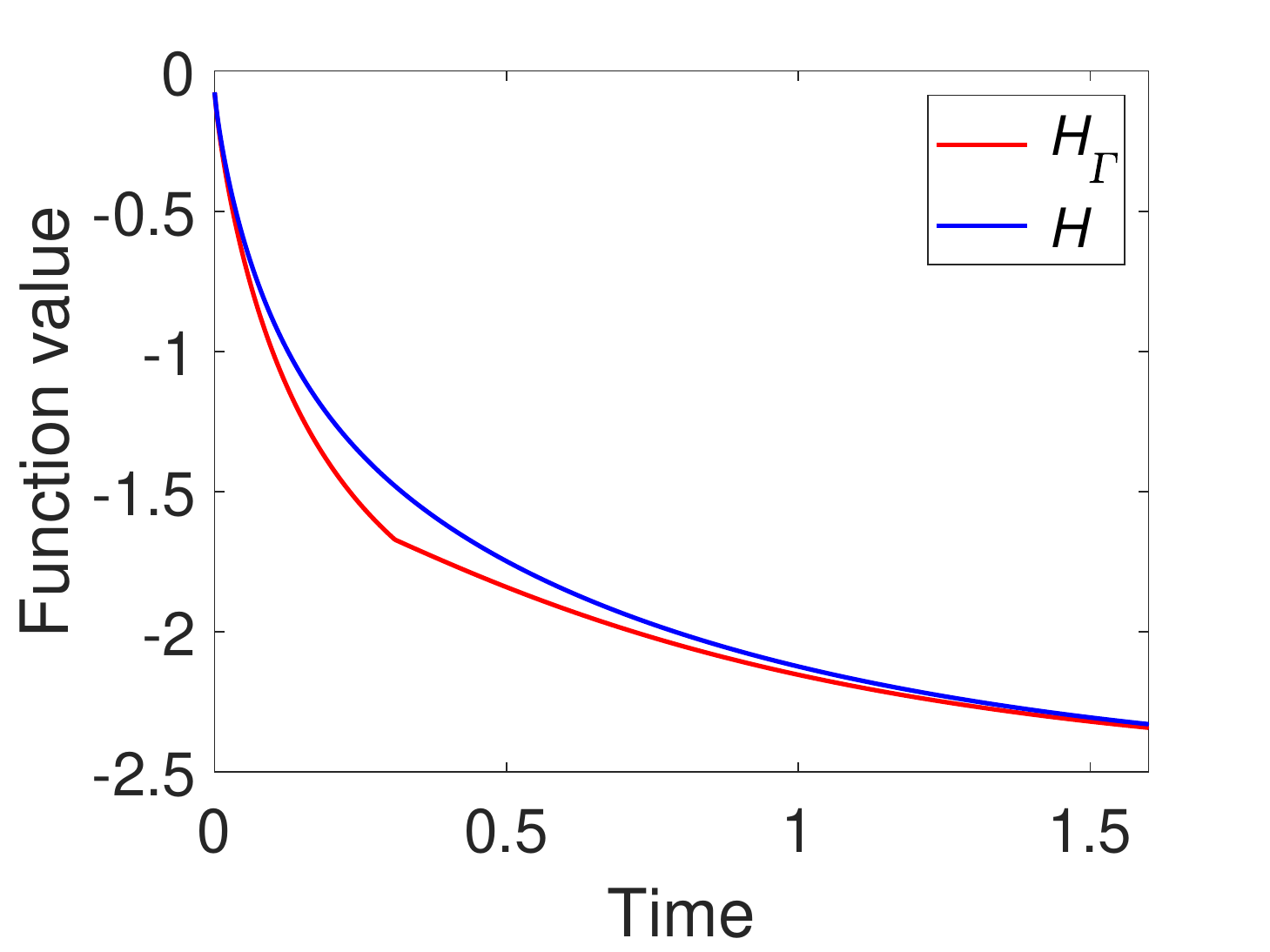}

\caption{The results of system (\ref{SystWGSAb}) simulation: (\textbf{a}) the trajectory (red line) in the phase plane (left), (\textbf{b}) the graphs of $H_{\gamma_i}$ and $H_\Gamma$ versus time, and (\textbf{c}) the graphs of Boltzmann's $H$ and Gorban's $H_\Gamma$ versus~time.}
\label{FigWGSAb}
\end{figure}

There are two pairs of identical equalities: the first equality coincides with the second one, and the third equality coincides with the fourth one. Moreover, the first and the third equalities are equivalent to the detailed balance conditions. This means that there is no difference between the detailed and complex balance conditions for system (\ref{EqWGSKin}). For simulation, we use equilibrium  concentrations $c^{\rm eq}=(0.25, 0.25, 0.5,0.5,0.25,0.25)$ and reaction rate constants of direct reactions $k^+=(1, 1)$. The simulation results are presented in Figure~\ref{FigWGSAb}. This figure clearly shows the difference between $H$ and $H_\Gamma$ functions and the switching from the $H_\Gamma = H_{\gamma_2}$ to $H_\Gamma=H_{\gamma_1}$ during dynamics.

Now we consider the real WGS reaction with the list of substances $\mathrm{H}_2\mathrm{O}, \mathrm{H}_2, \mathrm{CO}, \mathrm{CO}_2, \mathrm{red}, \mathrm{Ox}$ and reactions (\ref{SystWGS}).
All conservation laws are the same as for the abstract model, since these two models have the same structure. The coincidence of the complex balance condition with the detailed balance condition also takes place for WGS reaction. The WGS reaction parameters were found for the condition described in \cite{WGS3} ``a 1:1 molar feed ratio $[\mathrm{H}_2\mathrm{O}/\mathrm{CO}]$ and 220 $^\circ C$ the conversion reaches 70\%. The equilibrium conversion for these conditions is calculated as 87\%''. Additional parameters of reactor are described in \cite{WGS3}: ``... catalyst loading: 1.0 g; ... GHSV: 6100 $\mathrm{h}^{-1}$. Size of reactor is 1/2 inch in diameter and 12 inch long.'' 
From this information we can identify required values. Let us consider the case $b_\mathrm{H}=1$. Then from the equality of concentrations of $\mathrm{H}_2\mathrm{O}$ and $\mathrm{CO}$ and absence of all other gases in the original composition we can find $b_\mathrm{C}=b_\mathrm{H},\;\;b_\mathrm{O}=b_\mathrm{H}+b_\mathrm{C}$.
The time of a gas movement trough the reactor can be calculated as

\begin{equation*}
t_r=\frac{3600}{\mathrm{GHSV}}\approx0.59.
\end{equation*}

From the conversion 70\% we can require $c_3(t_r)=0.3b_\mathrm{C}$. From the equilibrium concentration of $\mathrm{CO}$ we can find $c_3^{\rm eq}=0.13b_\mathrm{C}$. From known values of $b_\mathrm{H},b_\mathrm{C},b_\mathrm{O},c^{\rm eq}_3$ and degree of conversion at time $t_r$ we can find $b_\mathrm{A},c^{\rm eq}_5,k^+_1,k^+_2$ by solving optimisation problem

$$\min_{b_\mathrm{A},c^{\rm eq}_5,k^+_1,k^+_2}\|c_3(t_r)-0.13\|.$$

\begin{figure}[tb]
\centering
(a)\includegraphics[width=0.45\textwidth]{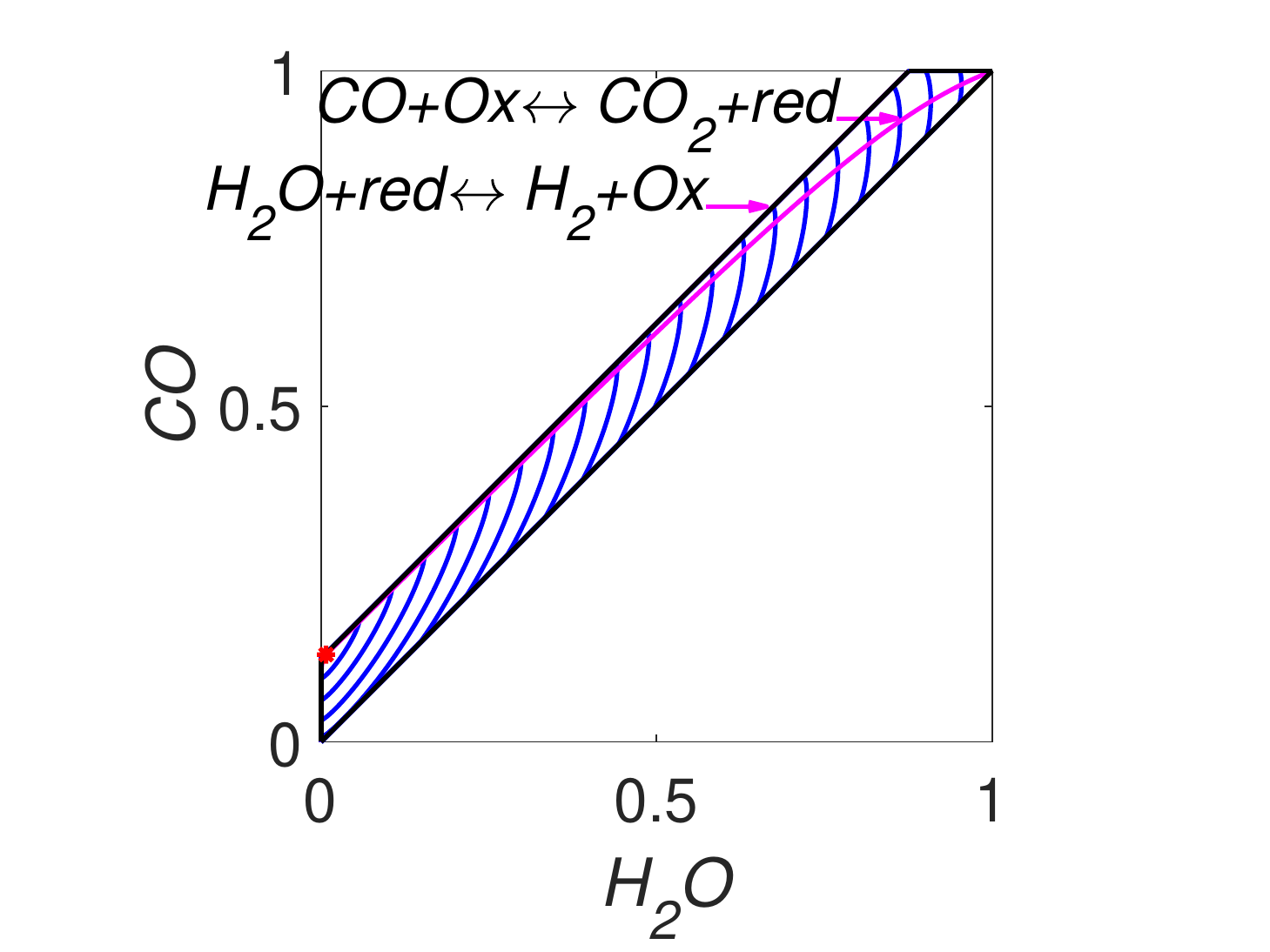}
(b)\includegraphics[width=0.45\textwidth]{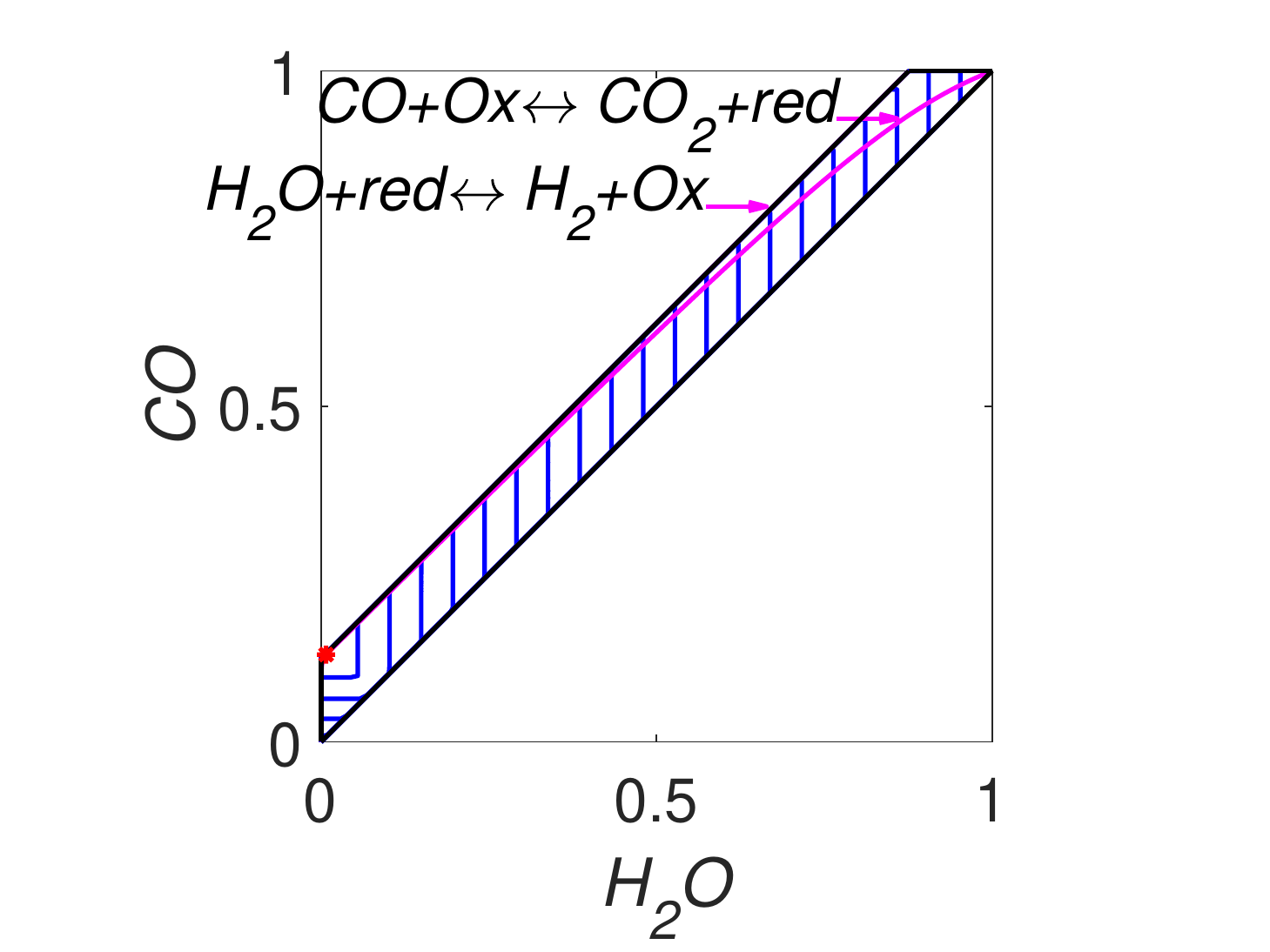}

\caption{Partial equilibrium lines (solid magenta lines) and level sets for: (\textbf{a}) Boltzmann's $H$ function and (\textbf{b}) Gorban's $H_\Gamma$ function.}
\label{FigWGSLev}
\end{figure}

\begin{figure}[tb]
\centering
(a)\includegraphics[width=0.30\textwidth]{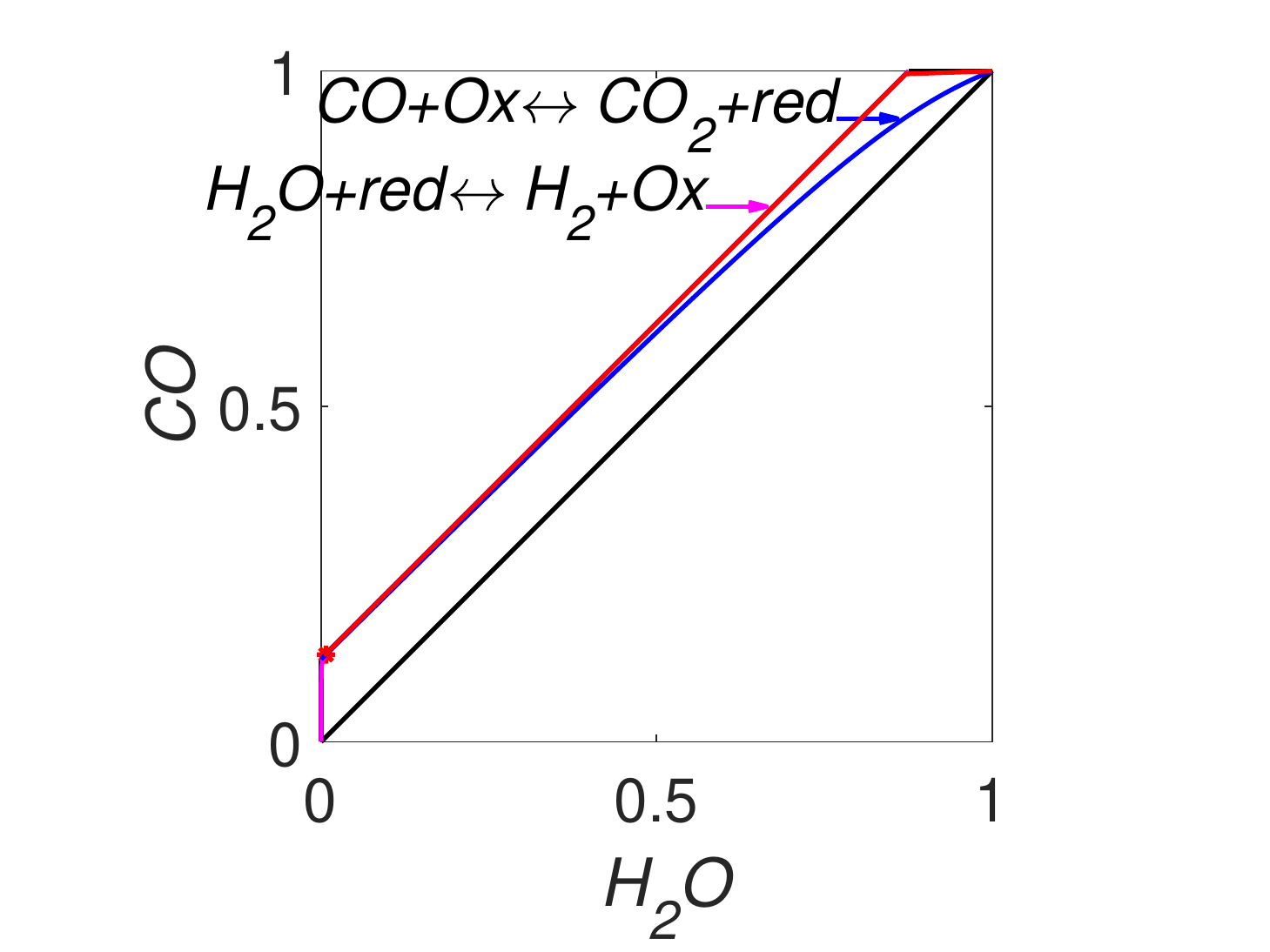}
(b)\includegraphics[width=0.30\textwidth]{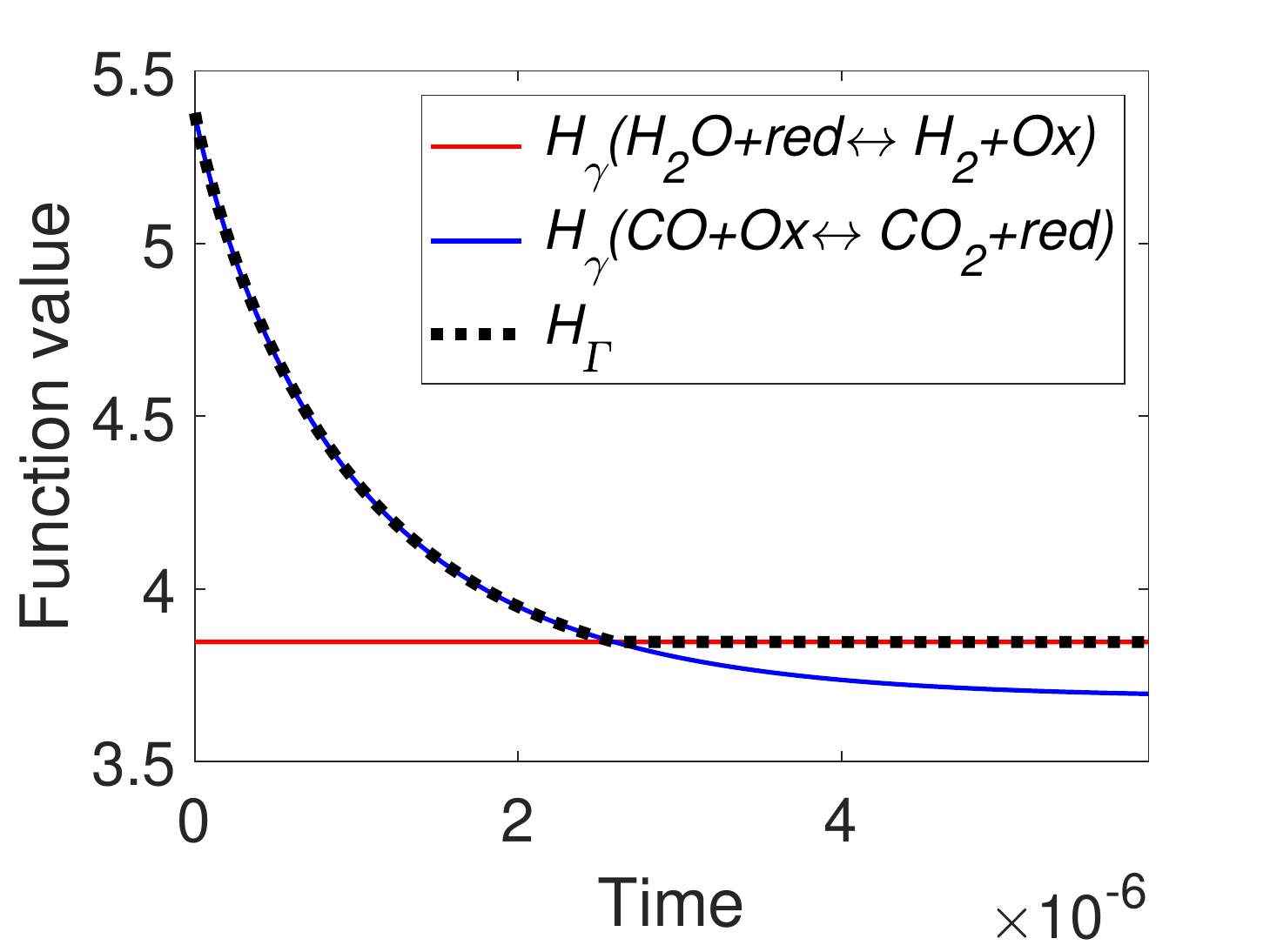}
(c)\includegraphics[width=0.30\textwidth]{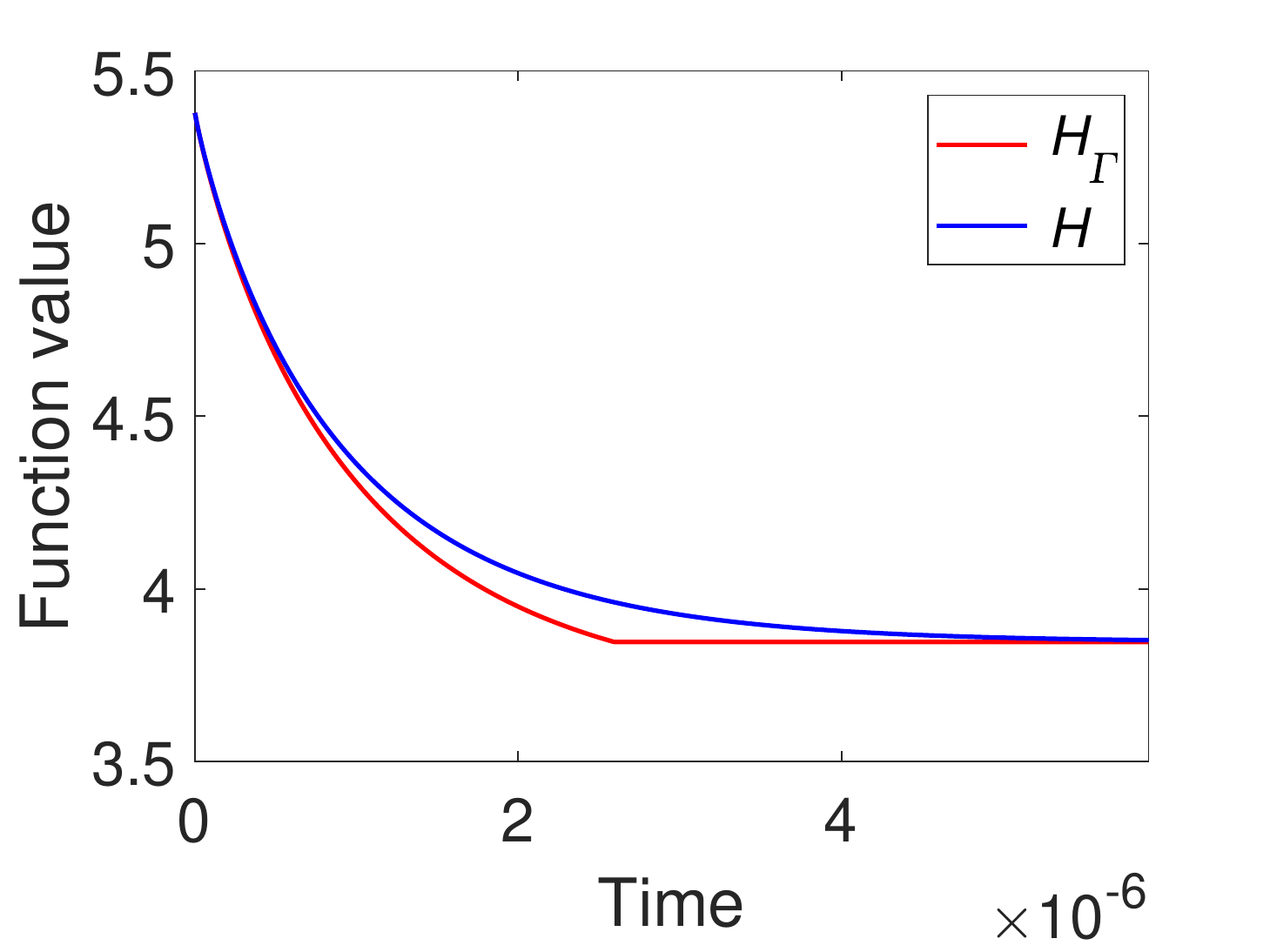}

\caption{The results of system (\ref{SystWGS}) simulation: (\textbf{a}) the trajectory in the phase plane (left), (\textbf{b}) the graphs of $H_{\gamma_i}$ and $H_\Gamma$ versus time, and (\textbf{c}) the graphs of Boltzmann's $H$ and Gorban's $H_\Gamma$ versus time.}
\label{FigWGS}
\end{figure}

The found parameters used in simulation are: the reaction rate constants of direct reactions $k^+=(80.53,146.31)$ and the equilibrium point $c^{\rm eq}=(0.0073,0.9927,0.13,0.87,0.0015,0.1227)$.
The lines of partial equilibrium and the level sets for Boltzmann's $H$-function and Gorban's $H_\Gamma$ function are presented in Figure~\ref{FigWGSLev}. The results of simulation are presented in Figure~\ref{FigWGS}. We can see that for the real WGS reaction, the equilibrium is very close to the boundary. As a result, the line of partial equilibrium of the first reaction also almost coincides with two sides of boundary of the reaction polygon. The trajectory very quickly  achieved the vicinity of the partial equilibrium line of the first reaction and then moved along this line to equilibrium. The time of achieving of the vicinity of the partial equilibrium line of the first reaction could be easily evaluated by switching $H_\Gamma$ from $H_{\gamma_2}$ to $H_{\gamma_1}$ and was approximately 3 microseconds. It was a very short time compared to 0.59 seconds of the total process time in the reactor. The difference between $H$ and $H_\Gamma$ is obvious for a very short initial time interval. By the way, the behaviour of the abstract system (\ref{SystWGSAb}) qualitatively coincides with the behaviour of the real WGS system (\ref{SystWGS}).

\subsection{Hydrogen Chloride Reaction \label{SectionHCL}}

In this subsection we consider the reaction of hydrogen chloride ($\mathrm{HCl}$) production \cite{HCL1,HCL2,HCL3}.
This reaction mechanism includes five substances $\mathrm{H}_2, \mathrm{H}, \mathrm{Cl}_2, \mathrm{Cl}, \mathrm{HCl}$ and four reactions

\begin{equation}\label{SystHCl}
\begin{split}
\mathrm{H}_2&\rightleftharpoons 2\mathrm{H},\\
\mathrm{Cl}_2&\rightleftharpoons 2\mathrm{Cl},\\
\mathrm{H}+\mathrm{Cl}_2&\rightleftharpoons \mathrm{HCl}+\mathrm{Cl},\\
\mathrm{Cl}+\mathrm{H}_2&\rightleftharpoons \mathrm{HCl}+\mathrm{H}.
\end{split}
\end{equation}

In the first part of this subsection we consider reactions with arbitrary chosen reaction rate constants. To avoid confusion we call this reaction 'abstract $\mathrm{HCl}$ reaction'. For this reaction we used the following substances $A_1, A_2, A_3, A_4, A_5$ and reactions

\begin{equation}\label{SystHClAb}
\begin{split}
A_1&\rightleftharpoons 2A_2,\\
A_3&\rightleftharpoons 2A_4,\\
A_2+A_3&\rightleftharpoons A_5+A_4,\\
A_4+A_1&\rightleftharpoons A_5+A_2.
\end{split}
\end{equation}

There are two conservation laws in the mechanism (\ref{SystHClAb}): $2c_1+c_2+c_5=b_\mathrm{H}$ is the hydrogen conservation law and $2c_3+c_4+c_5=b_\mathrm{Cl}$ is the chlorine conservation law. This means that there are only three independent variables in the system (\ref{SystHClAb}). For this study we selected the variables $A_1, A_3, A_5$ ($\mathrm{H}_2, \mathrm{Cl}_2, \mathrm{HCl}$ for system (\ref{SystHCl})) as independent and all the figures are presented in this space. The reaction polyhedron for this system can be found from the condition that all concentrations are nonnegative. The partial equilibrium surfaces of the first two reactions are defined by (\ref{GenDiss}). For example, for the first reaction, the partial equilibrium is

\begin{equation}\label{EqPart}
\begin{split}
c^*_1&=\frac{4(b_H-c_5)+k-\sqrt{8k(b_\mathrm{H}-c_5)+k^2}}{8},\\
c^*_2&=\frac{-k+\sqrt{8k(b_\mathrm{H}-c_5)+k^2}}{4},\\
c^*_3&=c^{}_3,\\
c^*_4&=c^{}_4,\\
c^*_5&=c^{}_5,
\end{split}
\end{equation}
where

$$ k = \frac{\big(c_2^{\rm eq})^2}{c_1^{\rm eq}}.$$

For the last two reactions, the surfaces of partial equilibrium are defined by (\ref{Gen1111}). For example, for the third reaction, this surface is

\begin{equation*}
\begin{split}
c^*_1&=c^{}_1,\\
c^*_2&=\frac{b_3+k(b_3-b_1)-\sqrt{(k+1)b_3^2+2k^2b_2^2-kb_1^2}}{2k},\\
c^*_3&=\frac{b_3+k(b_3+b_1)-\sqrt{(k+1)b_3^2+2k^2b_2^2-kb_1^2}}{2k},\\
c^*_4&=\frac{-b_3+kb_2+\sqrt{(k+1)b_3^2+2k^2b_2^2-kb_1^2}}{2k},\\
c^*_5&=\frac{-b_3-kb_2+\sqrt{(k+1)b_3^2+2k^2b_2^2-kb_1^2}}{2k},
\end{split}
\end{equation*}
where

$$ k = \frac{c_2^{\rm eq}c_3^{\rm eq}}{c_4^{\rm eq}c_5^{\rm eq}}-1, b_1=c_3-c_2, b_2=c_4-c_5, b_3=c_2+c_3+c_4+c_5.$$

The kinetic equations for the system (\ref{SystHClAb}) are:

\begin{equation}\label{EqHClKin}
\begin{split}
\frac{\mathrm{d}c_1}{\mathrm{d}t}&=-k^+_1c_1+k^-_1c_2^2-k_4^+c_1c_4+k^+_4c_2c_5,\\
c_2&=b_\mathrm{H}-2c_1-c_5,\\
\frac{\mathrm{d}c_3}{\mathrm{d}t}&=-k^+_2c_3+k^-_2c_4^2-k^+_3c_2c_3+k^-_3c_4c_5,\\
c_4&=b_\mathrm{Cl}-2c_3-c_5,\\
\frac{\mathrm{d}c_5}{\mathrm{d}t}&=k_4^+c_1c_4-k^+_4c_2c_5+k^+_3c_2c_3-k^-_3c_4c_5.
\end{split}
\end{equation}

For system (\ref{SystHClAb}) with  detailed balance, the conditions for the reaction rate constants are:

$$k^+_1c_1^{\rm eq}=k^-_1(c_2^{\rm eq})^2,\;\;k^+_2c_3^{\rm eq}=k^-_2(c_4^{\rm eq})^2,\;\;
k^+_3c_2^{\rm eq}c_3^{\rm eq}=k^-_3c_5^{\rm eq}c_4^{\rm eq},\;\;
k^+_4c_1^{\rm eq}c_4^{\rm eq}=k^-_4c_5^{\rm eq}c_2^{\rm eq}.$$

The system can be completely parametrised by six equilibrium concentrations $c^{\rm eq}_i$ and two reaction rate constants, for example, by the constants $k^+_1, k^+_2$.
To obtain the complex balance condition it is necessary to list all the different stoichiometric vectors $\alpha_\rho$ and $\beta_\rho$:

\begin{equation*}
\begin{split}
\alpha_{1}=\beta_{-1}&=(1,0,0,0,0),\\
\alpha_{-1}=\beta_{1}&=(0,2,0,0,0),\\
\alpha_{2}=\beta_{-2}&=(0,0,1,0,0),\\
\alpha_{-2}=\beta_{2}&=(0,0,0,2,0),\\
\alpha_{3}=\beta_{-3}&=(0,1,1,0,0),\\
\alpha_{-3}=\beta_{3}&=(0,0,0,1,1),\\
\alpha_{4}=\beta_{-4}&=(1,0,0,1,0),\\
\alpha_{-4}=\beta_{4}&=(0,1,0,0,1).
\end{split}
\end{equation*}

The conditions of  complex balance are

\begin{equation*}
\begin{split}
k^+_1c_1^{\rm eq}&=k^-_1(c_2^{\rm eq})^2,\\
k^-_1(c_2^{\rm eq})^2&=k^+_1c_1^{\rm eq},\\
k^+_2c_3^{\rm eq}&=k^-_2(c_4^{\rm eq})^2,\\
k^-_2(c_4^{\rm eq})^2&=k^+_2c_3^{\rm eq},\\
k^+_3c_2^{\rm eq}c_3^{\rm eq}&=k^-_3c_5^{\rm eq}c_4^{\rm eq},\\
k^-_3c_5^{\rm eq}c_4^{\rm eq}&=k^+_3c_2^{\rm eq}c_3^{\rm eq},\\
k^+_4c_1^{\rm eq}c_4^{\rm eq}&=k^-_4c_5^{\rm eq}c_2^{\rm eq},\\
k^-_4c_5^{\rm eq}c_2^{\rm eq}&=k^+_4c_1^{\rm eq}c_4^{\rm eq}.
\end{split}
\end{equation*}

We can see four pairs of identical equalities: the first and the second equalities, the third and the fourth equalities, the fifth and the sixth equalities, and the seventh and the eighth equalities. Moreover, the first, the third, the fifth and the seventh equalities are equivalent to the detailed balance conditions. This means that for system (\ref{SystHClAb}), the complex balance conditions are equivalent to the detailed balance~conditions.

\begin{figure}[H]
\centering
\includegraphics[width=0.45\textwidth]{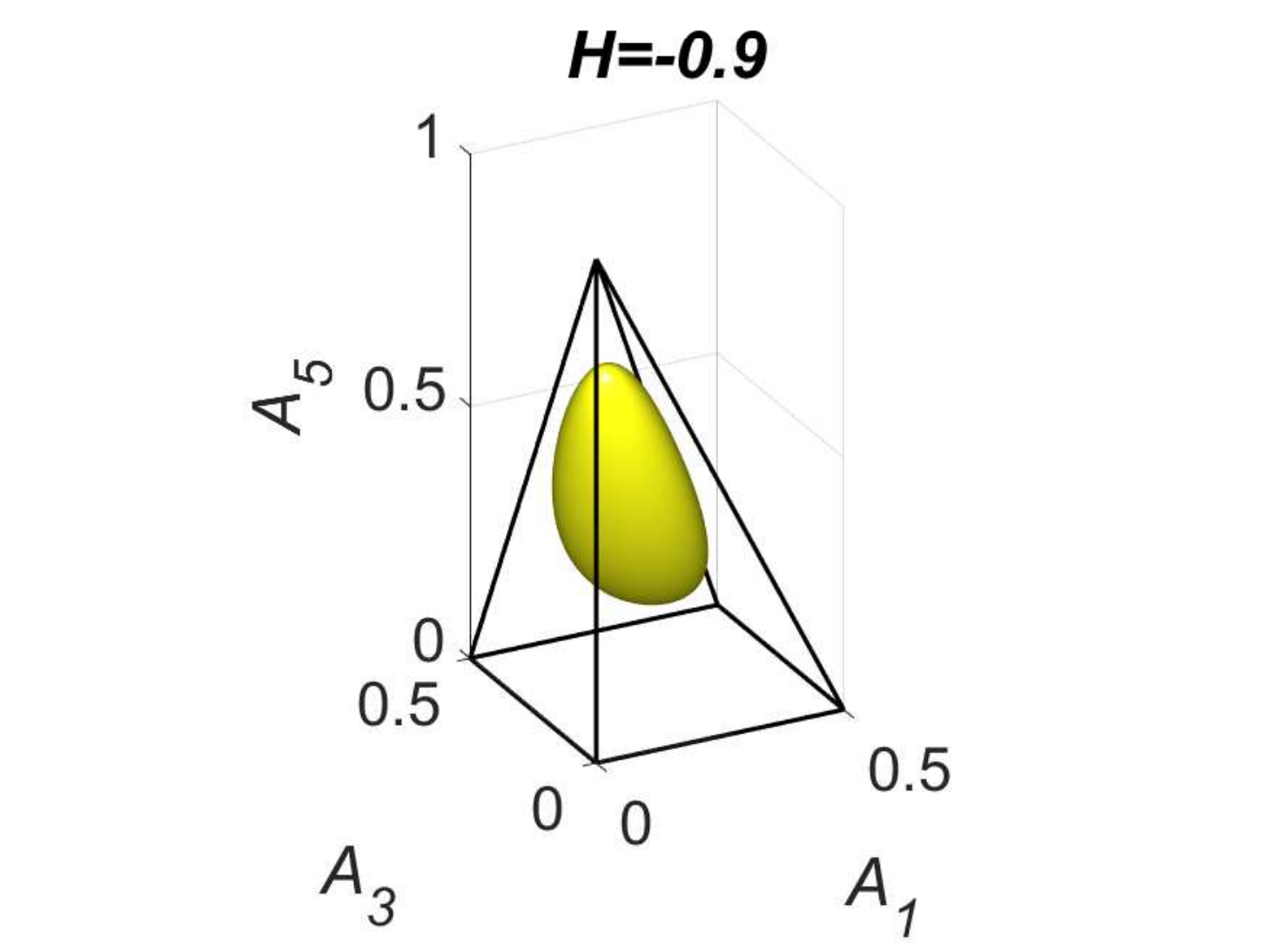}
\includegraphics[width=0.45\textwidth]{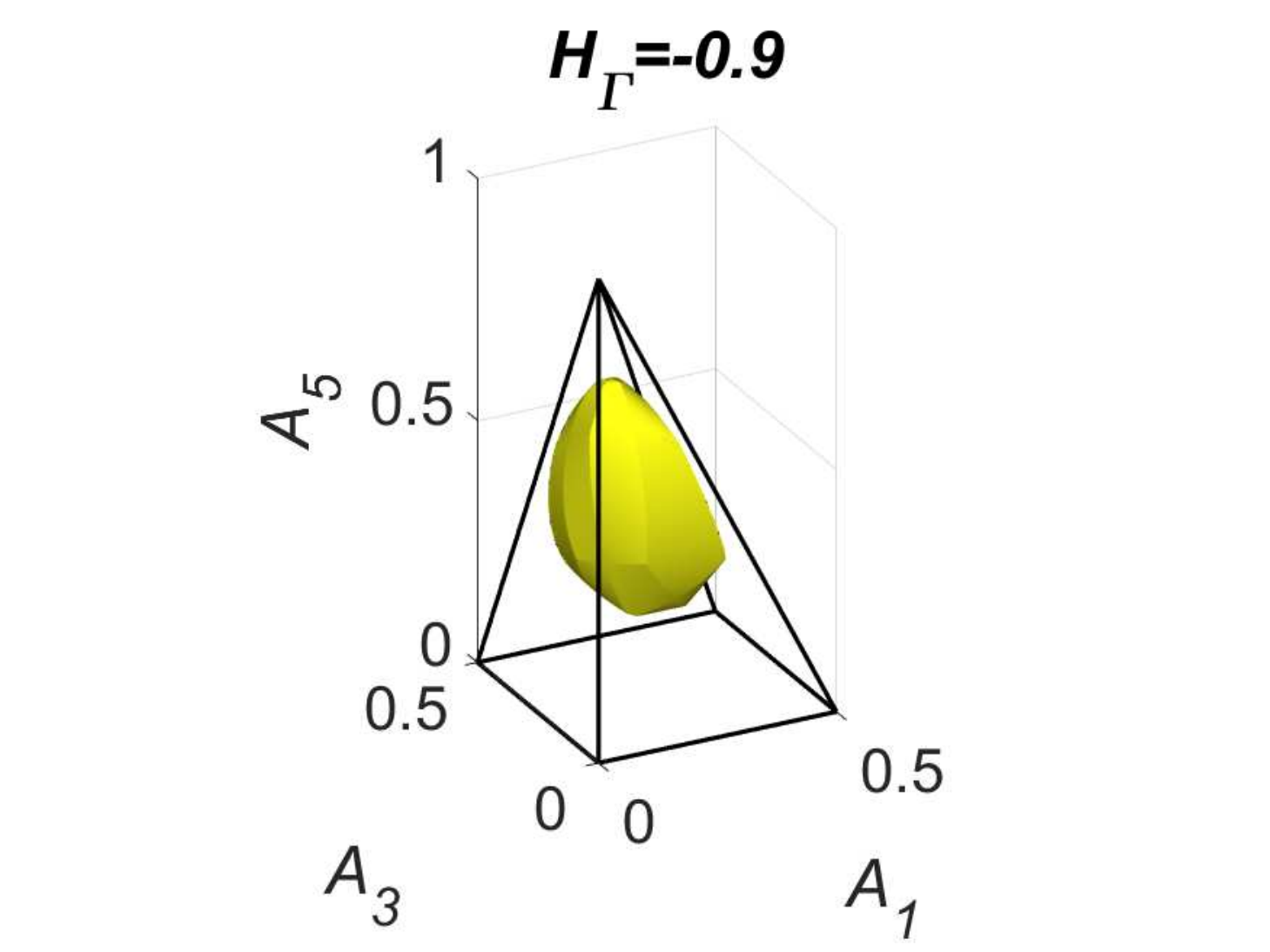}

\caption{The level sets for system (\ref{SystHClAb}): (\textbf{a}) $H=-0.9$ and (\textbf{b}) $H_\Gamma=-0.9$.}
\label{FigHClAbLevel}
\end{figure}

\begin{figure}[H]
\centering
\includegraphics[width=0.24\textwidth]{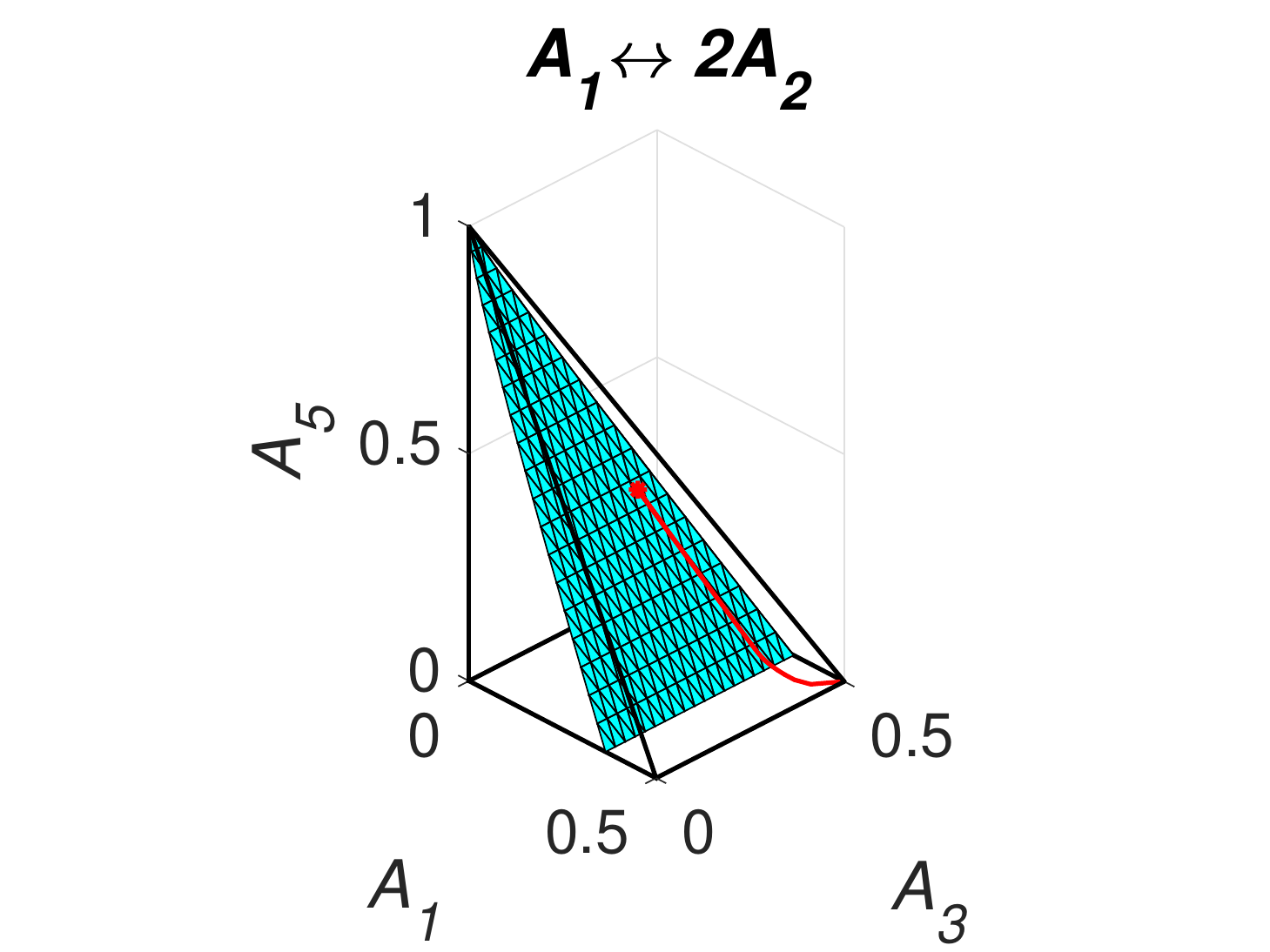}
\includegraphics[width=0.24\textwidth]{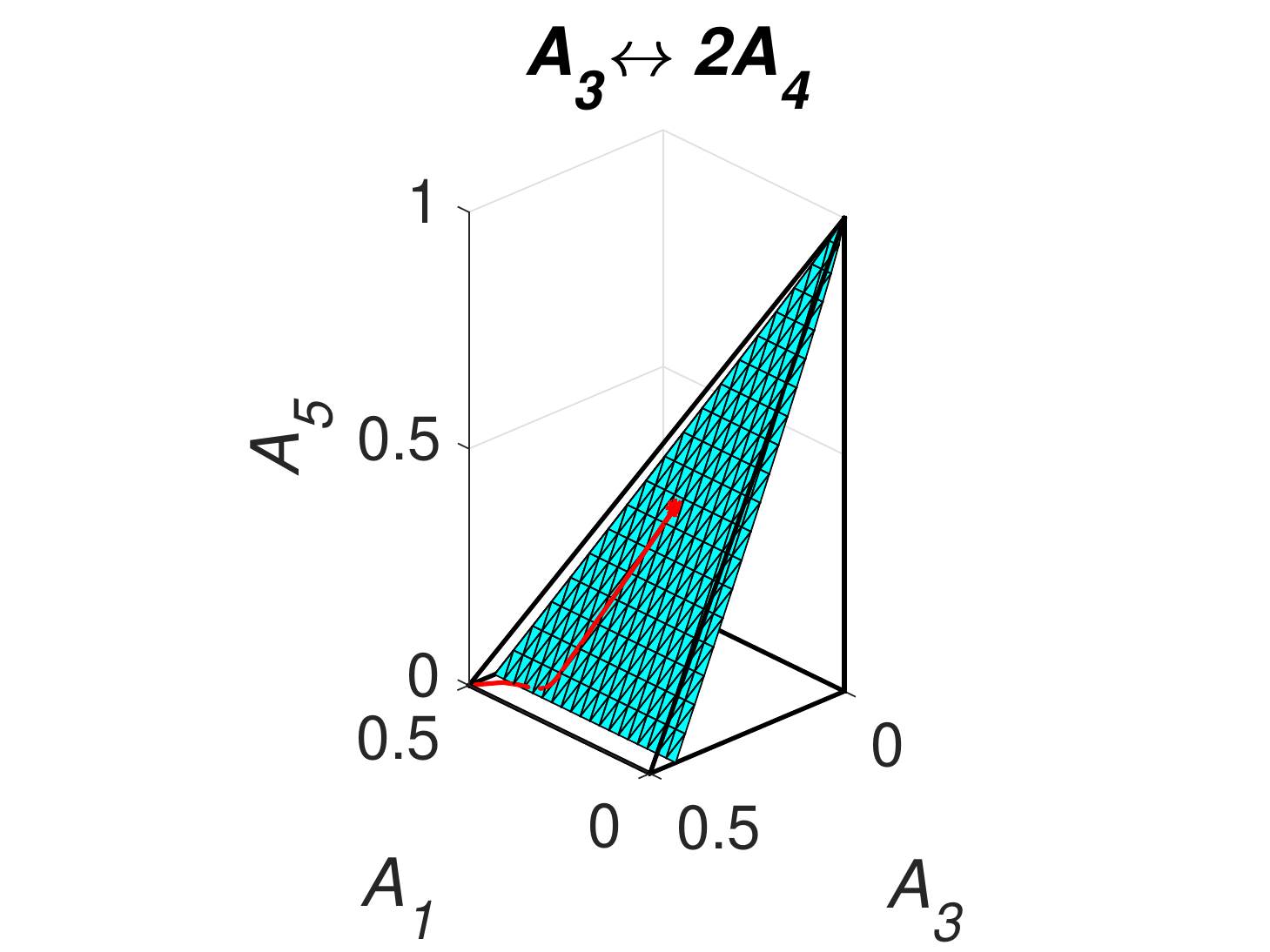}
\includegraphics[width=0.24\textwidth]{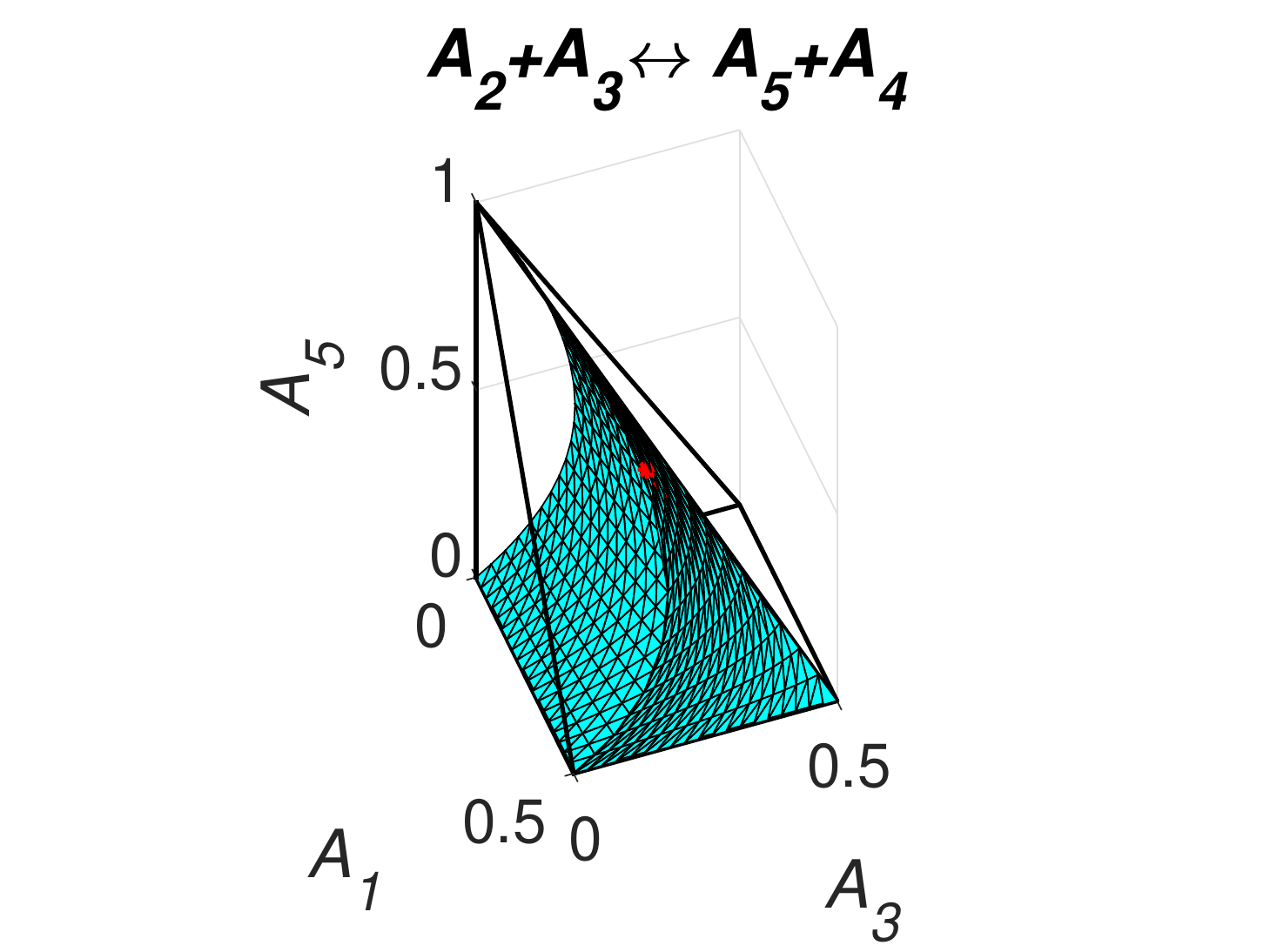}
\includegraphics[width=0.24\textwidth]{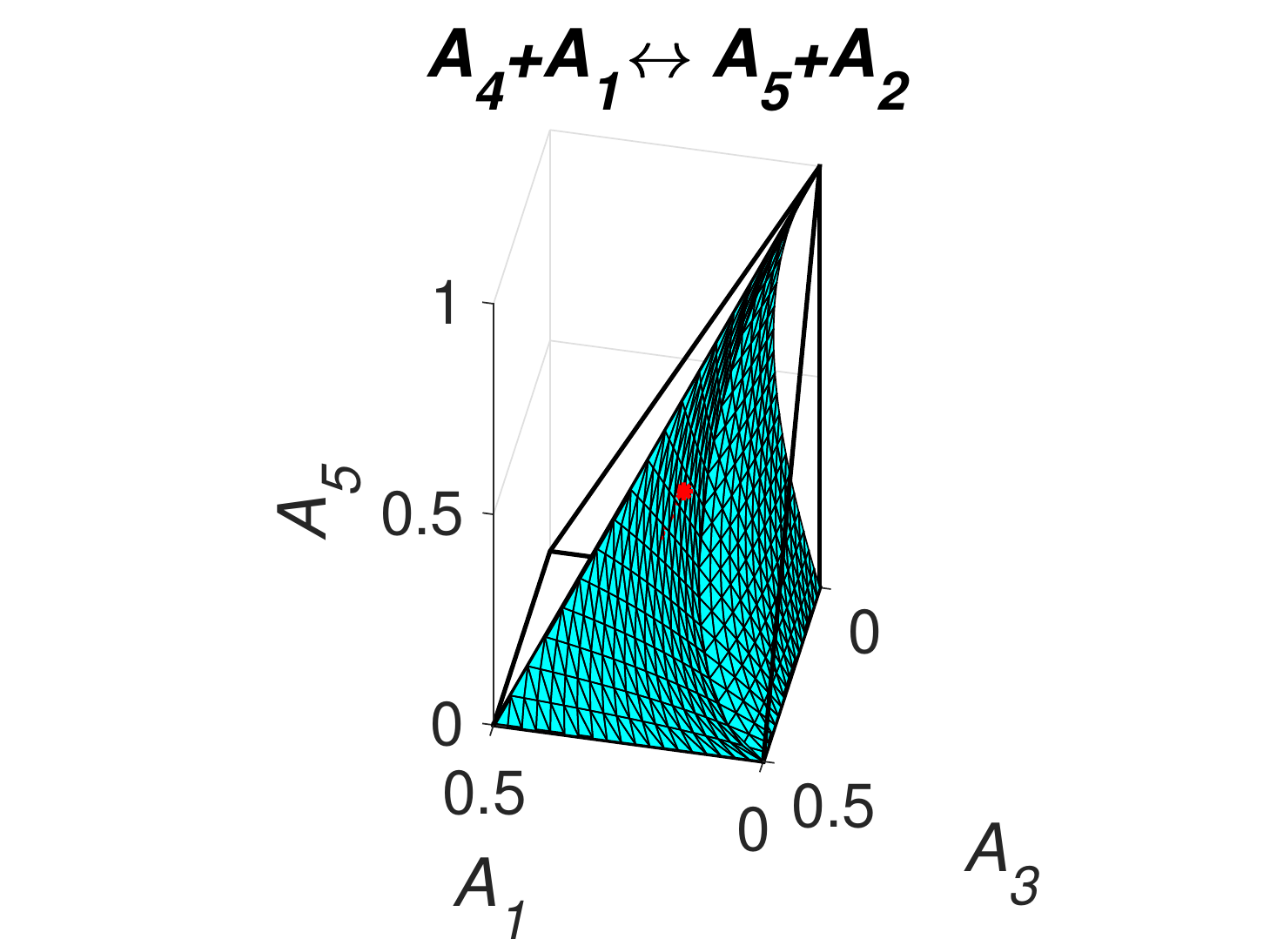}

\caption{Partial equilibrium surfaces and trajectory for system (\ref{SystHClAb}).}
\label{FigHClAbPart}
\end{figure}

The level sets for $H=-0.9$ and $H_\Gamma=-0.9$ are presented in Figure~\ref{FigHClAbLevel}. We can see that the level set $H=-0.9$ is smooth and the level set of $H_\Gamma=-0.9$ contains edges and faces. 
Partial equilibrium surfaces for system (\ref{SystHClAb}) with the kinetic curve (the trajectory) are presented for the equilibrium $c^{\rm eq}=(0.2, 0.2, 0.25, 0.1, 0.4)$ and the reaction rate constants of direct reactions $k^+=(5, 10, 2, 1)$ in Figure~\ref{FigHClAbPart}. It should be emphasised that the surfaces of partial equilibrium for the first two reactions only look like planes, but in fact they have square root type nonlinearity (see (\ref{EqPart}) for the surface of partial equilibrium of the first reaction). At the beginning of motion, the trajectory quickly (at about 0.03 seconds) achieved the partial equilibrium surface of the second reaction, then along this surface the trajectory reached (at about 0.18 seconds) the intersection of the partial equilibrium surfaces of the first two reactions and then moved along this intersection to the equilibrium (approximately 15 seconds for the tolerance level 0.0001).

\begin{figure}[H]
\centering
\includegraphics[width=0.544\textwidth]{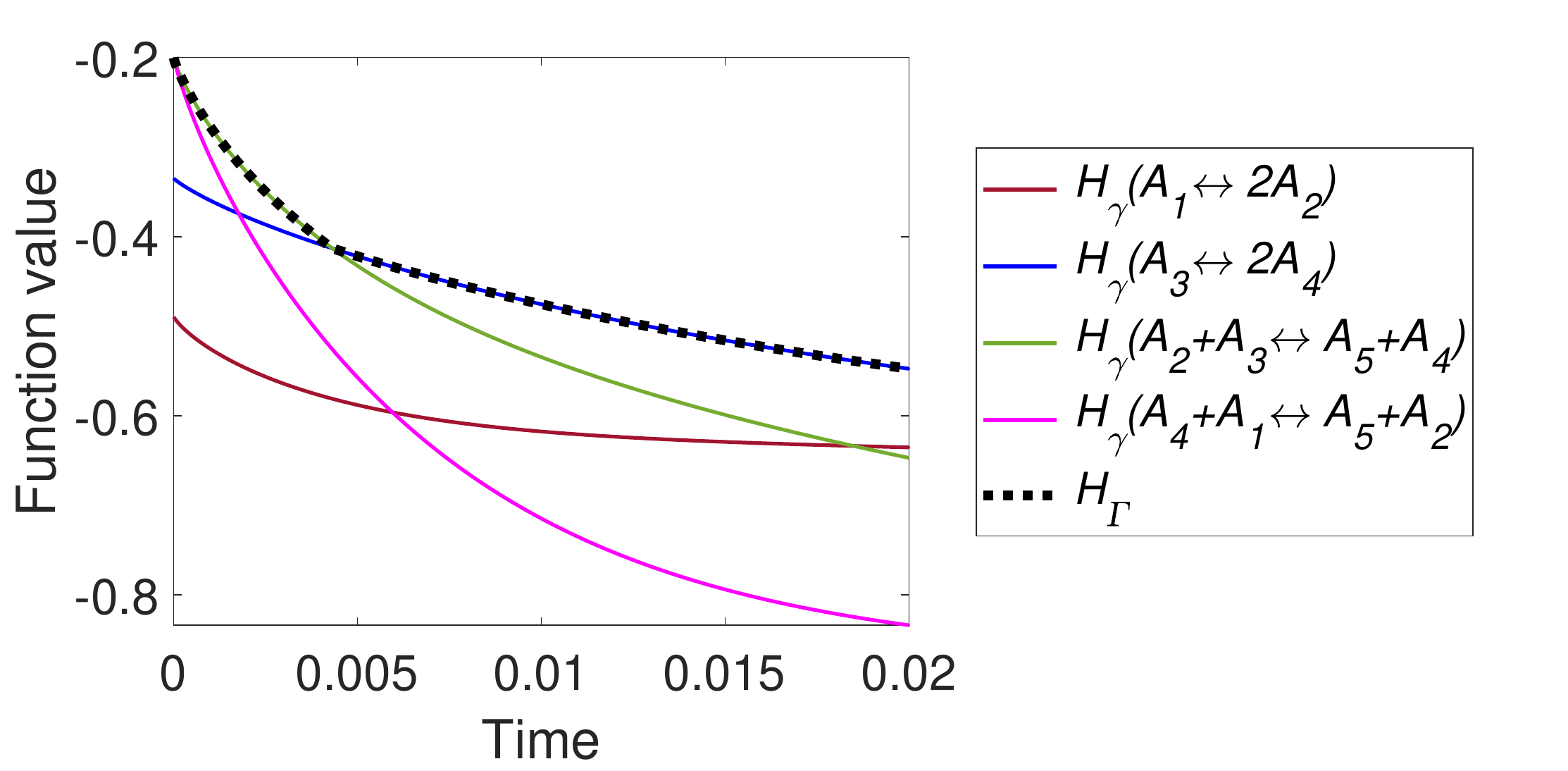}
\includegraphics[width=0.356\textwidth]{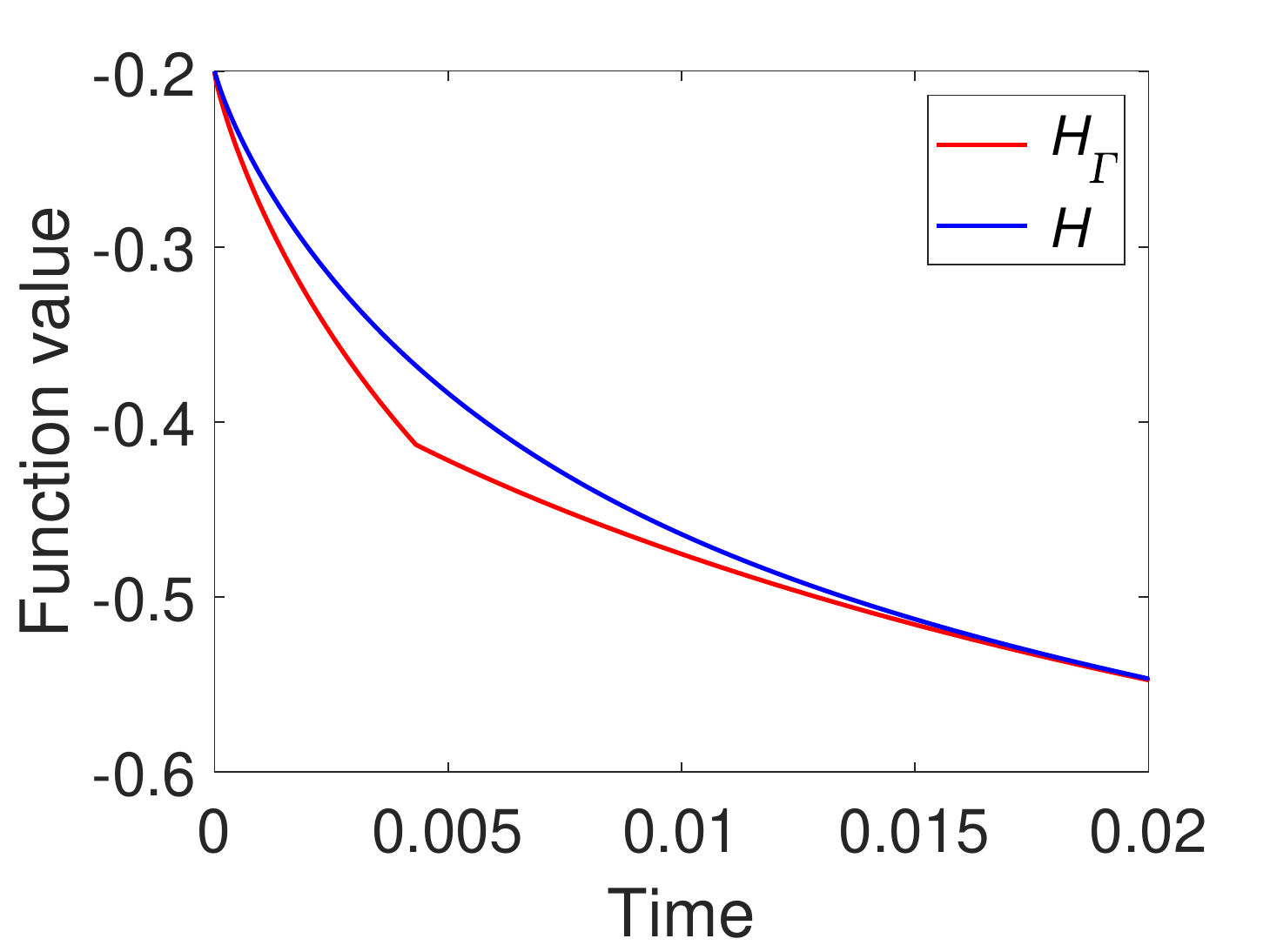}

\caption{The results of system (\ref{SystHClAb}) simulation: (\textbf{a}) the graphs of $H_{\gamma_i}$ and $H_\Gamma$ versus time, and (\textbf{b}) the graphs of Boltzmann's $H$ and Gorban's $H_\Gamma$ versus time.}
\label{FigHClAbH}
\end{figure}

The graphs of $H$ and $H_\Gamma$ are presented in Figure~\ref{FigHClAbH}. There is a difference between $H$ and $H_\Gamma$ at the initial stage of the reaction (during the first 0.02 seconds from the approximately 15 seconds of the full process). We also can observe two switches of $H_\Gamma$: from $H_{\gamma_4}$ to $H_{\gamma_3}$ at the first few microseconds and then from $H_{\gamma_3}$ to $H_{\gamma_2}$ in about 5 milliseconds after the start of the process.

\begin{figure}[htb]
\centering
\includegraphics[width=0.24\textwidth]{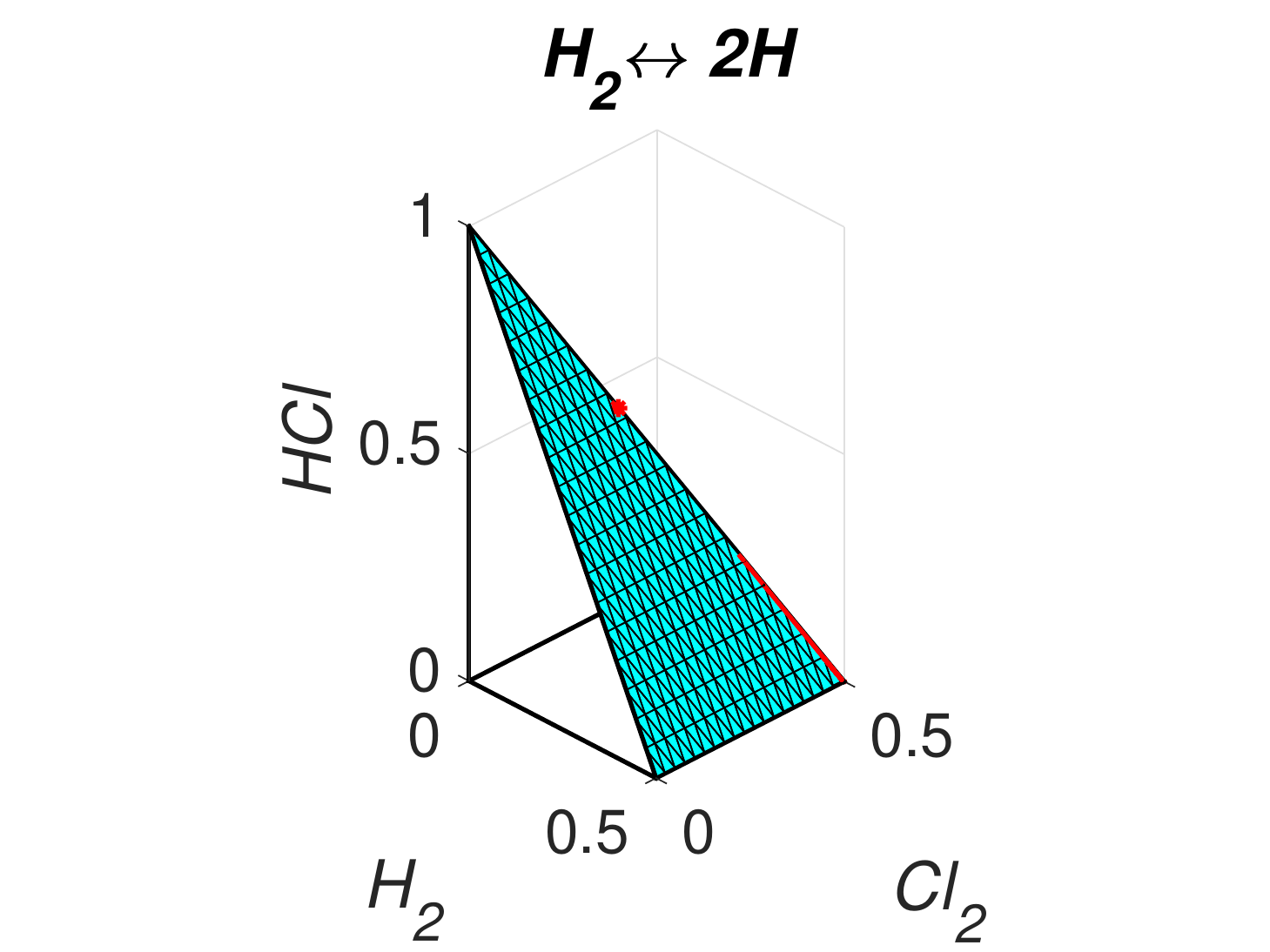}
\includegraphics[width=0.24\textwidth]{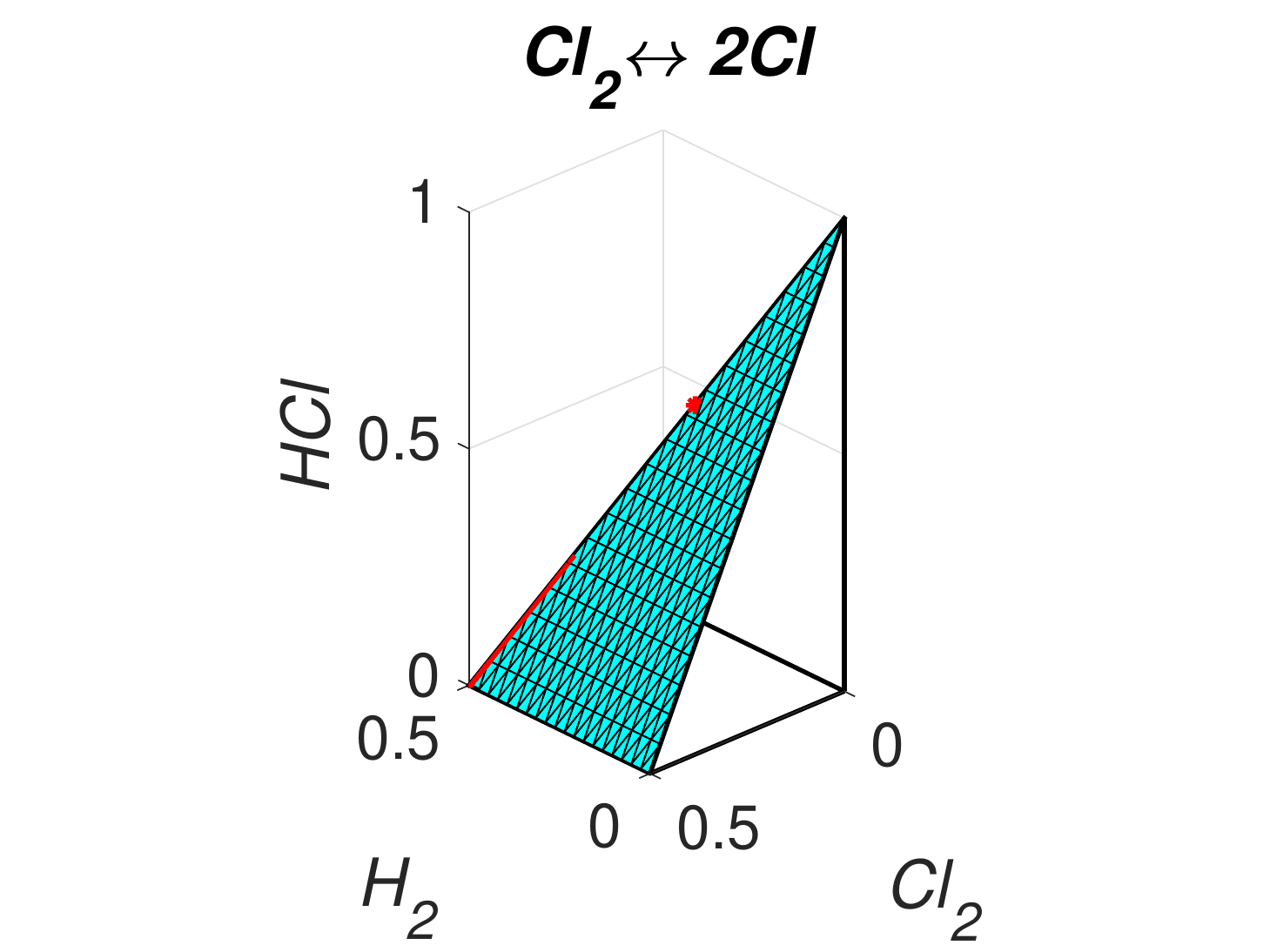}
\includegraphics[width=0.24\textwidth]{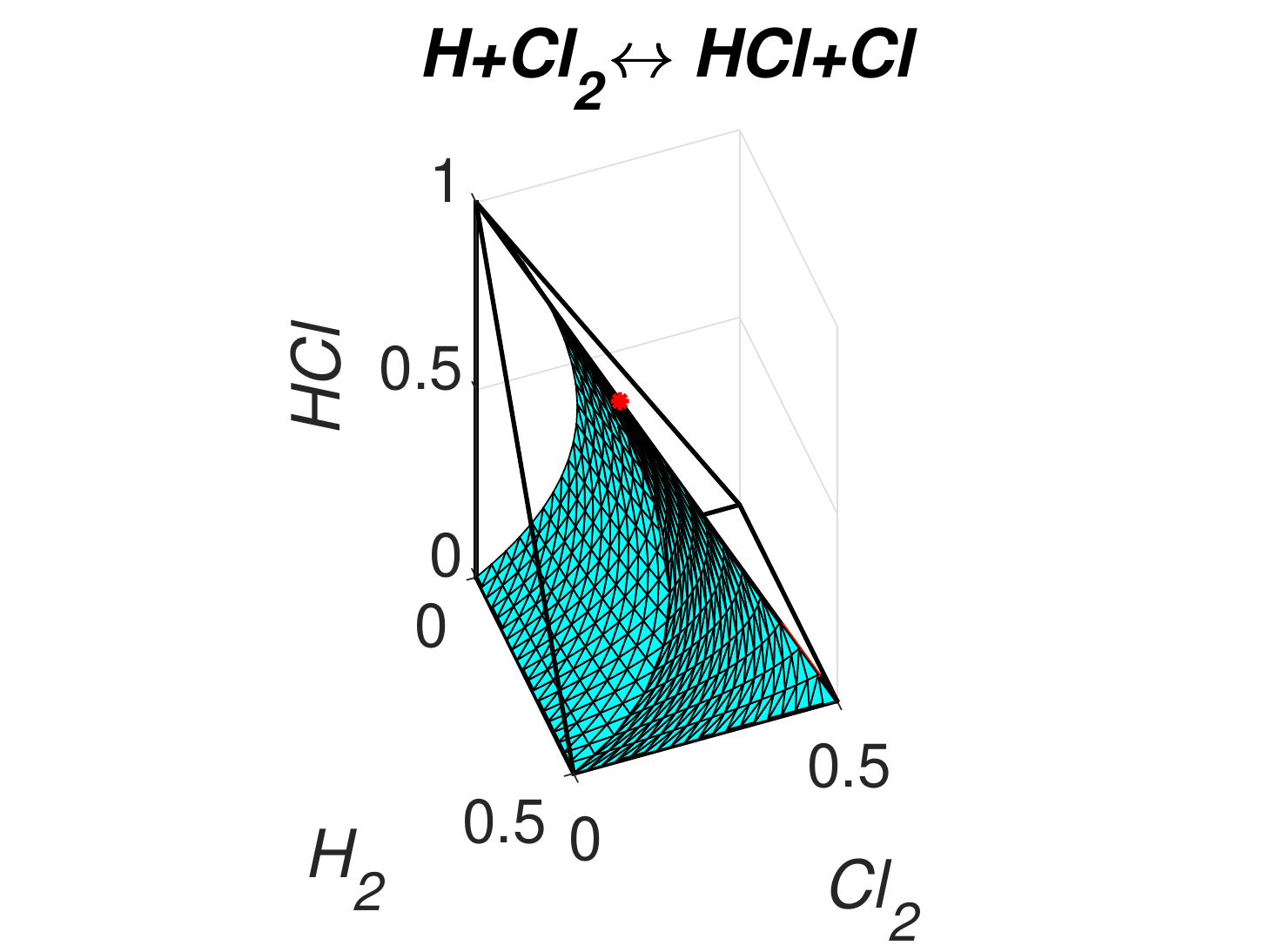}
\includegraphics[width=0.24\textwidth]{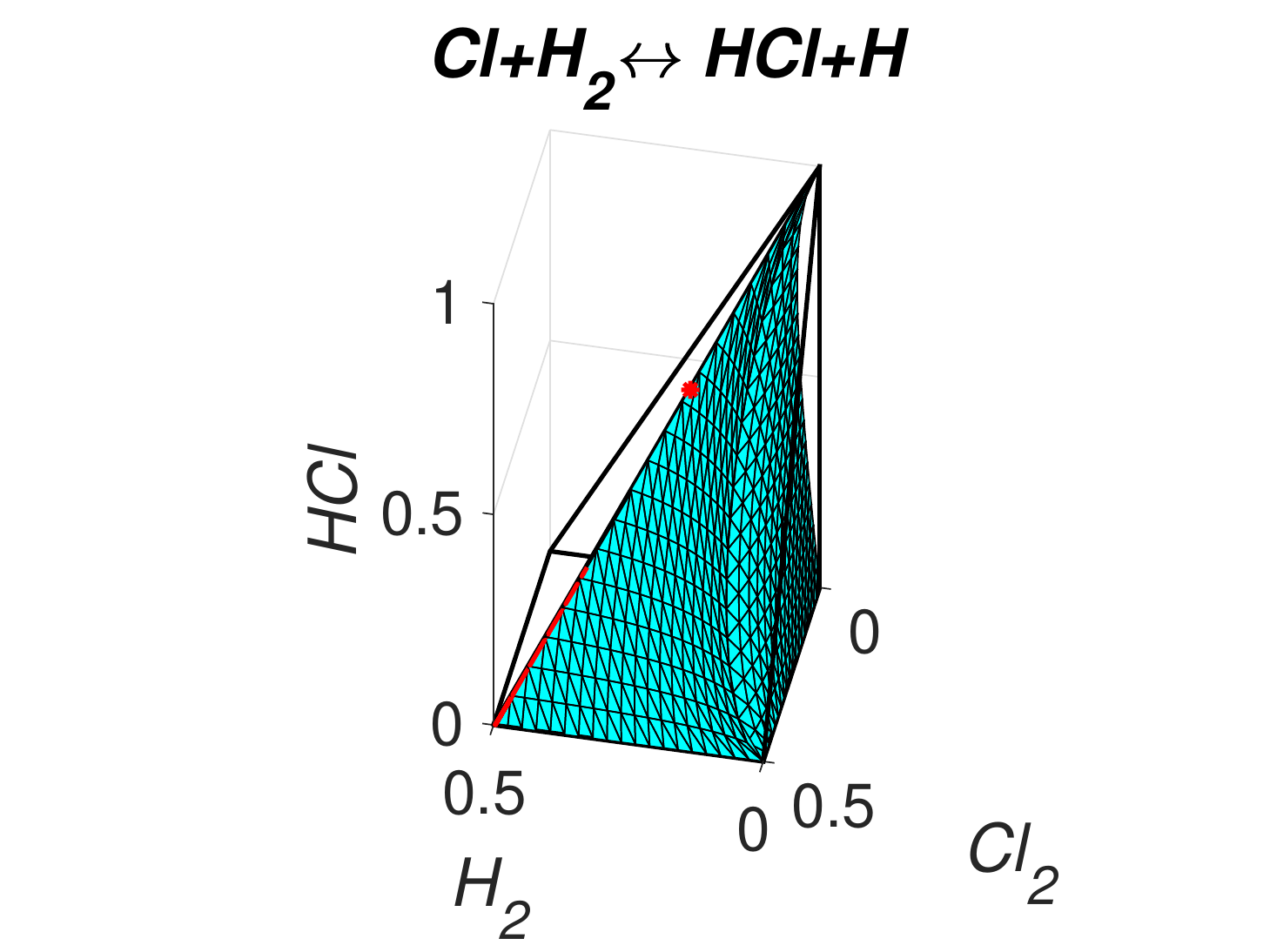}

\caption{Partial equilibrium surfaces and the trajectory for system (\ref{SystHCl}).}
\label{FigHClPart}
\end{figure}

Now we consider the real $\mathrm{HCl}$ reaction (\ref{SystHCl}). For the simulation, we used the information from~\cite{HCL1, HCL2, HCL3}: the equilibrium point was $c^{\rm eq}=(0.198, 0.004, 0.1995, 0.001, 0.6)$ and the reaction rate constants of the direct reactions were $k^+=(10^{16}, 10^{16}, 1.7\times 10^{11}, 1.59\times 10^8)$. This reaction system is very stiff and the equilibrium point is almost on the edge between the vertices $(0,0,0,0,1)$ and $(0.5,0,0.5,0,0)$. This means that the graphs of the level sets are uninformative and we omit them. Images of the level sets for system (\ref{SystHCl}) can be found in \cite{MirkesGit}.
The partial equilibrium surfaces for this system are presented in Figure~\ref{FigHClPart}. It should be emphasized that the partial equilibrium surfaces for the first two reactions only look like planes, but actually have a square root nonlinearity (see (\ref{EqPart}) for the partial equilibrium surface of the first reaction).

\begin{figure}[htb]
\centering
\includegraphics[width=0.544\textwidth]{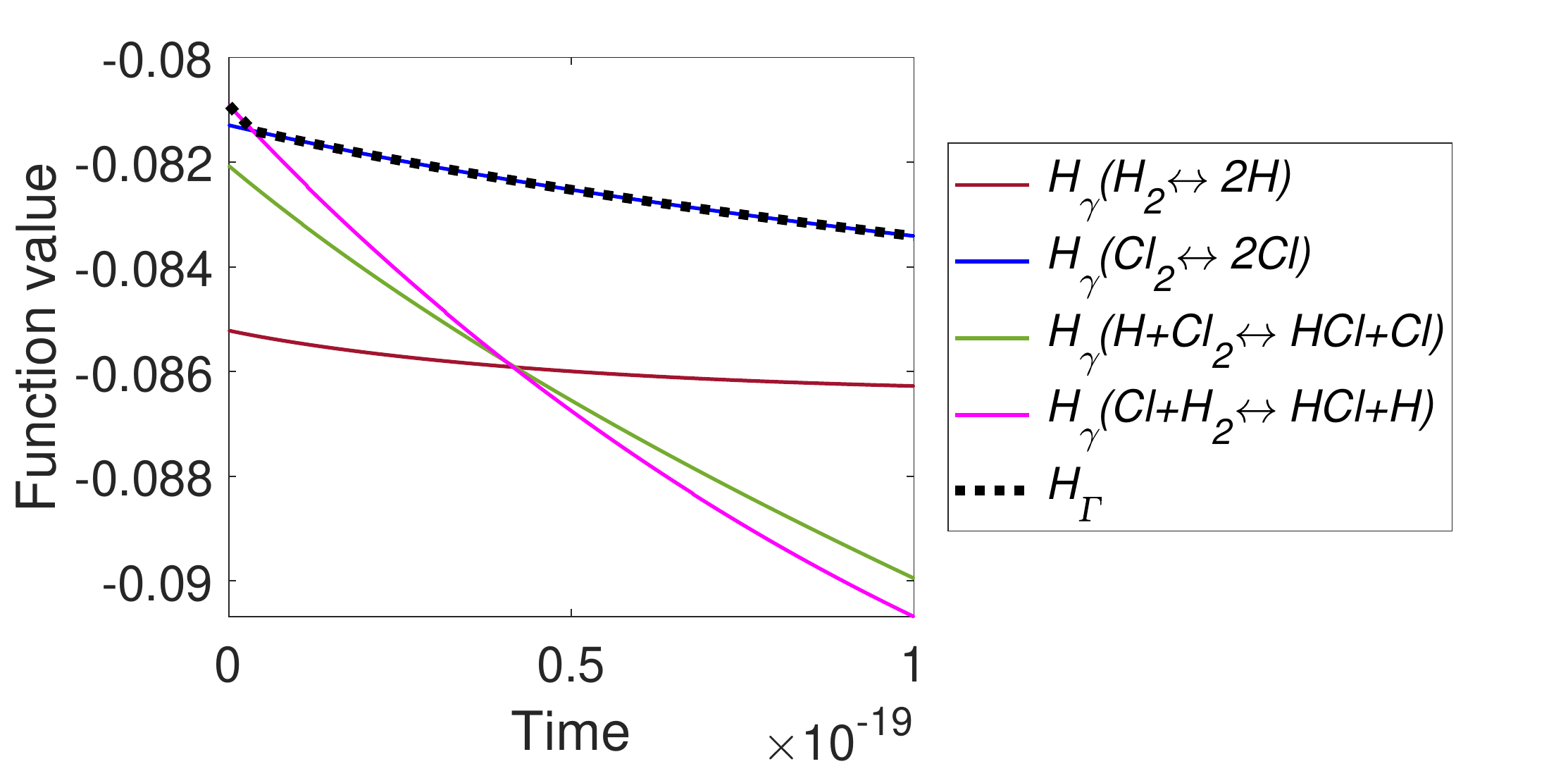}
\includegraphics[width=0.356\textwidth]{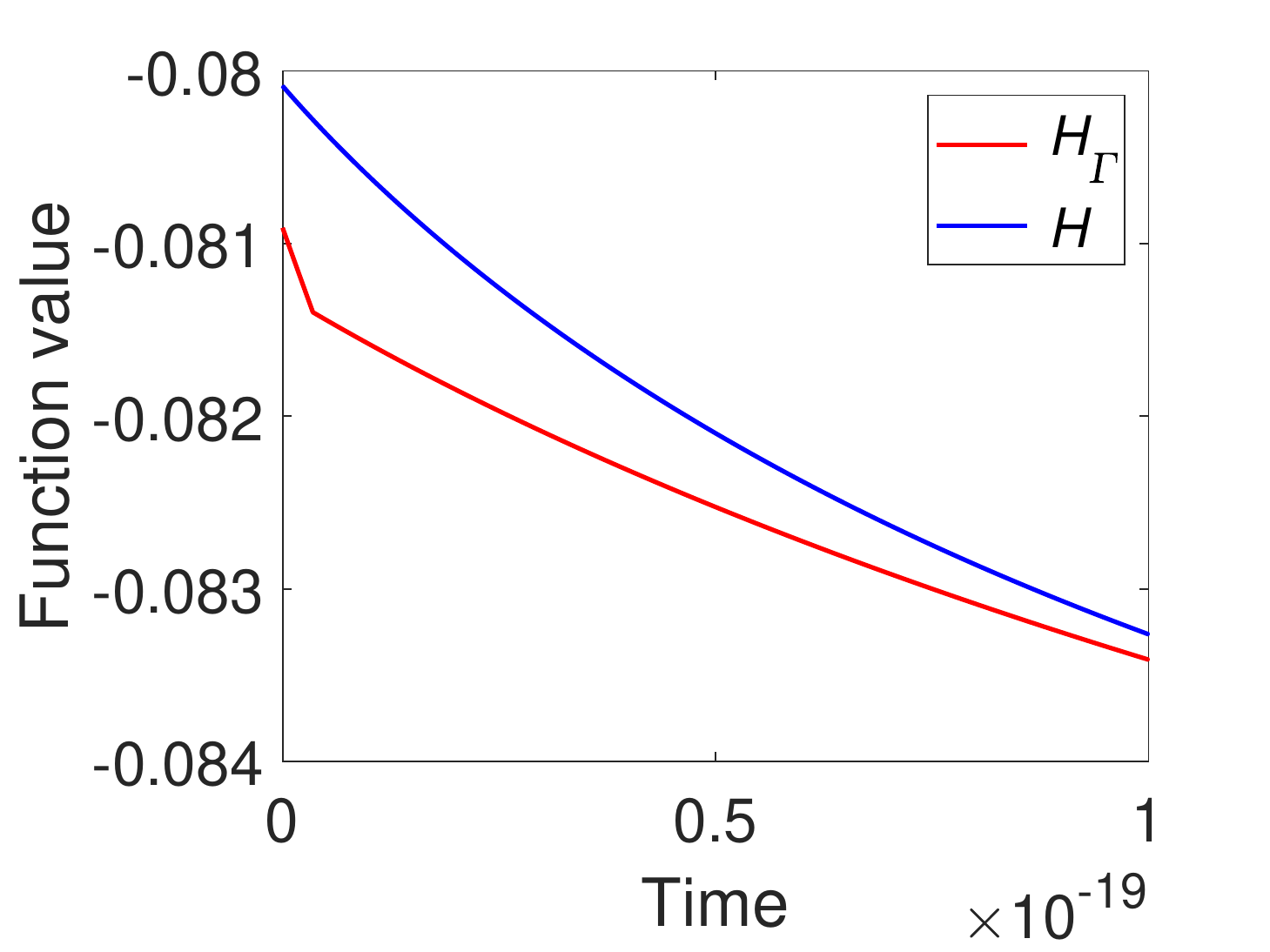}

\caption{The results of system (\ref{SystHCl}) simulation: (\textbf{a}) the graphs of $H_{\gamma_i}$ and $H_\Gamma$ versus time, and (\textbf{b}) the graphs of Boltzmann's $H$ and Gorban's $H_\Gamma$ versus time.}
\label{FigHClH}
\end{figure}

The graphs of $H$ and $H_\Gamma$ are presented in Figure~\ref{FigHClH}. It can be seen that there is a difference between $H$ and $H_\Gamma$ at the initial stage of reaction (during the first $10^{-19}$ seconds from the approximately $10^{-9}$ seconds of full process). There is one switch of $H_\Gamma$: from $H_{\gamma_4}$ to $H_{\gamma_2}$ approximately at the time moment $3.6\times 10^{-21}$ seconds.

\section{Conclusion and Outlook}

For each reaction mechanism, there exists an infinite family of Gorban's conditionally universal Lyapunov functions $H_{\Gamma}$ (\ref{GorbanFunction}) indexed by a finite set of $n$-dimensional vectors $\Gamma$, which should include all the stoichiometric vectors of the elementary reactions but may also include arbitrary vectors with at least one positive and at least one negative element. In all the cases, the level sets for $H_{\Gamma}$ were found significantly different from the level sets of the classical thermodynamic Lyapunov function $H(N)$ (\ref{LyapFreeEN}) (see Figures~\ref{FigLinHBHG}, \ref{FigNLinHBHG}, \ref{FigWGSAbLev} and \ref{FigHClAbLevel}).

The comparison of time dependences of $H_{\Gamma}$ and $H$ along kinetic trajectories gave more tricky results (Figures ~\ref{FigLin}, \ref{FigNLin}, \ref{FigWGSAb}, \ref{FigWGS}, \ref{FigHClAbH}, and \ref{FigHClH}). Of course, both functions decreased in time. Their values and the rates of descent were different if all elementary reactions were far from their partial equilibria, but if at least one reaction with the stoichiometric vector $\gamma$ approached closely its partial equilibrium then $H_{\Gamma}(c)\approx H_{\gamma}(c)\approx H(c)$ and the difference vanished. Nevertheless, if the kinetic trajectory leaved the small vicinity of the partial equilibrium, then the dynamics of $H_{\Gamma}(c)$ and $H(c)$ became  again different.

The new family of the conditionally universal Lyapunov functions gives the answer to an intriguing question about existence of such functions for non-linear reaction mechanisms (for linear reactions, the answer was done by R\'{e}nyi \cite{Renyi1961} and elaborated further by several authors \cite{Csiszar1963,Morimoto1963,ENTR3,Amari2009}). In addition to this theoretical value, we can expect some new fields of applications for these functions.

There may be many applications of the new conditionally universal Lyapunov functions. We can compare this situation to applications of many different divergences in the applied statistical inference problem \cite{Pardo2018,Judge2011}. Moreover, it is possible to use families of different entropies together and find Maximal Entropy (MaxEnt) sets of distributions instead of single distributions. This is the so-called Maximum of All Entropies (MaxAllEnt) approach that takes into account uncertainty in selection of the measure of uncertainty in the inference problem \cite{GorbanMaxAllEnt2013}.

Another application is the evaluation of the attainability regions. Each Lyapunov function can serve as a tool for evaluation (from above) the region attainable for kinetic curves  because the value of this function should decrease in time \cite{Horn1964,Feinberg1997,GorbanSIADS2013}.

There exists an obvious necessary condition of attainability of a state $y$ from the state $x$, $H(x)\geq H(y)$, but it is not sufficient for attainability by a continuous path, along which $H$ decreases monotonically. For example, a 1D system (with $n$ components and $n-1$ conservation laws) cannot come from a state $x$ to a state $y$ if they are on the opposite sides of the equilibrium even if $H(x)> H(y)$. Detailed analysis of attainability in several dimensions led to a beautiful chapter of computational convex combinatorial geometry (for more detailed review we refer to \cite{GorbanSIADS2013}). These results and their generalisations are proved to be useful in optimisation of chemical reactors and related problems~\cite{Glasser1987,Hildebrandt1990,GorbKagan2006}.

There remain also some problems. It was mentioned that all the $H_{\Gamma}$ should have an equivalent $f$-divergence form (\ref{F-div}) (possibly, after a monotonic transformation) and this form is still unknown \cite{Gorban2019}. From the application perspectives, the following question seems to be even more important: are there other families of the conditionally universal Lyapunov functions for non-linear reaction mechanisms? For linear mechanisms, such a question is fully resolved: any conditionally universal Lyapunov function for linear kinetics has the form of  $f$-divergence (or can be produced from an $f$-divergence by a  monotonic transformation) \cite{ENTR3,Amari2009,Gorban2Judge2010}.


\funding{This research was supported in part by the Ministry of Science and Higher Education of the Russian Federation, project number 14.Y26.31.0022.}

\conflictsofinterest{The author declares no conflict of interest.}

\abbreviations{The following abbreviations are used in this manuscript:\\

\noindent
\begin{tabular}{@{}ll}
GMAL & Generalised Mass Action Law 	\\
WGS & Water Gas Shift reaction\\
$\mathrm{HCl}$ & Reaction of hydrogen chloride production
\end{tabular}}

\reftitle{References}



\end{document}